\documentclass[
 reprint,
 amsmath,amssymb,
aps,prd,
floatfix,longbibliography,
]{revtex4-2}

\usepackage{threeparttable,multirow,booktabs,subfigure,color,xcolor,float}
\usepackage{graphicx}
\usepackage{dcolumn}
\usepackage{bm}

\begin{document}

\title{Non-collinear density functional theory}
\author{Zhichen Pu}
\affiliation{College of Chemistry and Molecular Engineering, Peking University,	Beijing 100871, the People's Republic of China}
\author{Hao Li}
\affiliation{College of Chemistry and Molecular Engineering, Peking University,	Beijing 100871, the People's Republic of China}
\author{Qiming Sun}
\affiliation{Axiomquant Investment Management LLC, Rong Ke Zi Xun building C 1211, Beijing 100086, the People's Republic of China}
\author{Ning Zhang}
\affiliation{College of Chemistry and Molecular Engineering, Peking University,	Beijing 100871, the People's Republic of China}
\author{Yong Zhang}
\affiliation{Qingdao Institute for Theoretical and Computational Sciences, Shandong University, Qingdao, Shandong 266237, the People's Republic of China}
\author{Sihong Shao}
\affiliation{CAPT, LMAM and School of Mathematical Sciences, Peking University, Beijing 100871, the People's Republic of China}
\author{Hong Jiang}
\affiliation{College of Chemistry and Molecular Engineering, Peking University,	Beijing 100871, the People's Republic of China}
\author{Yiqin Gao}
\affiliation{College of Chemistry and Molecular Engineering, Peking University,	Beijing 100871, the People's Republic of China}
\author{Yunlong Xiao}
\email{xiaoyl@pku.edu.cn}
\affiliation{College of Chemistry and Molecular Engineering, Peking University, Beijing 100871, the People's Republic of China}

\date{\today}

\begin{abstract}
    An approach to generalize any kind of collinear functionals in density functional theory to non-collinear functionals is proposed. This approach, for the very first time, satisfies the correct collinear limit for any kind of functionals, guaranteeing that the exact collinear functional after generalized is still exact for collinear spins. Besides, it has well-defined and numerically stable functional derivatives, a desired feature for non-collinear and spin-flip time-dependent density functional theory.
Furthermore, it provides local torque, hinting at its applications in spin dynamics.
\end{abstract}

\maketitle

\section{Background \label{background}}
Density functional theory (DFT) \cite{kohn1965self} has been widely and successfully applied in calculating electronic structures in molecules and materials. Spin-DFT \cite{von1972local} was developed to treat spin-polarized systems, generally in terms of the density $n(\boldsymbol{r})$ and the spin magnetization vector $ \boldsymbol{m}(\boldsymbol{r}) $, indicating the exchange-correlation energy depending on $n$ and $ \boldsymbol{m} $, \textit{i.e.}, $ E^{\mathrm{xc}}=E^{\mathrm{xc}}[n,\boldsymbol{m}] $.
However, widely used functionals, known as collinear functionals $E^{\mathrm{col}}$, do not depend on $n$ and $\boldsymbol{m}$ but on the spin-up density $n_{\uparrow}=\frac{1}{2}(n+m_z)$ and spin-down density $n_{\downarrow} = \frac{1}{2} (n-m_{z})$, or equivalently on $n$ and $m_{z}$. Collinear functionals can only handle a special spin configuration, known as collinear spin with $\boldsymbol{m} = (0,0,m_{z})$. The generalization of collinear functionals to non-collinear functionals, which can handle any spin configuration, is important in both theory and applications.

The first generalization is credited to K{\"u}bler \textit{et al.} in 1988 \cite{kubler1988density}, who suggested (with $m$ the norm of $\boldsymbol{m}$)
\begin{eqnarray}
	E^{\mathrm{xc}}[n,\boldsymbol{m}]
	= E^{\mathrm{col}}[n,\boldsymbol{m}\cdot\frac{\boldsymbol{m}}{m}]
    = E^{\mathrm{col}}[n,m].\label{lc} 
\end{eqnarray}
Because $\boldsymbol{m}$ is projected into its local direction $\frac{\boldsymbol{m}}{m}$, equation (\ref{lc}) is referred to as the \emph{locally collinear approach}.

In this work, four criteria for generalizations from collinear functionals to non-collinear functionals are addressed. It is interesting to investigate if they can be satisfied by the locally collinear approach, which has been widely used nowadays \cite{nordstrom1996noncollinear,oda1998fully,van2002spin,wang2004time,gao2005time,bast2009relativistic,bulik2013noncollinear,egidi2017two,liu2018relativistic,li2020real}.
The four criteria are as follows:
\begin{enumerate}
    \item[(\uppercase\expandafter{\romannumeral1})] \textit{Correct collinear limit} 
    \begin{eqnarray}
    	E^{\mathrm{xc}}[n,(0,0,m_z)]&=& E^{\mathrm{col}}[n,m_z].\label{equ:Exccollinear}
    \end{eqnarray}
    From the view of math, it respects that any extension of a functional domain should not change its values on the original domain. 
    The correct collinear limit is an important physical condition.
    Supposing we have known the exact collinear functional, an approach without the correct collinear limit will generalize it to a \emph{incorrect} functional for collinear spin states. This is bad, because the collinear spin is the most important spin configuration, as limits of non-collinear spins in the absence of spin-orbit couplings and magnetic field.
    Desmarais and co-workers pointed out in Ref. \citenum{desmarais2021spin} that the locally collinear approach Eq. (\ref{lc}) has the correct collinear limit for LSDA (local spin density approximation) \cite{vosko1980accurate} and GGA (generalized gradient approximation)\cite{perdew1992atoms,becke1988density} functionals. 
    However, for non-local functionals\cite{gritsenko1993weighted} of $n$ and $\boldsymbol{m}$ (having nothing to do with the exact Hartree-Fock exchange, which is a non-local functional of orbitals), the locally collinear approach does not satisfy the correct collinear limit condition Eq. (\ref{equ:Exccollinear}). 
    Indeed, consider the case where spins at two spatial positions point in the same direction. 
    Flipping the spin at one position results in different energies evaluated by collinear functionals while the same by the locally collinear functionals Eq. (\ref{lc}).
    
    \item[(\uppercase\expandafter{\romannumeral2})] \textit{Being invariant to the global rotation while sensitive to the local rotation of spin magnetization vector}. The former reveals that the functional should not depend on the choice of spin axes, while the latter enables the functional to distinguish essentially different states, which are connected by local spin rotations.
    The locally collinear approach is invariant to global rotation while not sensitive to local rotation.
    \item[(\uppercase\expandafter{\romannumeral3})] \textit{Well-defined functional derivatives}. In applications, functional derivatives are needed, such as the potential (the first-order derivative) in self-consistent field calculations, and the kernel (the second-order derivative) in LR-TDDFT \cite{casida1995time,TDDFT_1984} (linear response time-dependent density functional theory).
    In the locally collinear approach, the direction of $ \boldsymbol{m} $ at $ \boldsymbol{m}(\boldsymbol{r}) = \boldsymbol{0} $ is ill-defined, causing numerical singularities in functional derivatives. 
    Discussions on this issue can be found in Refs. \citenum{desmarais2021spin,bast2009relativistic,egidi2017two,komorovsky2019four,li2012theoretical,liu2018relativistic}, yet a general cure within the locally collinear approach is unknown and most likely does not exist.
    \item[(\uppercase\expandafter{\romannumeral4})] \textit{Providing global zero torque but non-vanishing local torque}. The global zero torque reflects the fact that the self-consistent exchange-correlation magnetic field ($ \boldsymbol{B}^{\mathrm{xc}} $) should not exert a net torque on the whole system, known as the zero torque theorem \cite{capelle2001spin}. Meanwhile, local torque ($\boldsymbol{m} \times \boldsymbol{B}^{\mathrm{xc}}$), which reflects the internal interactions and plays a crucial role in spin dynamics \cite{sharma2007first}, should not vanish.
    The locally collinear approach satisfies the zero torque theorem \cite{capelle2001spin}, but cannot provide local torque because the calculated $ \boldsymbol{B}^{\mathrm{xc}} $ is always parallel to $ \boldsymbol{m} $\cite{sharma2007first}. To obtain a non-vanishing local torque, considerable efforts have been made \cite{scalmani2012new,sharma2007first,eich2013transverse,eich2013transverse2}, such as the modified version of the locally collinear approach by Scalmani and Frisch \cite{scalmani2012new}, and exact exchange combined with optimized effective potential \cite{talman1976optimized} by Sharma \textit{et al.} \cite{sharma2007first}.
\end{enumerate}

 Considering that the locally collinear approach does not fully satisfy any criterion listed above, a new approach is proposed that fully satisfies them all. 

\section{Theory}\label{sec:theory}

\subsection{The establishment of the theory}
Considering that collinear functionals can only handle scalar functions, the vector function $ \boldsymbol{m} $ needs to be firstly projected to a given direction $\Omega$ (such as $ z $-direction in Eq. (\ref{equ:Exccollinear})),
\begin{eqnarray}
m_{\Omega}&=&\boldsymbol{m}\cdot\boldsymbol{e}_{\Omega},
\end{eqnarray}
with $\boldsymbol{e}_{\Omega}$ the unit vector representing the solid angle $\Omega$ in spin space,
\begin{eqnarray}
\boldsymbol{e}_{\Omega}=\left(e_{\Omega x},e_{\Omega y},e_{\Omega z}\right) 
=\left(\sin\theta\cos\phi,\sin\theta\sin\phi,\cos\theta\right).\label{equ:u_ome}
\end{eqnarray}
In the locally collinear approach, $ \boldsymbol{e} $ is chosen as a $ \boldsymbol{r} $-dependent function, $ \boldsymbol{e}(\boldsymbol{r}) = \frac{\boldsymbol{m}(\boldsymbol{r})}{m(\boldsymbol{r})} $, which is ill-defined at $\boldsymbol{m} = \boldsymbol{0}$. To avoid this ill-definition issue, we suggest choosing $ \boldsymbol{e}_\Omega $ as a constant unit vector ($\Omega$ as its parameter), which does not depend on $ \boldsymbol{r} $ or any other physical quantities. 
A single direction for projection will certainly break spin rotation invariance, but an average over all directions will not.

Inspired by this, we propose non-collinear functionals in the form of 
\begin{eqnarray}
	E^{\mathrm{xc}}[n, \boldsymbol{m}] =  \overline{E^{\mathrm{eff}}[n,m_\Omega]},\label{equ:un}
\end{eqnarray}
with the overline denoting the average over all directions $ \Omega $. The key of our approach Eq. (\ref{equ:un}) is introducing the effective collinear functional $E^{\mathrm{eff}}$, but not naively defining $E^{\mathrm{xc}}[n, \boldsymbol{m}] =  \overline{E^{\mathrm{col}}[n,m_\Omega]}$, which apparently ruins the correct collinear limit. Instead, the correct collinear limit condition Eq. (\ref{equ:Exccollinear}) is taken as a \emph{prerequisite} and is used to determine $E^{\mathrm{eff}}$ in Eq. (\ref{equ:un}) as shown below.

Considering a collinear spin state $\boldsymbol{m}=(0,0,m_z)$, equation (\ref{equ:un}) becomes
\begin{eqnarray}
E^{\mathrm{xc}}[n,(0,0,m_z)]=\overline{E^{\mathrm{eff}}[n,m_z\cos\theta]},
\label{equ:coll_tmp}
\end{eqnarray}
which, compared with Eq. (\ref{equ:Exccollinear}), leads to
\begin{eqnarray}
\overline{E^{\mathrm{eff}}[n,m_z\cos\theta]}=E^{\mathrm{col}}[n,m_z].
\label{equ:connect}
\end{eqnarray}
First, we replace $ m_z(\boldsymbol{r}) $ with the short notation $ s(\boldsymbol{r}) $ for spin density, \textit{i.e.}, 
\begin{eqnarray}
 \overline{E^{\mathrm{eff}}[n,s\cos\theta]}=E^{\mathrm{col}}[n,s].\label{equ:tempa8} 
\end{eqnarray}
The left-hand side of Eq. (\ref{equ:tempa8}), involving the average over solid angles, is evaluated as
\begin{eqnarray}
 \overline{E^{\mathrm{eff}}[n,s\cos\theta]} &=&\frac{
 \int_{\theta=0}^{\pi}\int_{\phi=0}^{2\pi} E^{\mathrm{eff}}[n,s\cos\theta] \sin\theta \mathrm{d}\phi \mathrm{d}\theta }
 {\int_{\theta=0}^{\pi} \int_{\phi=0}^{2\pi} \sin\theta  \mathrm{d}\phi \mathrm{d}\theta} \label{equ:tmra6}\\ 
&=& \frac{1}{2}\int_{-1}^{1} E^{\mathrm{eff}}[n,st] \mathrm{d} t \quad (t=\cos \theta) \label{tmr1}.
\end{eqnarray}
$E^{\mathrm{eff}}[n,s]$ can always be expressed as a summation of an odd functional and an even functional with respect to $s$, but the former does not contribute to the integral in Eq. (\ref{tmr1}). Without loss of generality, $E^{\mathrm{eff}}$ is assumed to be an even functional, leading to
\begin{eqnarray}
    \overline{E^{\mathrm{eff}}[n,s\cos\theta]}  = \int_{0}^{1} E^{\mathrm{eff}}[n,st] \mathrm{d} t.\label{tmr2}
\end{eqnarray}
By comparing Eq. (\ref{equ:tempa8}) and (\ref{tmr2}), one immediately obtains
\begin{eqnarray}
 \int_0^{1} E^{\mathrm{eff}}[n,st] \mathrm{d} t=  E^{\mathrm{col}}[n,s].\label{int-t}
\end{eqnarray}

To solve Eq. (\ref{int-t}), we regard scalar function $n$ and $s$ as ``parameters", and introduce univariate functions $F^{\mathrm{col}}_{[n,s]}(t)$ and $F^{\mathrm{eff}}_{[n,s]}(t)$, as
\begin{eqnarray}
F^{\mathrm{col/eff}}_{[n,s]}(t)&=&  E^{\mathrm{col/eff}}[n,ts].
\label{equ:tmrunif1}\end{eqnarray}
Equation (\ref{int-t}) is rewritten as, using the notations in Eq. (\ref{equ:tmrunif1}),
\begin{eqnarray}
 \int_0^{1} F^{\mathrm{eff}}_{[n,s]}(t) \mathrm{d} t=  F^{\mathrm{col}}_{[n,s]}(1) \label{equ:tmrunif2}.
\end{eqnarray}
Equation (\ref{equ:tmrunif2}) validates for any function $s$, therefore, also for $\lambda s$  ($\lambda > 0$), \textit{i.e.},
\begin{eqnarray}
 \int_0^{1} F^{\mathrm{eff}}_{[ n,\lambda s]}(t) \mathrm{d} t=  F^{\mathrm{col}}_{[n,\lambda s]}(1).
\end{eqnarray}
Further replacing $t$ by $ \frac{t}{\lambda}$, 
\begin{eqnarray}
 \int_{\frac{t}{\lambda} =0}^{\frac{t}{\lambda}=1} F^{\mathrm{eff}}_{[n,\lambda s]}\left(\frac{t}{\lambda}\right)  \mathrm{d} \left( \frac{t}{\lambda}\right)=  F^{\mathrm{col}}_{[n,\lambda s]}(1),
\end{eqnarray}
leads to
\begin{eqnarray}
\int_{0}^{\lambda} F^{\mathrm{eff}}_{[n,s]}( t) \mathrm{d} t= \lambda F^{\mathrm{col}}_{[n,s]}(\lambda),
\end{eqnarray}
whose derivative with respect to $\lambda$ is
\begin{eqnarray}
  F^{\mathrm{eff}}_{[n,s]}( \lambda)   =  F^{\mathrm{col}}_{[n,s]}(\lambda)+\lambda
  \frac{\mathrm{d} F^{\mathrm{col}}_{[n,s]}(\lambda)}{\mathrm{d} \lambda}.
  \label{equ:function_eff}
\end{eqnarray}
At $\lambda=1$, in virtue of
\begin{eqnarray}
 \left.\frac{\mathrm{d} F^{\mathrm{col}}_{[n,s]}(\lambda)}{\mathrm{d} \lambda}\right|_{\lambda=1} &=& \lim_{\Delta\rightarrow 0} \frac{F^{\mathrm{col}}_{[n,s]}(1+\Delta)-F^{\mathrm{col}}_{[n,s]}(1)}{\Delta} \nonumber \\
  &=& \left. \frac{\mathrm{d}}{\mathrm{d} x} E^{\mathrm{col}} [n,xs] \right|_{x=1}\label{equ:tmrunif3},
\end{eqnarray}
equation (\ref{equ:function_eff}) leads to 
\begin{eqnarray}
  E^{\mathrm{eff}}[n,s]  &=&  E^{\mathrm{col}}[n,s]
  +\left. \frac{\mathrm{d}}{\mathrm{d} x} E^{\mathrm{col}} [n,xs] \right|_{x=1},
  \label{equ:un-col_f}
\end{eqnarray}
which can be further written as
\begin{eqnarray}
	E^{\mathrm{eff}}[n,s]=E^{\mathrm{col}}[n,s]
	+ \int \frac{\delta E^{\mathrm{col}}[n,s]}{\delta s(\boldsymbol{r})} s(\boldsymbol{r})\mathrm{d} \boldsymbol{r}.\label{equ:UN-col}
\end{eqnarray}
Considering the fact that simply taking $E^{\mathrm{eff}} = E^{\mathrm{col}}$ ruins the correct collinear limit, the second term of Eq. (\ref{equ:un-col_f}) or (\ref{equ:UN-col}) is crucial to restoring the correct collinear limit.
Its physical meaning is clear (seeing from Eq. (\ref{equ:un-col_f})), describing the response of the collinear functional to the squeezing factor $x$ of the spin density $s$.

\begin{figure}
	\centering
	\includegraphics[width=8.5cm]{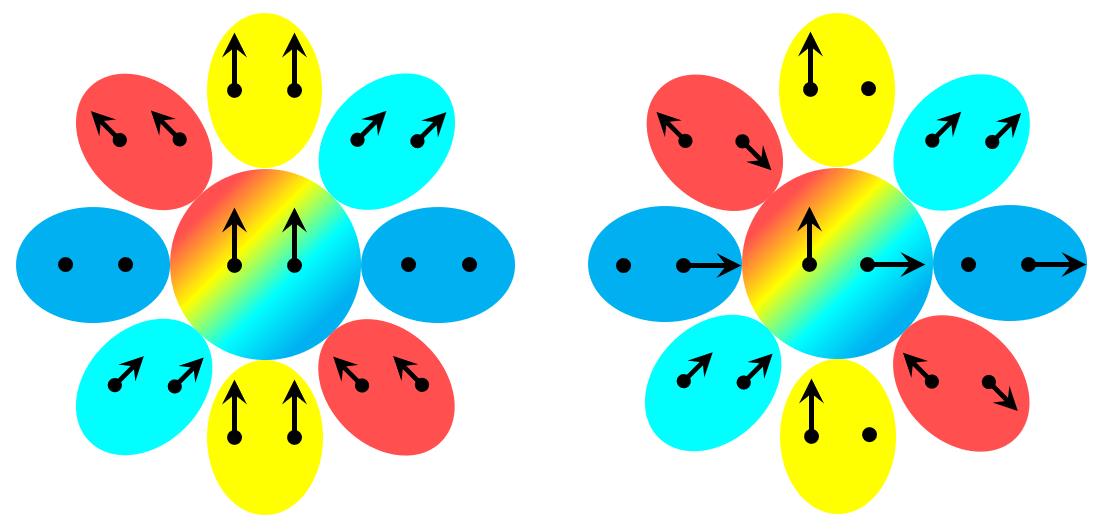}
	\caption{Schematic representation of the multi-collinear approach. 
    Consider a toy system containing only two spatial grids represented by two dots, with magnetization vectors represented by arrows.
    Two spin configurations of this toy system are shown, a collinear spin state (the left colorful circle), and a non-collinear spin state (the right colorful circle).
    They are projected in all directions (eight in this example) to obtain projected collinear states (represented by ellipses).  
	}
	\label{fig1}
\end{figure}

Substituting Eq. (\ref{equ:UN-col}) into Eq. (\ref{equ:un}) leads to the final expression of the exchange-correlation functional,
\begin{eqnarray}
	E^{\mathrm{xc}}[n,\boldsymbol{m}] =  \overline{E^{\mathrm{col}}[n,m_\Omega] }
 +\int\overline{ \frac{\delta E^{\mathrm{col}}[n,m_\Omega]}{\delta m_\Omega(\boldsymbol{r})} m_\Omega(\boldsymbol{r})}\mathrm{d} \boldsymbol{r},\nonumber\\
    \label{equ:MC2}
\end{eqnarray}
\textit{i.e.}, energies plus their responses of globally projected collinear states, followed by an average.
Since multiple projected collinear states are used to represent the original non-collinear state, we call Eq. (\ref{equ:MC2}) the multi-collinear (MC) approach. 
Because no specific form of collinear functional is assumed in our deduction, the multi-collinear functionals satisfy the correct collinear limit for all kinds of collinear functionals, such as LSDA, GGA, meta-GGA\cite{becke1988correlation,tschinke1989shape,neumann1997higher}, hybrid\cite{becke1993new} and non-local functionals.

An illustrative example is shown in FIG. \ref{fig1}, in which two target states, a collinear spin state (the left colorful circle) and a non-collinear spin state (the right colorful circle), are exhibited.
Although treated by the multi-collinear approach in a uniform way, their projected collinear states are different, indicating that the multi-collinear approach can distinguish essentially different states connected by local spin rotations.
It seems that for the collinear spin state (left in FIG. \ref{fig1}), projections in all directions are superfluous, since only the collinear energy of the yellow ellipse (projected to its spin-polarized direction) matters.
Its explanation has two folds.
First, it is necessary to project the target spin state in all directions, if we 
want to treat collinear and non-collinear spin states in a uniform way. 
Secondly, all the projected states do contribute to the multi-collinear energy, but through their effective collinear energies instead of collinear energies. If we count contributions from their collinear energies, only the state projected along the spin-polarization axis (yellow ellipse) contributes. 

\subsection{Some properties of multi-collinear approach}

In this subsection, some properties of the multi-collinear approach are addressed for a deeper understanding of this approach.

\subsubsection{For closed-shell systems}

Let us first consider the simplest case, a closed-shell system with $ \boldsymbol{m}(\boldsymbol{r}) = \boldsymbol{0} $. In this case, the first-order derivative $ \frac{\delta E^{\mathrm{col}}[n,m_\Omega]}{\delta m_\Omega(\boldsymbol{r})} $ in Eq. (\ref{equ:MC2}) vanishes, in virtue of $ E^{\mathrm{col}} $ being an even functional, and the multi-collinear functional Eq. (\ref{equ:MC2}) becomes
\begin{eqnarray}
    E^{\mathrm{MC}}[n,\boldsymbol{0}]=E^{\mathrm{col}}[n,0] \label{equ:AB2}.
\end{eqnarray}
Equation (\ref{equ:AB2}) indicates that for a spin-unpolarized system, the exchange-correlation energy evaluated by the multi-collinear functional goes back to the energy by the collinear functional. This is not surprising, because the closed-shell system is just a special case of collinear spins and the multi-collinear approach has the correct collinear limit. 

\subsubsection{For spin-independent functionals}

Another easily obtained property is that for collinear functionals depending only on density $n$, the multi-collinear functional is just the original collinear functional 
\begin{eqnarray}
    E^{\mathrm{col}}[n,s]=E^{\mathrm{col}}[n] 
    \Rightarrow
    E^{\mathrm{MC}}[n,\boldsymbol{m}]=E^{\mathrm{col}}[n] \label{equ:property_2_new}.
\end{eqnarray}

\subsubsection{Linearity}

The third property of multi-collinear functionals is their linear dependence on collinear functionals. That is to say, if a collinear functional $E^{\mathrm{col}}_\mathrm{C}$ is a linear combination of two collinear functionals $E^{\mathrm{col}}_\mathrm{A}$ and $E^{\mathrm{col}}_\mathrm{B}$, their multi-collinear counterparts retain the same relation, \textit{i.e.}, 
\begin{eqnarray}
    E^{\mathrm{col}}_\mathrm{C}[n,s]&=& aE^{\mathrm{col}}_\mathrm{A}[n,s]+bE^{\mathrm{col}}_\mathrm{B}[n,s] \nonumber \\ \Rightarrow E^{\mathrm{MC}}_\mathrm{C}[n,\boldsymbol{m}]&=&aE^{\mathrm{MC}}_\mathrm{A}[n,\boldsymbol{m}]+bE^{\mathrm{MC}}_\mathrm{B}[n,\boldsymbol{m}].\label{equ:AB4}
\end{eqnarray}

A straightforward result by combining Eq. (\ref{equ:property_2_new}) and Eq. (\ref{equ:AB4}) is   
\begin{eqnarray}
    E^{\mathrm{col}}_{\mathrm{B}}[n,s]&=& E^{\mathrm{col}}_{\mathrm{A}}[n,s]+C \nonumber\\
    \Rightarrow
    E^{\mathrm{MC}}_{\mathrm{B}}[n,\boldsymbol{m}]&=& E^{\mathrm{MC}}_{\mathrm{A}}[n,\boldsymbol{m}]+C, \label{shift}
\end{eqnarray}   
with $C$ a constant.
Equation (\ref{shift}) exhibits a good property of the multi-collinear approach that a shift in the collinear energy by a constant leads to the same shift in the multi-collinear energy. 
Equation (\ref{shift}) is satisfied because the arithmetic average is adopted in Eq. (\ref{equ:un}), which leads to the linearity property Eq. (\ref{equ:AB4}) of the multi-collinear approach. Imagining other kinds of averages were adopted in the multi-collinear approach, such as
 \begin{eqnarray}
 E^{\mathrm{MC}}[n,\boldsymbol{m}]
 =\pm \sqrt{\overline{|E^{\mathrm{eff}}[n,m_{\Omega}]|^2}},
\end{eqnarray}
the good property Eq. (\ref{shift}) would be ruined. 

\subsection{Multi-collinear approach for LSDA functionals} \label{sec:LSDA}

The multi-collinear functional Eq. (\ref{equ:MC2}) and locally collinear functional Eq. (\ref{lc}) look quite different, but for LSDA functionals, they are the same. 

To see this, consider a collinear LSDA functional
\begin{eqnarray}
    E^{\mathrm{col}}[n,s] = \int f^{\mathrm{col}}(n,s)  \mathrm{d} \boldsymbol{r},\label{equ:AC1}
\end{eqnarray}
whose corresponding effective collinear functional $ E^{\mathrm{eff}}$ is calculated according to Eq. (\ref{equ:UN-col}), with integrand
\begin{eqnarray}
f^{\mathrm{eff}}(n,s)&=&f^{\mathrm{col}}(n,s)+s
\frac{\partial f^{\mathrm{col}}(n,s)}{\partial s}.\label{equ:AC2}
\end{eqnarray}
The integrand of the multi-collinear functional, according to Eq. (\ref{equ:un}), reads
\begin{eqnarray}
f^{\mathrm{MC}}(n,\boldsymbol{m})=\overline{f^{\mathrm{eff}}(n,m_{\Omega})},  
\label{equ:AC4}
\end{eqnarray}
which hints that $f^{\mathrm{MC}}(n,\boldsymbol{m})$ is independent of the direction of $\boldsymbol{m}$, without loss of generality, assuming $\boldsymbol{m}$ points in the positive direction of the $z$-axis,
\begin{eqnarray}
    f^{\mathrm{MC}}(n,\boldsymbol{m})
    =f^{\mathrm{MC}}(n,(0,0,m)) \label{equ:AC5}.
\end{eqnarray}
By noting that the multi-collinear functionals have the correct collinear limit Eq. (\ref{equ:Exccollinear}), 
\begin{eqnarray}
    f^{\mathrm{MC}}(n,(0,0,m)) = f^{\mathrm{col}}(n,m) \label{equ:AC_limit_re},
\end{eqnarray}
one immediately obtains
\begin{eqnarray}
    f^{\mathrm{MC}}(n,\boldsymbol{m}) = f^{\mathrm{col}}(n,m),\label{equ:AC_limit}
\end{eqnarray}
which validates the equivalence between the multi-collinear approach and the locally collinear approach for LSDA functionals.

In the following, another proof of Eq. (\ref{equ:AC_limit}) is given without using the property that the multi-collinear approach has the correct collinear limit.
We calculate the average over solid angles in Eq. (\ref{equ:AC4}) directly, with the simplification given by Eq. (\ref{equ:AC5}),
\begin{eqnarray}
f^{\mathrm{MC}}(n,\boldsymbol{m})    
&=&\frac{
 \int_{\theta=0}^{\pi}\int_{\phi=0}^{2\pi} f^{\mathrm{eff}}(n,m\cos\theta)  \sin\theta \mathrm{d}\phi \mathrm{d}\theta }
 {\int_{\theta=0}^{\pi} \int_{\phi=0}^{2\pi} \sin\theta  \mathrm{d}\phi \mathrm{d}\theta}   \nonumber  
 \\ 
&=&\int_0^1  f^{\mathrm{eff}}(n,mt) \mathrm{d} t. \label{equ:AC6}
\end{eqnarray}
If $m=0$, Eq. (\ref{equ:AC_limit}) holds apparently.
Otherwise $m > 0$, equation (\ref{equ:AC6}) becomes
\begin{eqnarray}
f^{\mathrm{MC}}(n,\boldsymbol{m})
&=&\frac{1}{m} \int_0^m   \left[f^{\mathrm{col}}(n,s)+s\frac{\partial f^{\mathrm{col}}(n,s)}{\partial s}\right] \mathrm{d}s \nonumber    \\
&=&\frac{1}{m} \left[\int_0^m f^{\mathrm{col}}(n,s)\mathrm{d}s+\int_{s=0}^{s=m} s \mathrm{d}f^{\mathrm{col}}(n,s)\right], \nonumber \\ \label{equ:ibp}
\end{eqnarray}
where the first term is the contribution of collinear energy (from the first term in Eq. (\ref{equ:MC2})), and the second term is the contribution of functional derivative (from the second term in Eq. (\ref{equ:MC2})). With the help of integration by parts, equation (\ref{equ:ibp}) turns to
\begin{eqnarray}
f^{\mathrm{MC}}(n,\boldsymbol{m})
&=&\frac{1}{m} \int_0^m f^{\mathrm{col}}(n,s)\mathrm{d}s
+ \frac{1}{m} \left.\left[s f^{\mathrm{col}}(n,s)\right]\right|_{s=0}^{s=m}   \nonumber   \\
&&- \frac{1}{m} \int_{0}^{m} f^{\mathrm{col}}(n,s) \mathrm{d} s,\label{equ:ibp2}
\end{eqnarray}
directly leading to Eq. (\ref{equ:AC_limit}).
In Eq. (\ref{equ:ibp2}), $s$ represents the squeezed spin of projected collinear states (the norm of spin is squeezed from $m$ to $s$). The cancelation between the first and third term in Eq. (\ref{equ:ibp2}) suggests no contributions from those squeezed states if collinear energies are counted, thanks to the integration by parts.
In the end, only the second term of Eq. (\ref{equ:ibp2}), the boundary condition term, survives.

The equivalence between the multi-collinear approach and the locally collinear approach for LSDA functionals is expected, because LSDA functionals, seeing only local spin density, are not sensitive to local spin rotations.

\subsection{Multi-collinear approach for toy GGA, meta-GGA and non-local functionals} \label{sec:toys}
Although for LSDA functionals, multi-collinear and the locally collinear approach are equivalent,
for functionals beyond LSDA, they are essentially different. To see this, we start with the investigation of several toy functionals.

In TABLE \ref{table1}, five toy non-collinear functionals are displayed, which are determined intuitively, including GGA, meta-GGA and non-local functionals, all with simple forms. First, restricting $\boldsymbol{m}$ to collinear case $\boldsymbol{m} = (0,0,m_z)$, their corresponding toy collinear functionals are obtained. 
We pretend that we do not know the original non-collinear functionals, but only the collinear functionals, which are further generalized to non-collinear functionals in the multi-collinear or locally collinear approach, and finally compared with the original non-collinear functionals.
As shown in TABLE \ref{table1}, the first four toy functionals are successfully reproduced by the multi-collinear approach, while the locally collinear approach reproduces none of them and fails to satisfy the correct collinear limit for functional No. 4. 

The deduction of TABLE \ref{table1} is displayed below. 

\begin{table*}
  \centering
  \caption{Multi-collinear approach for toy functionals.$^\mathrm{(a)}$}
  \begin{threeparttable}
    \begin{tabular}{lcccc}
    \toprule
    No. of functionals  & $f^{\mathrm{NC}}$                                                                                     & $f^{\mathrm{col}}$                                        & $ f^{\mathrm{MC}}$                                                                                    & $f^{\mathrm{LC}}$                                         \\
    \midrule
    1 (GGA)             & $ \boldsymbol{\nabla} \boldsymbol{m} \cdot \circ  \boldsymbol{\nabla} \boldsymbol{m}^\mathrm{(b)}$    & $\boldsymbol{\nabla} s \cdot \boldsymbol{\nabla} s$       & $ \boldsymbol{\nabla} \boldsymbol{m} \cdot \circ   \boldsymbol{\nabla} \boldsymbol{m} ^\mathrm{(b)}$  &$\boldsymbol{\nabla} m \cdot \boldsymbol{\nabla} m$        \\
    2 (GGA)             & $ \boldsymbol{m} \cdot (\boldsymbol{\nabla} n \cdot \boldsymbol{\nabla} \boldsymbol{m}) $             & $s  (\boldsymbol{\nabla} n \cdot \boldsymbol{\nabla} s)$  & $ \boldsymbol{m} \cdot (\boldsymbol{\nabla} n \cdot \boldsymbol{\nabla} \boldsymbol{m}) $             &$m  (\boldsymbol{\nabla} n \cdot \boldsymbol{\nabla} m)$   \\
    3 (meta-GGA)        & $\boldsymbol{m} \cdot  \nabla^2 \boldsymbol{m}$                                                       & $s \nabla^2s$                                             & $\boldsymbol{m} \cdot \nabla^2 \boldsymbol{m}$                                                        & $m \nabla^2m$ $^{\mathrm{(c)}}$                           \\
    4 (non-local)       & $\boldsymbol{m_1} \cdot  \boldsymbol{m_2} $                                                           & $ s_1 s_2$                                                & $\boldsymbol{m_1} \cdot            \boldsymbol{m_2}$                                                  & $m_1 m_2$                                                 \\
    5 (non-local)       & $(\boldsymbol{m_1} \cdot  \boldsymbol{m_2})^2 $                                                       & $ s_1^2 s_2^2$                                            & $[2(\boldsymbol{m_1} \cdot  \boldsymbol{m_2})^2 + m_1^2 m_2^2]/3$                                     & $m_1^2 m_2^2$                                             \\
    \bottomrule
    \end{tabular}
    	\begin{tablenotes}
    	    \item[(a)] Five toy non-collinear functionals ($f^{\mathrm{NC}}$ as integrands) are given intuitively, which become collinear functionals ($f^{\mathrm{col}}$ as integrands) when applied to collinear spins. Assuming only collinear functionals are known, their generalizations to non-collinear functionals are obtained in the multi-collinear approach ($f^{\mathrm{MC}}$ as integrands) and locally collinear approach ($f^{\mathrm{LC}}$ as integrands).
			\item[(b)] $\boldsymbol{\nabla} \boldsymbol{m} \cdot \circ  \boldsymbol{\nabla} \boldsymbol{m} = \sum\limits_{\alpha,\beta = x,y,z}  \nabla_\alpha m_\beta \nabla_\alpha m_\beta $.
            \item[(c)] Another interpretation of the locally collinear approach for meta-GGA gives $f^{\mathrm{LC}} = \boldsymbol{m} \cdot  \nabla^2 \boldsymbol{m}$.
		\end{tablenotes}
		\end{threeparttable}
  \label{table1}
\end{table*}

\subsubsection{Functional No. 1}

Applying a toy non-collinear GGA functional,
\begin{eqnarray}
E^{\mathrm{NC}}[n, \boldsymbol{m}]=\int \sum_{\alpha=x,y,z} \boldsymbol{\nabla} m_{\alpha} \cdot \boldsymbol{\nabla} m_{\alpha} \mathrm{d}\boldsymbol{r},\label{equ:toy_gga}
\end{eqnarray}
to a collinear spin system leads to 
\begin{eqnarray}
E^{\mathrm{NC}}[n, (0,0,m_z)]=\int \boldsymbol{\nabla} m_z \cdot \boldsymbol{\nabla} m_z \mathrm{d}\boldsymbol{r},\label{equ:toy_gga_mz}
\end{eqnarray}
hinting that the corresponding collinear functional reads
\begin{eqnarray}
E^{\mathrm{col}}[n,s]=\int \boldsymbol{\nabla} s \cdot \boldsymbol{\nabla} s \mathrm{d}\boldsymbol{r}.\label{equ:toy_gga_col}
\end{eqnarray}

Now assuming that only the collinear functional Eq. (\ref{equ:toy_gga_col}) is known, we generalize it to the non-collinear functional in the multi-collinear approach and see whether the original toy non-collinear functional Eq. (\ref{equ:toy_gga}) can be reproduced. 

According to Eq. (\ref{equ:UN-col}), the effective collinear functional reads
\begin{eqnarray}
E^{\mathrm{eff}}[n,s]&=& \int \boldsymbol{\nabla} s \cdot \boldsymbol{\nabla} s \mathrm{d}\boldsymbol{r} - 2\int (\nabla^2 s)s \mathrm{d}\boldsymbol{r}  \nonumber   \\
&=& 3 \int \boldsymbol{\nabla} s \cdot \boldsymbol{\nabla} s \mathrm{d}\boldsymbol{r}.
\end{eqnarray}
The non-collinear functional in the multi-collinear approach, according to Eq. (\ref{equ:un}), reads
\begin{eqnarray}
&&E^{\mathrm{MC}}[n,\boldsymbol{m}] \nonumber\\
&=&3 \int \overline{\boldsymbol{\nabla} m_{\Omega} \cdot \boldsymbol{\nabla} m_{\Omega}} \mathrm{d}\boldsymbol{r}\nonumber \\
&=&3 \sum\limits_{\alpha,\beta=x,y,z}\overline{e_{\Omega \alpha} e_{\Omega \beta}}\int \boldsymbol{\nabla} m_{\alpha} \cdot \boldsymbol{\nabla} m_{\beta}  \mathrm{d}\boldsymbol{r}.
\label{equ:DMC}
\end{eqnarray}
Using the explicit form of $\boldsymbol{r}$-independent vector $\boldsymbol{e}_{\Omega}$ in Eq. (\ref{equ:u_ome}), the average on the right-hand side of Eq. (\ref{equ:DMC}) is easy to calculate
\begin{eqnarray}
    \overline{e_{\Omega \alpha} e_{\Omega \beta}} = \frac{1}{3}\delta_{\alpha \beta},\label{equ:uab}
\end{eqnarray}
leading to
\begin{eqnarray}
E^{\mathrm{MC}}[n,\boldsymbol{m}]    
&=& \sum_{\alpha=x,y,z} \int \boldsymbol{\nabla} m_{\alpha} \cdot \boldsymbol{\nabla} m_{\alpha} \mathrm{d}\boldsymbol{r}, \label{equ:toy_GGA_MC}
\end{eqnarray}
nothing but the original toy non-collinear functional Eq. (\ref{equ:toy_gga}).

However, the generalization of Eq. (\ref{equ:toy_gga_col}) to non-collinear functional in the locally collinear approach is
\begin{eqnarray}
E^{\mathrm{LC}}[n,\boldsymbol{m}]=\int \boldsymbol{\nabla} |\boldsymbol{m}| \cdot \boldsymbol{\nabla} |\boldsymbol{m}| \mathrm{d}\boldsymbol{r},\label{equ:toy_GGA_LC}
\end{eqnarray}
failing to reproduce the original toy non-collinear functional Eq. (\ref{equ:toy_gga}).

\subsubsection{Functional No. 2}

Consider another toy non-collinear GGA functional,
\begin{align}
	E^{\mathrm{col}}[n,\boldsymbol{m}] = \boldsymbol{m} \cdot \left(  \boldsymbol{\nabla} n \cdot \boldsymbol{\nabla} \boldsymbol{m} \right) , \label{equ:tl_diff_nc}
\end{align}
whose corresponding collinear functional reads
\begin{align}
	E^{\mathrm{col}}[n,s] = s \boldsymbol{\nabla} n \cdot \boldsymbol{\nabla} s. \label{equ:tl_diff_col}
\end{align}
In the multi-collinear approach, the effective collinear and multi-collinear functionals are ready to be obtained
\begin{eqnarray}
E^{\mathrm{eff}}[n,s] &=& 3 E^{\mathrm{col}}[n,s] ,\label{equ:tl_diff_eff}\\
E^{\mathrm{MC}}[n,\boldsymbol{m}] &=& \boldsymbol{m} \cdot \left(  \boldsymbol{\nabla} n \cdot \boldsymbol{\nabla} \boldsymbol{m} \right) .
	\label{equ:tl_diff_mc}
\end{eqnarray}

On the other hand, in the locally collinear approach, the generalized non-collinear functional reads
\begin{eqnarray}
E^{\mathrm{LC}}[n,\boldsymbol{m}]=\int  |\boldsymbol{m}| \nabla n \cdot \nabla |\boldsymbol{m}| \mathrm{d}\boldsymbol{r} \label{equ:tl_diff_LC}.
\end{eqnarray}

\subsubsection{Functional No. 3}
Consider a toy non-collinear meta-GGA functional,
\begin{eqnarray}
E^{\mathrm{NC}}[n,\boldsymbol{m}]=\int \boldsymbol{m} \cdot \nabla^2 \boldsymbol{m} \mathrm{d}\boldsymbol{r},\label{equ:toy_mGGA_NC}
\end{eqnarray}
whose corresponding collinear functional reads
\begin{eqnarray}
E^{\mathrm{col}}[n,s]=\int s \nabla^2 s \mathrm{d}\boldsymbol{r}. \label{equ:mgga_toy_col}
\end{eqnarray}
In the multi-collinear approach, the effective collinear and multi-collinear functionals are ready to be obtained
\begin{eqnarray}
E^{\mathrm{eff}}[n,s]&=&3 E^{\mathrm{col}}[n,s],\\
E^{\mathrm{MC}}[n,\boldsymbol{m}]
&=&\int \boldsymbol{m} \cdot \nabla^2 \boldsymbol{m} \mathrm{d}\boldsymbol{r}.
\label{equ:mgga_toy_mc}
\end{eqnarray}

On the other hand, in the locally collinear approach, the generalized non-collinear functional reads
\begin{eqnarray}
E^{\mathrm{LC}}[n,\boldsymbol{m}]=\int  |\boldsymbol{m}| \nabla^2 |\boldsymbol{m}| \mathrm{d}\boldsymbol{r}.
\end{eqnarray}

It is worth noting that the locally collinear approach was originally proposed for LSDA functionals only \cite{kubler1988density}. There are two intuitive interpretations \cite{kurz2004ab,sjostedt2002noncollinear} to include the derivatives of $\boldsymbol{m}$. In the first interpretation, $\boldsymbol{m}$ is projected to its local direction, and then the derivatives are calculated, taking first and second-order derivatives as examples, reading
\begin{eqnarray}
    \nabla_{\alpha} s(\boldsymbol{r}) &\leftarrow& \nabla_{\alpha} \left(\boldsymbol{m}(\boldsymbol{r}) \cdot \boldsymbol{e}(\boldsymbol{r})\right), \label{equ:kubler_u1} \\
    \nabla^2 s(\boldsymbol{r}) &\leftarrow& \nabla^2 \left(\boldsymbol{m}(\boldsymbol{r}) \cdot \boldsymbol{e}(\boldsymbol{r})\right),  \label{equ:kubler_u3}
\end{eqnarray}
with $\alpha=x,y,z$ and $ \boldsymbol{e}(\boldsymbol{r}) = \frac{\boldsymbol{m}(\boldsymbol{r})}{m(\boldsymbol{r})} $.
In the second interpretation, the derivatives of $\boldsymbol{m}$ are calculated before the projection, reading
\begin{eqnarray}
    \nabla_{\alpha} s(\boldsymbol{r}) &\leftarrow&  \boldsymbol{e}(\boldsymbol{r}) \cdot \nabla_{\alpha} \boldsymbol{m}(\boldsymbol{r}) , \label{equ:kubler_u2} \\
    \nabla^2 s(\boldsymbol{r}) &\leftarrow& \boldsymbol{e}(\boldsymbol{r}) \cdot \nabla^2 \boldsymbol{m}(\boldsymbol{r}). \label{equ:kubler_u4}
\end{eqnarray}

For GGA functionals, the two interpretations are equivalent, because equations (\ref{equ:kubler_u1}) and (\ref{equ:kubler_u2}) are the same, in virtue of $ \nabla_{\alpha} \boldsymbol{e}(\boldsymbol{r}) \perp \boldsymbol{m}(\boldsymbol{r}) $. Direct implementations according to Eqs. (\ref{equ:kubler_u1}) and (\ref{equ:kubler_u2}) in Ref. \cite{kurz2004ab} provided different results, probably caused by the numerical singularities of the locally collinear approach.

However, for meta-GGA functionals depending on $\nabla^2 \boldsymbol{m}$, the two interpretations are different, because equations (\ref{equ:kubler_u3}) and (\ref{equ:kubler_u4}) are not the same. Nevertheless, they have been both used, such as the first interpretation Eq. (\ref{equ:kubler_u3}) by Peralta \textit{et al.} in Ref. \citenum{peralta2007noncollinear} and the second interpretation Eq. (\ref{equ:kubler_u4}) by Kn{\"o}pfle \textit{et al.} in Ref. \citenum{knopfle2000spin}.
In this work, the first interpretation is adopted, because it provides a compact and simple form for non-collinear functional, Eq. (\ref{lc}). Besides, $\nabla^2 s(\boldsymbol{r})$ calculated according to Eq. (\ref{equ:kubler_u3}), does equal to $ \nabla^2 $ performing on $ s(\boldsymbol{r}) $, a kind of self-consistency, which is absent in Eq. (\ref{equ:kubler_u4}). 
However, the first interpretation involves the derivatives of $ \boldsymbol{e}(\boldsymbol{r}) $, exacerbating numerical singularities, and even ruining the correct collinear limit.
To see this, applying a collinear functional,
$E^{\mathrm{col}}[n, s] = \int |\nabla^2 s| \mathrm{d} \boldsymbol{r}$, on a simple collinear spin system $\boldsymbol{m} = (0,0,z)$ within the cube $-\frac{1}{2} \le x,y,z \le \frac{1}{2}$, provides $E^{\mathrm{col}}[n, m_z] = \int |\nabla^2 z| \mathrm{d} \boldsymbol{r} = 0$. However, the first interpretation, Eq. (\ref{equ:kubler_u3}), leads to $E^{\mathrm{xc}}[n, \boldsymbol{m}] = \int |\nabla^2 m| \mathrm{d} \boldsymbol{r} = \int |\nabla^2 |z|| \mathrm{d} \boldsymbol{r} = 2 \ne 0$, ruining the correct collinear limit.

Nevertheless, according to the second interpretation of the locally collinear approach, the non-collinear generalization of collinear functional Eq. (\ref{equ:mgga_toy_col}) reads
\begin{eqnarray}
    E^{\mathrm{LC}}[n,\boldsymbol{m}]
&=&\int \boldsymbol{m} \cdot \nabla^2 \boldsymbol{m} \mathrm{d}\boldsymbol{r},
\end{eqnarray}
the same as the multi-collinear functional Eq. (\ref{equ:mgga_toy_mc}).

\subsubsection{Functional No. 4}

Consider a toy non-collinear non-local functional
\begin{eqnarray}
E^{\mathrm{NC}}[n,\boldsymbol{m}]&=&\iint \boldsymbol{m}_1\cdot\boldsymbol{m}_2 \mathrm{d}\boldsymbol{r}_1 \mathrm{d}\boldsymbol{r}_2,\label{equ:nc1_NC} 
\end{eqnarray}
whose corresponding collinear functional reads
\begin{eqnarray}
E^{\mathrm{col}}[n,s]&=&\iint s_1s_2 \mathrm{d}\boldsymbol{r}_1 \mathrm{d}\boldsymbol{r}_2.
\end{eqnarray}
In the multi-collinear approach, the effective collinear and multi-collinear functionals are ready to be obtained
\begin{eqnarray}
E^{\mathrm{eff}}[n,s]&=&3E^{\mathrm{col}}[n,s], \\
E^{\mathrm{MC}}[n,\boldsymbol{m}]
&=& \iint \boldsymbol{m}_1 \cdot \boldsymbol{m}_2  \mathrm{d}\boldsymbol{r}_1 \mathrm{d}\boldsymbol{r}_2.  \label{equ:nc1_MC} 
\end{eqnarray}
Actually, not only the non-collinear functional in the form of Eq. (\ref{equ:nc1_NC}), but also functionals in more complicated forms, such as
\begin{eqnarray}
E^{\mathrm{NC}}[n,\boldsymbol{m}] = \iint \frac{\boldsymbol{m}_1\cdot\boldsymbol{m}_2}{|\boldsymbol{r}_{1}-\boldsymbol{r}_{2}|^3}  \mathrm{d}\boldsymbol{r}_1 \mathrm{d}\boldsymbol{r}_2,\label{equ:nc1_diss}
\end{eqnarray}
can be reproduced by the multi-collinear approach.

On the other hand, in the locally collinear approach, the generalized non-collinear functional reads
\begin{eqnarray}
E^{\mathrm{LC}}[n,\boldsymbol{m}]
&=& \iint |\boldsymbol{m}_1|  |\boldsymbol{m}_2| \mathrm{d}\boldsymbol{r}_1 \mathrm{d}\boldsymbol{r}_2,
\end{eqnarray}
which does not satisfy the correct collinear limit Eq. (\ref{equ:Exccollinear}) when $\boldsymbol{m}_1$ and $\boldsymbol{m}_2$ pointing in the opposite directions.

\subsubsection{Functional No. 5}

Consider another toy non-collinear non-local functional
\begin{eqnarray}
E^{\mathrm{NC}}[n, \boldsymbol{m}]&=&\iint (\boldsymbol{m}_1 \cdot \boldsymbol{m}_2)^2 \mathrm{d}\boldsymbol{r}_1 \mathrm{d}\boldsymbol{r}_2,\label{equ:nc2_NC}
\end{eqnarray}
whose corresponding collinear functional reads
\begin{eqnarray}
E^{\mathrm{col}}[n, s]&=&\iint (s_1 s_2)^2 \mathrm{d}\boldsymbol{r}_1 \mathrm{d}\boldsymbol{r}_2. \label{equ:col_NC_2}
\end{eqnarray}
In the multi-collinear approach, the effective collinear and multi-collinear functionals are ready to be obtained
\begin{eqnarray}
E^{\mathrm{eff}}[n, s]&=&5E^{\mathrm{col}}[n,s], \\ \label{equ:eff_NC_2}
E^{\mathrm{MC}}[n, \boldsymbol{m}]
&=& \iint \frac{2(\boldsymbol{m}_1\cdot\boldsymbol{m}_2)^2+m_1^2m_2^2}{3}  \mathrm{d}\boldsymbol{r}_1 \mathrm{d}\boldsymbol{r}_2.\nonumber \\\label{equ:nc2_MC}
\end{eqnarray}
The multi-collinear functional Eq. (\ref{equ:nc2_MC}) does not reproduce the non-collinear toy functional Eq. (\ref{equ:nc2_NC}). It is expected, because another non-collinear functional,
\begin{eqnarray}
E^{\mathrm{NC}}[n, \boldsymbol{m}]
&=& \iint m_1^2  m_2^2 \mathrm{d}\boldsymbol{r}_1 \mathrm{d}\boldsymbol{r}_2,\label{equ:noncollinear2_nc}
\end{eqnarray}
provides the same collinear functional Eq. (\ref{equ:col_NC_2}), such that no approach can distinguish those two non-collinear functionals Eq. (\ref{equ:nc2_NC}) and Eq. (\ref{equ:noncollinear2_nc}). 

On the other hand, in the locally collinear approach, the generalized non-collinear functional reads
\begin{eqnarray}
E^{\mathrm{LC}}[n, \boldsymbol{m}]
&=& \iint m_1^2  m_2^2 \mathrm{d}\boldsymbol{r}_1 \mathrm{d}\boldsymbol{r}_2.
\end{eqnarray}

The deduction of TABLE \ref{table1} is finished.

\subsection{Multi-collinear approach  satisfying four criteria}\label{sec:four}

In this subsection, we revisit the four criteria proposed in Section \ref{background} to show that they can be fully satisfied by the multi-collinear approach.

\subsubsection{Correct collinear limit}

In constructing the multi-collinear functionals, the correct collinear limit Eq. (\ref{equ:Exccollinear}) is applied as a prerequisite. Therefore, the correct collinear limit is always satisfied. It should be noted that our deduction (from Eq. (\ref{equ:un}) to Eq. (\ref{equ:MC2})) is completely general, without any constraint for the form of collinear functionals. As a result, the multi-collinear functionals satisfy the correct collinear limit for all functionals, including but not limited to LSDA, GGA, meta-GGA, hybrid and non-local functionals.

\subsubsection{Being invariant to the global rotation while sensitive to the local rotation of spin magnetization vector}

\begin{figure}
	\centering
	\includegraphics[width=8.6cm]{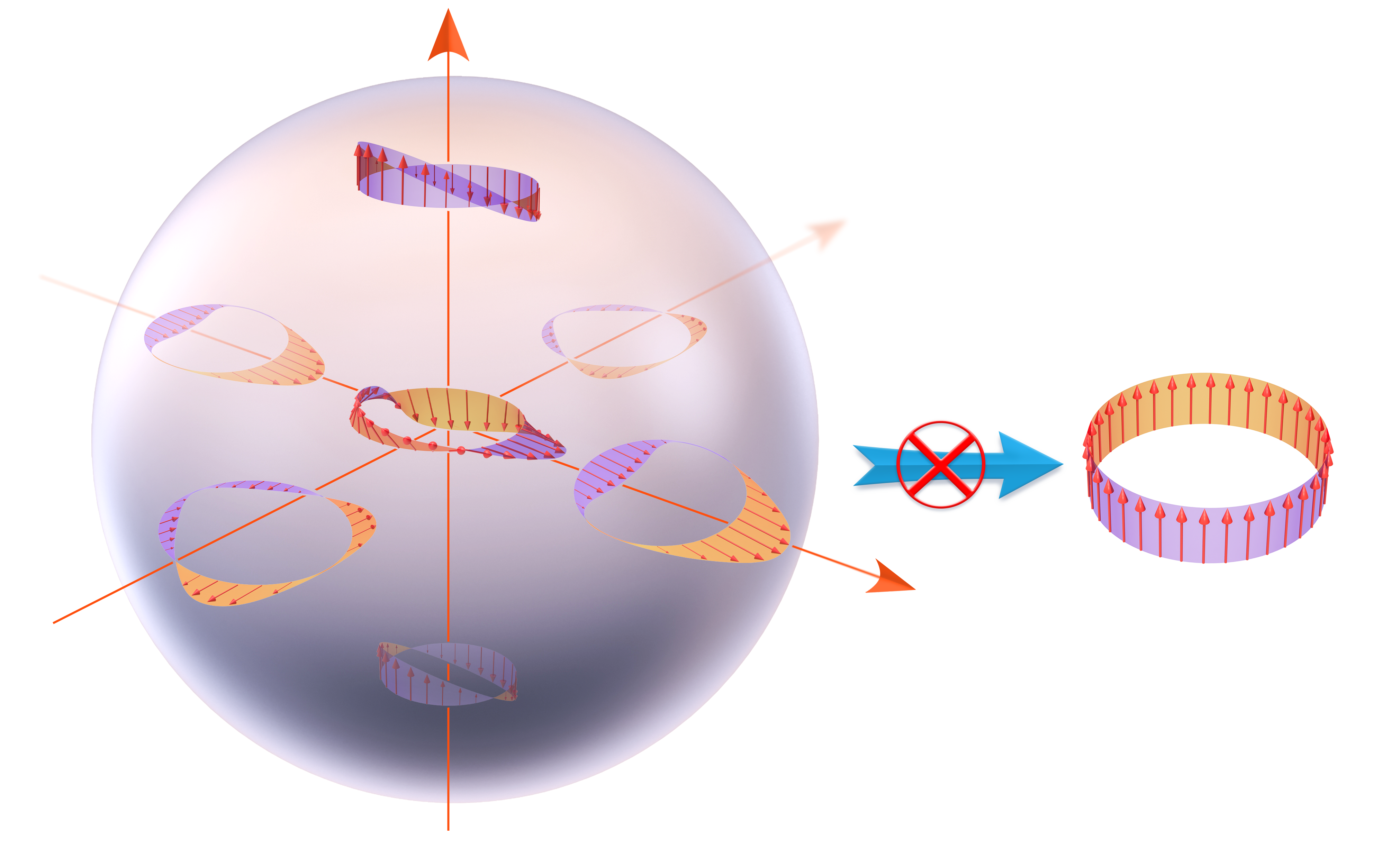}
	\caption{A spin spiral state in the multi-collinear approach. The twisted band in the sphere's center denotes a spin spiral state, with red arrows indicating its $\boldsymbol{m}$. Multiple projected collinear states (six bands on the sphere's surface) represent the original non-collinear spin spiral state, exhibiting its non-trivial magnetic structure. 
	}
	\label{fig2}
\end{figure}

The average over all directions in Eq. (\ref{equ:un}) indicates that the multi-collinear approach must satisfy the global rotation invariance in the property (\uppercase\expandafter{\romannumeral2}).
The sensitivity to local spin rotations of the multi-collinear approach has been shown by the illustrative example in FIG. \ref{fig1}.
A more complicated example is shown in FIG. \ref{fig2}, where a non-collinear spin state (a spin spiral state \cite{overhauser1962spin} represented by the band at the sphere's center) is represented by a set of projected collinear states (six bands on the sphere's surface). 
This sensitivity endows multi-collinear functionals with dependence on the directions of $\boldsymbol{m}$ on the whole spatial space. Such dependence is useful in studying the topological structures of magnetic materials, such as skyrmions \cite{skyrme1962unified}.

\subsubsection{Well-defined functional derivatives}

In the locally collinear approach, numerical singularities result from the ill-definition of the direction of $\boldsymbol{m}$ when $\boldsymbol{m}\to\boldsymbol{0}$. Taking the locally collinear GGA functionals in TABLE \ref{table1} as an example, the numerical singularities root in $\nabla_{\alpha}m = \frac{\boldsymbol{m} \cdot\nabla_{\alpha} \boldsymbol{m}}{m}$ ($\alpha = x,y,z$), which goes to $\frac{0}{0}$ at $ m \to 0 $.

In the multi-collinear approach, $ \boldsymbol{m} $ is projected to constant global directions $ \boldsymbol{e}_\Omega $.
Since the $ \boldsymbol{e}_\Omega $ does not depend on spatial positions or any physical quantity, its arbitrary derivatives are zero, avoiding numerical singularities. Therefore, the multi-collinear approach satisfies criterion (\uppercase\expandafter{\romannumeral3}). 

\subsubsection{Providing global zero torque but non-vanishing local torque}

As pointed out by Capelle and co-workers, any static functional that is invariant under the infinitesimal global spin rotation provides global zero torque\cite{capelle2001spin}. Thus, the multi-collinear approach apparently provides global zero torque, satisfying the zero torque theorem.

The local torque is defined as $\boldsymbol{m} \times \boldsymbol{B}^{\mathrm{xc}}$, with $\boldsymbol{B}^{\mathrm{xc}}$ the exchange-correlation magnetic field
\begin{eqnarray}
     \boldsymbol{B}^{\mathrm{xc}} &=&- \frac{\delta E^{\mathrm{xc}}[n,\boldsymbol{m}]}{\delta \boldsymbol{m}}.\label{equ:bxc_define}
\end{eqnarray}
In the multi-collinear approach, $\boldsymbol{B}^{\mathrm{xc}}$ and $\boldsymbol{m}$ are generally not parallel, providing non-vanishing local torque, with an explicit example given below. Consider the toy multi-collinear GGA functional, functional No. 1 in TABLE \ref{table1}, with the exchange-correlation magnetic field 
\begin{eqnarray}
   \boldsymbol{B}^{\mathrm{xc}}
   = \sum_{\alpha,\beta,\gamma = x,y,z} \nabla_{\gamma} \frac{\partial [(\nabla_{\alpha} m_{\beta})(\nabla_{\alpha} m_{\beta})]}{\partial(\nabla_\gamma \boldsymbol{m})}
    = 2 \nabla^2 \boldsymbol{m}, 
\end{eqnarray}
and a model system, with $\boldsymbol{m}=(x^2,0,1)$. The local exchange-correlation magnetic field $\boldsymbol{B}^{\mathrm{xc}}$ is $(4,0,0)$, apparently not parallel to $\boldsymbol{m}$, providing non-vanishing local torque.

Eich and co-workers\cite{eich2013transverse2} showed that for non-collinear GGA functionals, non-vanishing local torque arises from the dependence of functionals on transverse gradients of spin density, and suggested that non-collinear GGA functionals should depend on both transverse and longitude gradients, in an unequal way. In the same work, they pointed out\cite{eich2013transverse2} that the modified locally collinear approach by Scalmani and Frisch\cite{scalmani2012new} does depend on transverse gradients of spin density, but in the same way as longitudinal gradients.
It is easy to see that the multi-collinear functional depends on both transverse and longitudinal gradients, in an unequal way. 

\subsection{Exchange-correlation potential of multi-collinear approach}

In applications, not only the exchange-correlation energy but also its derivatives with respect to density matrix $ \boldsymbol{D} = D_{pq} $ ($p,q$ for orbital indexes) are needed.
Its first-order derivative is exchange-correlation potential $V^{\mathrm{MC}}$,
\begin{align}
    V^{\mathrm{MC}}_{qp} = \frac{\partial E^{\mathrm{MC}}}{\partial D_{pq}}= \frac{\partial \overline{E^{\mathrm{eff}}}}{\partial D_{pq}} = \frac{\overline{\partial E^{\mathrm{eff}}}}{\partial D_{pq}} = \overline{V^{\mathrm{eff}}_{qp}}.\label{equ:interchange}
\end{align}
In deriving Eq. (\ref{equ:interchange}), the fact that two operations, averaging over $\Omega$ and calculating the first-order derivative with respect to the density matrix, are interchangeable has been used (since the projection directions do not depend on the density matrix).
It should be noted that such interchangeability also holds for higher-order derivatives.

With the help of Eq. (\ref{equ:interchange}), the calculation of $V^{\mathrm{MC}}_{qp}$ is transformed into $V^{\mathrm{eff}}_{qp}$, which can be further evaluated using the chain rule
\begin{eqnarray}
  V^{\mathrm{eff}}_{qp}[n,m_{\Omega}] &=& \int  E^{\mathrm{eff}}_{10}[n,m_{\Omega}] \psi_q^{\dagger}(\boldsymbol{r}_1)\psi_p(\boldsymbol{r}_1) 
  \mathrm{d} \boldsymbol{r}_1  +  \nonumber   \\
  && \int  E^{\mathrm{eff}}_{01}[n,m_{\Omega}] \psi_q^{\dagger}(\boldsymbol{r}_1){\sigma_{\Omega}}\psi_p(\boldsymbol{r}_1) \mathrm{d} \boldsymbol{r}_1.
  \label{equ:potential_gga_original}
\end{eqnarray}
For convenience, in Eq. (\ref{equ:potential_gga_original}), we have introduced the short notations for $(N+M)$-th order functional derivatives ($N$ for $n$, $M$ for $s$) of collinear and effective collinear functionals, 
\begin{eqnarray}
  &&E^{\mathrm{col/eff}}_{NM}[n,s]\nonumber\\ &= &  \frac{\delta^{N+M} E^{\mathrm{col/eff}}[n,s]}{\delta n({\boldsymbol{r}_{1}}) \delta n({\boldsymbol{r}_{2}}) \cdots \delta n({\boldsymbol{r}_{N}}) \delta s({\boldsymbol{r}_{N+1}})\cdots \delta s({\boldsymbol{r}_{N+M}})}, \nonumber \\
  \label{equ:nm_definite}
\end{eqnarray}
which are related via, after simple calculations,
\begin{eqnarray}
  E^{\mathrm{eff}}_{NM}&=& (1+M)E^{\mathrm{col}}_{NM} \nonumber \\
  &&+ \int  E^{\mathrm{col}}_{NM+1} s(\boldsymbol{r}_{N+M+1}) \mathrm{d} \boldsymbol{r}_{N+M+1}.
  \label{equ:eff_derivative}
\end{eqnarray}
For $N=M=0$, equation (\ref{equ:eff_derivative}) is nothing but Eq. (\ref{equ:UN-col}). For $N=1,M=0$ and $N=0,M=1$, equation (\ref{equ:eff_derivative}) reads
\begin{equation}
    \left\{
    \begin{aligned}
    E^{\mathrm{eff}}_{10} & = 
    \frac{\delta E^{\mathrm{col}}}{\delta n({\boldsymbol{r}_{1}})} + \int  \frac{\delta^2 E^{\mathrm{col}}}{\delta n({\boldsymbol{r}_{1}})\delta s({\boldsymbol{r}_{2}})} s(\boldsymbol{r}_{2}) \mathrm{d} \boldsymbol{r}_{2}  \\
    E^{\mathrm{eff}}_{01} & = 2\frac{\delta E^{\mathrm{col}}}{\delta s({\boldsymbol{r}_{1}})} + \int   \frac{\delta^2 E^{\mathrm{col}}}{\delta s({\boldsymbol{r}_{1}})\delta s({\boldsymbol{r}_{2}})}s(\boldsymbol{r}_{2}) \mathrm{d} \boldsymbol{r}_{2}
    \end{aligned}
    \right..
\end{equation}

\subsection{Multi-collinear potential for collinear spin systems} \label{sec:potential}

The general expression of the multi-collinear potential has been given by Eq. (\ref{equ:interchange}) and Eq. (\ref{equ:potential_gga_original}).
The question we want to address in this subsection is, for collinear spin systems, what the relation between the multi-collinear potential and collinear potential is.
The answer is that they are identical, with the proof given below.

Let us consider a collinear spin system with $\boldsymbol{m} = (0,0,m_z)$, whose multi-collinear potential $V^{\mathrm{MC}}$ is evaluated via
\begin{eqnarray}
&&V_{qp}^{\mathrm{MC}}[n, (0,0,m_z)] \nonumber \\
&=&\int\frac{\delta E^{\mathrm{MC}}[n,(0,0,m_z)]}{\delta n(\boldsymbol{r})}\psi_q^{\dagger}(\boldsymbol{r})\psi_p(\boldsymbol{r}) \mathrm{d}\boldsymbol{r} \nonumber\\
&+&
\int\frac{\delta E^{\mathrm{MC}}[n,(0,0,m_z)]}{\delta m_z(\boldsymbol{r})} \psi_q^{\dagger}(\boldsymbol{r})\sigma_{z}\psi_p(\boldsymbol{r})  \mathrm{d}\boldsymbol{r}\nonumber\\
&+&\int \left.\frac{ \delta E^{\mathrm{MC}}[n,(m_x,m_y,m_z)]}{\delta m_x(\boldsymbol{r})}\right|_{m_x= m_y=0} \psi_q^{\dagger}(\boldsymbol{r})\sigma_{x}\psi_p(\boldsymbol{r}) \mathrm{d}\boldsymbol{r}\nonumber\\
&+&\int \left. \frac{\delta E^{\mathrm{MC}}[n,(m_x,m_y,m_z)]}{\delta m_y(\boldsymbol{r})}\right|_{m_x= m_y=0} \psi_q^{\dagger}(\boldsymbol{r})\sigma_{y}\psi_p(\boldsymbol{r}) \mathrm{d}\boldsymbol{r}. \nonumber \\ \label{equ:col_potential_start} 
\end{eqnarray}
By noticing that the multi-collinear approach preserves the global rotation symmetry, 
\begin{eqnarray}
E^{\mathrm{MC}}[n,m_x,m_y,m_z] = E^{\mathrm{MC}}[n,m_x,-m_y,-m_z],
\end{eqnarray}
and the time-reversal symmetry,
\begin{eqnarray}
E^{\mathrm{MC}}[n,m_x,m_y,m_z] = E^{\mathrm{MC}}[n,-m_x,-m_y,-m_z],
\end{eqnarray}
one immediately obtains
\begin{eqnarray}
E^{\mathrm{MC}}[n,m_x,m_y,m_z] = E^{\mathrm{MC}}[n,-m_x,m_y,m_z],
\end{eqnarray}
hinting that the third term in Eq. (\ref{equ:col_potential_start}) vanishes (so does the fourth term). 
By further realizing the fact that the multi-collinear approach satisfies the correct collinear limit Eq. (2), equation (\ref{equ:col_potential_start}) is simplified into
\begin{eqnarray}
   && V_{qp}^{\mathrm{MC}}[n, (0,0,m_z)] \nonumber\\ &=&\int \frac{\delta E^{\mathrm{col}}[n,m_z]}{\delta n(\boldsymbol{r})}\psi_q^{\dagger}(\boldsymbol{r})\psi_p(\boldsymbol{r}) \mathrm{d}\boldsymbol{r} \nonumber \\
   &&+  \int \frac{\delta E^{\mathrm{col}}[n,m_z]}{\delta m_z(\boldsymbol{r})} \psi_q^{\dagger}(\boldsymbol{r})\sigma_{z}\psi_p(\boldsymbol{r})  \mathrm{d}\boldsymbol{r}, \label{V-MC}
\end{eqnarray}
which is nothing but $V^{\mathrm{col}}$, the potential of traditional collinear functional.

The multi-collinear approach satisfying the correct collinear limit indicates that the energies evaluated by multi-collinear functionals and traditional collinear functionals are identical for collinear spins. 
Discussions in this subsection show that their exchange-correlation potentials are also identical for collinear spins.
However, such equivalence cannot be extended to the second-order derivatives, the exchange-correlation kernel, which will be further discussed elsewhere.

\section{Implementation and benchmark tests}\label{sec:implementation}

For developers willing to implement the multi-collinear approach, we provide a package called MCfun, which transforms collinear functionals into multi-collinear functionals. MCfun, available on GitHub (\url{https://github.com/Multi-collinear/MCfun}) and PyPI (\url{https://pypi.org/project/MCfun}), now supports LSDA, GGA and meta-GGA functionals.

By calling MCfun, the multi-collinear approach has been implemented in the official version (2.1) of PySCF \cite{sun2015libcint,sun2018pyscf,sun2020recent}. Now it supports non-collinear DFT and TDDFT calculations, for LSDA, GGA, meta-GGA and hybrid functionals for molecules. People interested in applications can directly download PySCF and use it. All the tests in this work are carried out in a locally modified PySCF to support calculations of local torque, solid states and forces \textit{et al.}.

In the following, the implementation of the multi-collinear approach will be addressed, including energy Eq. (\ref{equ:un}) and potential Eq. (\ref{equ:interchange}), both in two steps. The first step is the calculation of the effective collinear energy and effective collinear potential. The second step is the average over all directions in spin space.

\subsection{The first step: the effective collinear energy and potential}

The expressions of effective collinear energies Eq. (\ref{equ:UN-col}) and effective collinear potentials Eq. (\ref{equ:potential_gga_original}) are general, independent of the form of collinear functionals.
In this subsection, their working equations for commonly used functionals, LSDA, GGA and meta-GGA, will be given. It is worth noting that when applying the multi-collinear approach to hybrid functionals, only their pure functional parts need to be treated, in virtue of Eq. (\ref{shift}). 

For LSDA, the integrand of the effective collinear energy Eq. (\ref{equ:UN-col}) is evaluated via
\begin{eqnarray}
    f^{\mathrm{eff}}(n,m_\Omega) = f^{\mathrm{col}}(n,m_\Omega) + m_\Omega\frac{\partial f^{\mathrm{col}}(n,m_\Omega)}{\partial m_\Omega}. \label{equ:eff_lda}
\end{eqnarray}
For GGA, it is evaluated via
\begin{eqnarray}
    f^{\mathrm{eff}}
    &=& f^{\mathrm{col}}(n,m_\Omega,\boldsymbol{\nabla}n,\boldsymbol{\nabla}m_\Omega) \nonumber    \\
    &&+  m_\Omega \frac{\partial f^{\mathrm{col}}(n,m_\Omega,\boldsymbol{\nabla}n,\boldsymbol{\nabla}m_\Omega)}{\partial m_\Omega} 
      \nonumber \\
    &&-  m_\Omega \boldsymbol{\nabla} \cdot \frac{\partial f^{\mathrm{col}}(n,m_\Omega,\boldsymbol{\nabla}n,\boldsymbol{\nabla}m_\Omega)}{\partial \boldsymbol{\nabla} m_\Omega},  
    \label{equ:gga_old}
\end{eqnarray}
or, in virtue of the integration by parts,
\begin{eqnarray}
    &&f^{\mathrm{eff}}(n,m_\Omega,\boldsymbol{\nabla}n,\boldsymbol{\nabla}m_\Omega) \nonumber \\
    &=& f^{\mathrm{col}}(n,m_\Omega,\boldsymbol{\nabla}n,\boldsymbol{\nabla}m_\Omega) \nonumber    \\
    &&+ m_\Omega \frac{\partial f^{\mathrm{col}}(n,m_\Omega,\boldsymbol{\nabla}n,\boldsymbol{\nabla}m_\Omega)}{\partial m_\Omega} 
     \nonumber \\
    &&+   \left(\boldsymbol{\nabla}  m_\Omega\right) \cdot \frac{\partial f^{\mathrm{col}}(n,m_\Omega,\boldsymbol{\nabla}n,\boldsymbol{\nabla}m_\Omega)}{\partial \boldsymbol{\nabla}  m_\Omega}.\label{equ:gga_ibp}
\end{eqnarray}
In our code, the effective collinear energy for GGA functional, as well as its derivatives, is implemented according to Eq. (\ref{equ:gga_ibp}), which has a simpler form compared with Eq. (\ref{equ:gga_old}).
For meta-GGA functionals depending on not only densities but also orbital kinetic energy densities, 
equation (\ref{equ:un}) needs to be generalized to
\begin{eqnarray}
	E^{\mathrm{MC}}[n, \boldsymbol{m}, \tau, \boldsymbol{u}] =  \overline{E^{\mathrm{eff}}[n,m_\Omega,\tau, u_{\Omega}]},\label{equ:un_mgga}
\end{eqnarray}
where 
\begin{eqnarray}
\tau(\boldsymbol{r})&=&\frac{1}{2} \sum_{i \in \mathrm{occ.}}\left[\boldsymbol{\nabla} \psi_{i}(\boldsymbol{r})\right]^{\dagger} \cdot \boldsymbol{\nabla} \psi_{i}(\boldsymbol{r}), \\
u_{\alpha}(\boldsymbol{r})&=&\frac{1}{2} \sum_{i \in \mathrm{occ.}}\left[\boldsymbol{\nabla} \psi_{i}(\boldsymbol{r})\right]^{\dagger} \sigma_{\alpha} \cdot \boldsymbol{\nabla} \psi_{i}(\boldsymbol{r}),
\end{eqnarray} 
with $\mathrm{occ.}$ for occupied orbitals and $\alpha$ for $x,y,z$. It is worth noting that the directions of $\boldsymbol{m}$ and $\boldsymbol{u}$ are generally not parallel. Projecting $\boldsymbol{u}$ to the local direction of $\boldsymbol{m}$, suggested by the locally collinear approach, seems not to be a good choice, especially for spatial grids with significant $\boldsymbol{u}$ but negligible $\boldsymbol{m}$. However, in the multi-collinear approach, $\boldsymbol{m}$ and $\boldsymbol{u}$ are projected to all global directions, regardless of their own directions, which is a fair and stable treatment.
To evaluate $E^{\mathrm{eff}}[n, m_\Omega, \tau, u_\Omega]$ in meta-GGA, equation (\ref{equ:UN-col}) needs to be extended to
\begin{eqnarray}
    E^{\mathrm{eff}}[n, m_\Omega, \tau, u_\Omega] 
    &=&  E^{\mathrm{col}}[n,m_\Omega, \tau, u_\Omega] \nonumber \\
	&&+\int \frac{\delta E^{\mathrm{col}}[n,m_\Omega, \tau, u_\Omega]}{\delta m_\Omega(\boldsymbol{r})} m_\Omega(\boldsymbol{r})  \mathrm{d} \boldsymbol{r}
	\nonumber \\
	&& + \int  \frac{\delta E^{\mathrm{col}}[n,m_\Omega, \tau,u_\Omega]}{\delta u_\Omega(\boldsymbol{r})}
	u_\Omega(\boldsymbol{r}) \mathrm{d} \boldsymbol{r}, \nonumber \\  \label{equ:mgga_1}
\end{eqnarray}
with the integrand
\begin{eqnarray}
    &&f^{\mathrm{eff}}(n,m_\Omega,\boldsymbol{\nabla}n,\boldsymbol{\nabla}m_\Omega,\nabla^2 n,\nabla^2 m_\Omega,\tau,u_\Omega)  \nonumber \\
    &=&  f^{\mathrm{col}}(n,m_\Omega,\boldsymbol{\nabla}n,\boldsymbol{\nabla}m_\Omega,\nabla^2 n,\nabla^2 m_\Omega,\tau,u_\Omega)  \nonumber
    \\
	&&+   m_{\Omega} \frac{\partial f^{\mathrm{col}}(n,m_\Omega,\boldsymbol{\nabla}n,\boldsymbol{\nabla}m_\Omega,\nabla^2 n,\nabla^2 m_\Omega,\tau,u_\Omega)}{\partial m_{\Omega}}  \nonumber \\
    &&+  \left(\boldsymbol{\nabla} m_{\Omega}\right) \cdot \frac{\partial f^{\mathrm{col}}(n,m_\Omega,\boldsymbol{\nabla}n,\boldsymbol{\nabla}m_\Omega,\nabla^2 n,\nabla^2 m_\Omega,\tau,u_\Omega)}{\partial \boldsymbol{\nabla}  m_{\Omega}} \nonumber   \\
	&&+  (\nabla^2 m_{\Omega}) \frac{\partial f^{\mathrm{co l}}(n,m_\Omega,\boldsymbol{\nabla}n,\boldsymbol{\nabla}m_\Omega,\nabla^2 n,\nabla^2 m_\Omega,\tau,u_\Omega)}{\partial \nabla^2 m_{\Omega}} \nonumber \\
 &&+  u_{\Omega} \frac{\partial f^{\mathrm{col}}(n,m_\Omega,\boldsymbol{\nabla}n,\boldsymbol{\nabla}m_\Omega,\nabla^2 n,\nabla^2 m_\Omega,\tau,u_\Omega)}{\partial u_{\Omega}}.
	\label{equ:MGGA_explicit}
\end{eqnarray}
It is worth noting that the integrands $f^{\mathrm{eff}}$ for LSDA Eq. (\ref{equ:eff_lda}), GGA Eq. (\ref{equ:gga_ibp}) and meta-GGA Eq. (\ref{equ:MGGA_explicit}) share a uniform form
\begin{eqnarray}
    &&f^{\mathrm{eff}}(\kappa_{1},\kappa_{2},\cdots,\kappa_{n},\chi_{1},\chi_{2},\cdots,\chi_{n}) \nonumber    \\
    &=& f^{\mathrm{col}}(\kappa_{1},\kappa_{2},\cdots,\kappa_{n},\chi_{1},\chi_{2},\cdots,\chi_{n}) \nonumber    \\
    &&+ \sum_{j=1}^{n} \chi_j \frac{\partial f^{\mathrm{col}}(\kappa_{1},\kappa_{2},\cdots,\kappa_{n},\chi_{1},\chi_{2},\cdots,\chi_{n})}{\partial \chi_j},
    \label{equ:eff_intgrand_new}
\end{eqnarray}
with $\kappa_i$ for time-reversal even variables ($n$, $\boldsymbol{\nabla} n $, $\nabla^2 n$, $\tau$), and $\chi_i$ for time-reversal odd variables ($m_\Omega$, $\boldsymbol{\nabla} m_\Omega $, $\nabla^2 m_\Omega$, $u_\Omega$). Equation (\ref{equ:eff_intgrand_new}) is the working equation for effective collinear energy.

With the help of the chain rule, the working equations of the effective collinear potential for LSDA, GGA and meta-GGA read
\begin{eqnarray}
    V^{\mathrm{eff}}_{qp}
  &&= \int \mathrm{d}\boldsymbol{r}  \frac{\partial f^{\mathrm{eff}}}{\partial n}
	\psi_{q}^{\dagger}\psi_{p}
	+ \int \mathrm{d}\boldsymbol{r} \frac{\partial f^{\mathrm{eff}}}{\partial m_{\Omega}}
	\psi_{q}^{\dagger}\sigma_{\Omega}\psi_{p} \label{equ:wk_lda},
 \end{eqnarray}
and
\begin{eqnarray}
    V^{\mathrm{eff}}_{qp}
  &=& \int \mathrm{d}\boldsymbol{r}  \frac{\partial f^{\mathrm{eff}}}{\partial n}
	\psi_{q}^{\dagger}\psi_{p}
	+ \int \mathrm{d}\boldsymbol{r} \frac{\partial f^{\mathrm{eff}}}{\partial m_{\Omega}} 	\psi_{q}^{\dagger}\sigma_{\Omega}\psi_{p}   \nonumber\\
	&&+
	\int \mathrm{d}\boldsymbol{r} \frac{\partial f^{\mathrm{eff}}}{\partial \boldsymbol{\nabla}n}
	 \cdot
	\boldsymbol{\nabla} \left( \psi_{q}^{\dagger}\psi_{p}\right)  \nonumber   \\
	&&+ \int \mathrm{d}\boldsymbol{r}\frac{\partial f^{\mathrm{eff}}}{\partial \boldsymbol{\nabla} m_{\Omega}}\cdot\boldsymbol{\nabla}
	\left( \psi_{q}^{\dagger}\sigma_{\Omega}\psi_{p}\right) ,\label{equ:veff_gga}
\end{eqnarray}
and
\begin{eqnarray}
	 &&V^{\mathrm{eff}}_{qp} \nonumber   \\
	 &=& \int \mathrm{d}\boldsymbol{r} \frac{\partial f^{\mathrm{eff}}}{\partial n}
	\psi_{q}^{\dagger}\psi_{p}
	+ \int \mathrm{d}\boldsymbol{r} \frac{\partial f^{\mathrm{eff}}}{\partial m_{\Omega}}
		\psi_{q}^{\dagger}\sigma_{\Omega}\psi_{p} \nonumber   \\
	&+&    
	\int \mathrm{d}\boldsymbol{r} \frac{\partial f^{\mathrm{eff}}}{\partial \boldsymbol{\nabla}n}
	 \cdot
	\boldsymbol{\nabla} \left( \psi_{q}^{\dagger}\psi_{p}\right) +
 \int \mathrm{d}\boldsymbol{r} \frac{\partial f^{\mathrm{eff}}}{\partial \boldsymbol{\nabla} m_{\Omega}} \cdot\boldsymbol{\nabla}
	\left( \psi_{q}^{\dagger}\sigma_{\Omega}\psi_{p}\right)   
\nonumber   \\	
 &+&  \int \mathrm{d}\boldsymbol{r} \frac{\partial f^{\mathrm{eff}}}{\partial \nabla^2n}
	 \nabla^2 \left( \psi_{q}^{\dagger}\psi_{p}\right)
  + \int \mathrm{d}\boldsymbol{r} \frac{\partial f^{\mathrm{eff}}}{\partial \nabla^2 m_{\Omega}} \nabla^2
	\left( \psi_{q}^{\dagger}\sigma_{\Omega}\psi_{p}\right) \nonumber  \\  
 &+&  \int \mathrm{d}\boldsymbol{r}
	\frac{\partial f^{\mathrm{eff}}}{\partial \tau}
	\frac{1}{2} \left(\boldsymbol{\nabla} \psi_{q}^{\dagger}\cdot\boldsymbol{\nabla}\psi_{p} \right) \nonumber  \\  
&+& \int \mathrm{d}\boldsymbol{r}
	\frac{\partial f^{\mathrm{eff}}}{\partial u_{\Omega}}
	\frac{1}{2} \left(\boldsymbol{\nabla} \psi_{q}^{\dagger}\cdot\sigma_{\Omega}\boldsymbol{\nabla}\psi_{p} \right),
	 \label{equ:potential_we} 
\end{eqnarray}
respectively.
The partial derivatives of $f^{\mathrm{eff}}$ appearing in Eq. (\ref{equ:wk_lda}), Eq. (\ref{equ:veff_gga}) and Eq. (\ref{equ:potential_we}) are obtained from Eq. (\ref{equ:eff_intgrand_new}),
\begin{equation}
    \left\{
    \begin{aligned}
    \frac{\partial f^{\mathrm{eff}}}{\partial \kappa_i} =& \frac{\partial f^{\mathrm{col}}}{\partial \kappa_i} + \sum_{j=1}^{n} \chi_j \frac{\partial^2 f^{\mathrm{col}}}{\partial \kappa_i \partial \chi_j}  \\
    \frac{\partial f^{\mathrm{eff}}}{\partial \chi_i} =& 2\frac{\partial f^{\mathrm{col}}}{\partial \chi_i} + \sum_{j=1}^{n} \chi_j \frac{\partial^2 f^{\mathrm{col}}}{\partial \chi_i \partial \chi_j}
    \end{aligned}
    \right. , \label{equ:pxy} 
\end{equation}
which can be easily extended to the second-order 
\begin{equation}
    \left\{
    \begin{aligned}
    \frac{\partial^2 f^{\mathrm{eff}}}{\partial \kappa_i \partial \kappa_k} =& \frac{\partial^2 f^{\mathrm{col}}}{\partial \kappa_i \partial \kappa_k} + \sum_{j=1}^{n} \chi_j \frac{\partial^3 f^{\mathrm{col}}}{\partial \kappa_i \partial \kappa_k \partial \chi_j}  \\
    \frac{\partial^2 f^{\mathrm{eff}}}{\partial \kappa_i \partial \chi_k} =& 2\frac{\partial^2 f^{\mathrm{col}}}{\partial \kappa_i \partial \chi_k} + \sum_{j=1}^{n} \chi_j \frac{\partial^3 f^{\mathrm{col}}}{\partial \kappa_i \partial \chi_j \partial \chi_k} \\
    \frac{\partial^2 f^{\mathrm{eff}}}{\partial \chi_i \partial \chi_k} =& 3\frac{\partial^2 f^{\mathrm{col}}}{\partial \chi_i \partial \chi_k} + \sum_{j=1}^{n} \chi_j \frac{\partial^3 f^{\mathrm{col}}}{\partial \chi_i \partial \chi_j \partial \chi_k}
    \end{aligned}
    \right. , \label{equ:pxyz} 
\end{equation}
and high-orders.
Equation (\ref{equ:pxy}) is the working equation in our code, whose explicit form for GGA reads
\begin{equation}
\left\{
\begin{aligned}
 \frac{\partial f^{\mathrm{eff}}}{\partial n}
 &= \frac{\partial 			  f^{\mathrm{col}}}{\partial n}
	+ \frac{\partial^2 f^{\mathrm{col}}}{\partial n \partial m_{\Omega}}m_{\Omega}
    +\frac{\partial^2 f^{\mathrm{col}}}{\partial n \partial \boldsymbol{\nabla} m_{\Omega}} \cdot \boldsymbol{\nabla} m_{\Omega}
 \\
 \frac{\partial f^{\mathrm{eff}}}{\partial m_{\Omega}}
 &= 2 \frac{\partial f^{\mathrm{col}}}{\partial m_{\Omega}}
    + \frac{\partial^2 f^{\mathrm{col}}}{\partial m_{\Omega}^2} m_{\Omega}
    + \frac{\partial^2 f^{\mathrm{col}}}{\partial m_{\Omega} \partial \boldsymbol{\nabla} m_{\Omega}} \cdot \boldsymbol{\nabla} m_{\Omega}
 \\
 \frac{\partial f^{\mathrm{eff}}}{\partial \boldsymbol{\nabla}n}
 &= \frac{\partial f^{\mathrm{col}}}{\partial \boldsymbol{\nabla} n}
	+ \frac{\partial^2 f^{\mathrm{col}}}{\partial \boldsymbol{\nabla} n \partial m_{\Omega}}m_{\Omega}  \\
	&+ \sum_{\alpha} \frac{\partial^2 f^{\mathrm{col}}}{\partial \boldsymbol{\nabla} n \partial \nabla_{\alpha} m_{\Omega}}  \nabla_{\alpha} m_{\Omega}
 \\
 \frac{\partial f^{\mathrm{eff}}}{\partial \boldsymbol{\nabla} m_{\Omega}}
 &= 2 \frac{\partial f^{\mathrm{col}}}{\partial \boldsymbol{\nabla} m_{\Omega}}
	+ \frac{\partial^2 f^{\mathrm{col}}}{\partial \boldsymbol{\nabla} m_{\Omega} \partial m_{\Omega}}m_{\Omega}  \\
	&+ \sum_{\alpha} \frac{\partial^2 f^{\mathrm{col}}}{\partial \boldsymbol{\nabla} m_{\Omega} \partial \nabla_{\alpha} m_{\Omega}}  \nabla_{\alpha} m_{\Omega}
  \label{equ:potential_gga}
  \end{aligned} \right. ,
\end{equation}
leading to the expression of the exchange-correlation magnetic field, needed in the calculations of local torque
\begin{align}
   \boldsymbol{B}^{\mathrm{xc}} 
   = &  -\overline{\left( 2 \frac{\partial f^{\mathrm{col}}}{\partial m_{\Omega}}
    + \frac{\partial^2 f^{\mathrm{col}}}{\partial m_{\Omega}^2} m_{\Omega}
    + 
	\frac{\partial^2 f^{\mathrm{col}}}{\partial m_{\Omega} \partial \boldsymbol{\nabla} m_{\Omega}} \cdot \boldsymbol{\nabla} m_{\Omega}
	 \right) \boldsymbol{e}_{\Omega}}  \nonumber\\
	&+ \overline{\boldsymbol{\nabla} \cdot\left( 2 \frac{\partial f^{\mathrm{col}}}{\partial \boldsymbol{\nabla} m_{\Omega}}
	+ \frac{\partial^2 f^{\mathrm{col}}}{\partial \boldsymbol{\nabla} m_{\Omega} \partial m_{\Omega}}m_{\Omega}
	\right) \boldsymbol{e}_{\Omega} } \nonumber \\
    &+ \overline{\boldsymbol{\nabla} \cdot\left( \sum_{\alpha} \frac{\partial^2 f^{\mathrm{col}}}{\partial \boldsymbol{\nabla} m_{\Omega} \partial \nabla_{\alpha} m_{\Omega}}  \nabla_{\alpha} m_{\Omega}
	\right) \boldsymbol{e}_{\Omega} }
 .
	  \label{equ:bxc}
\end{align}

\subsection{The second step: averages over solid angles in spin space}

The averages over projection directions for multi-collinear energy Eq. (\ref{equ:un}) and potential Eq. (\ref{equ:interchange}) are implemented in the same numerical way
\begin{eqnarray}
    E^{\mathrm{MC}}[n, \boldsymbol{m}] &=& 
    \sum_{\Omega} E^{\mathrm{eff}}[n,\boldsymbol{m} \cdot \boldsymbol{e}_{\Omega}] \omega_{\Omega} ,\\
    V^{\mathrm{MC}}[n, \boldsymbol{m}] &=&
    \sum_{\Omega} V^{\mathrm{eff}}[n,\boldsymbol{m} \cdot \boldsymbol{e}_{\Omega}] \omega_{\Omega},
\end{eqnarray}
with $\omega_{\Omega}$ the normalized weight of direction $\Omega$, satisfying $ \sum_{\Omega} \omega_{\Omega} = 1 $.
Three kinds of numerical strategies for solid angle distributions are tested: Lebedev quadrature\cite{Lebedev1975,lebedev1999quadrature}, Gauss-Legendre quadrature\cite{golub1969calculation}, and Fibonacci lattice\cite{gonzalez2010measurement}. All the tests in this section are performed under the default settings of PySCF, except a larger grid in real space (the default setting is level 3, we used level 5).

\begin{figure}
	\centering
	\subfigure[Errors of total energy in three numerical strategies.]{\includegraphics[width=4.2cm]{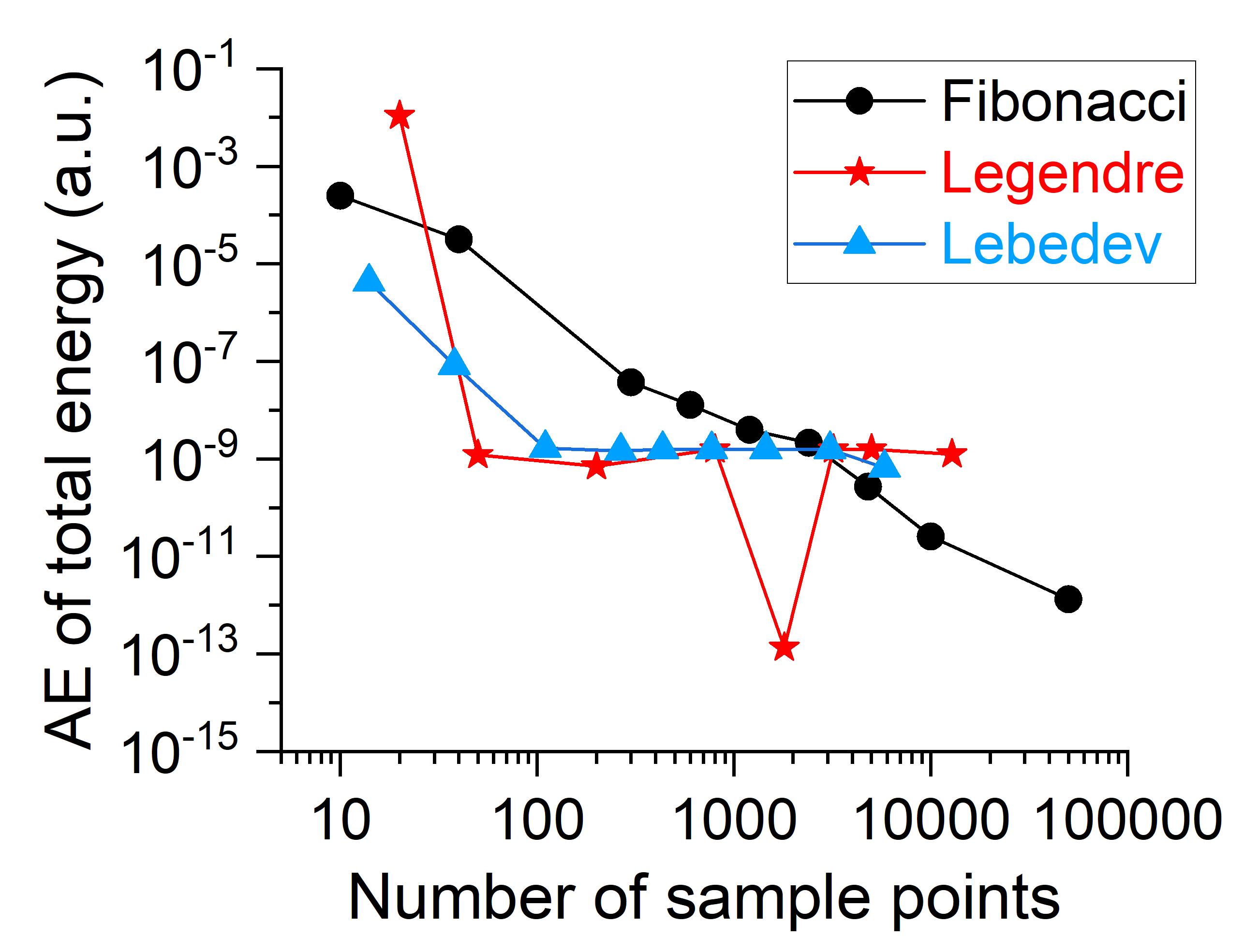}\label{fig3a}} 
	\subfigure[Errors of orbital energy in three numerical strategies.]{\includegraphics[width=4.2cm]{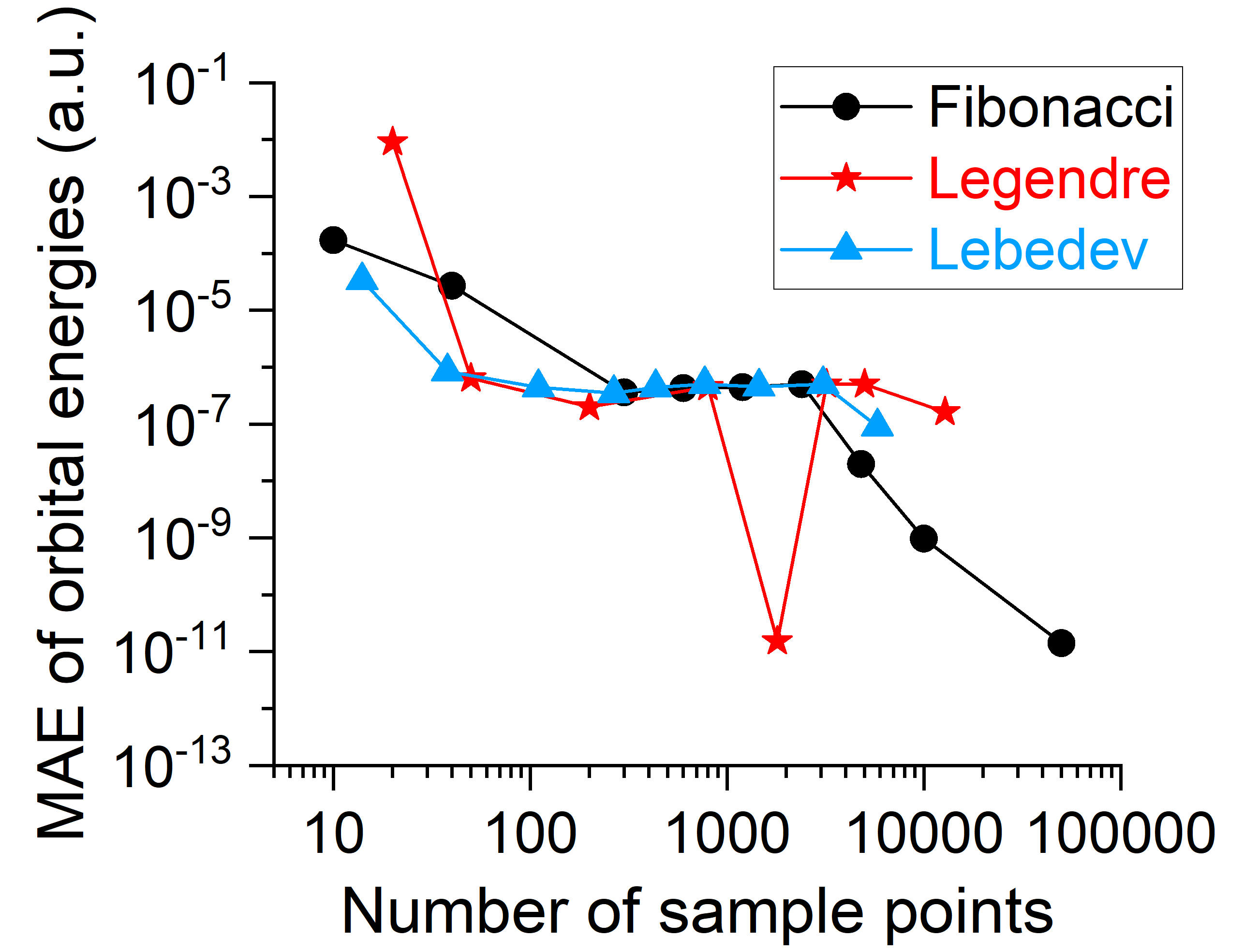}\label{fig3b}} \\
	\caption{
 Errors of the self-consistent field converged multi-collinear energy using three numerical strategies (the Lebedev quadrature, Gauss-Legendre quadrature and Fibonacci lattice) with respect to the number of solid angles in spin space. The non-collinear $\mathrm{Li_3}$ cluster ($\mathrm{D_{3h}}$ symmetry with bond length 4.0 Angstroms taken from Ref. \citenum{egidi2017two}) is tested using the SVWN5\cite{vosko1980accurate} functional and cc-pVTZ basis set\cite{balabanov2005systematically}.
	The AE (absolute error) of total energy and the MAE (mean absolute error) of orbital energies (including virtual orbitals) are plotted in (a) and (b), respectively. 
	 \label{fig3}
	}
\end{figure}

The fact that the multi-collinear and the locally collinear approaches are equivalent for LSDA functionals, as mentioned in Section \ref{sec:LSDA}, allows us to use the locally collinear LSDA as the benchmark to test the errors of the multi-collinear approach with respect to the solid angle distributions.
As plotted in FIG. \ref{fig3}, the Lebedev quadrature and Gauss-Legendre quadrature reach the accuracy of $10^{-9}$ a.u. in energies (the same magnitude as the energy convergence threshold in the self-consistent field iterations) and $10^{-6}$ a.u. in orbital energies with about 1000 sample points. This accuracy, which can be further improved with the energy convergence threshold decreasing, is enough for regular calculations, and the Lebedev quadrature is used in our following calculations.

\begin{figure}
	\centering
    \subfigure[Changes of energy under global spin rotations for doublet $\mathrm{H_2O^+}$ using SVWN5.]{\includegraphics[width=4.2cm]{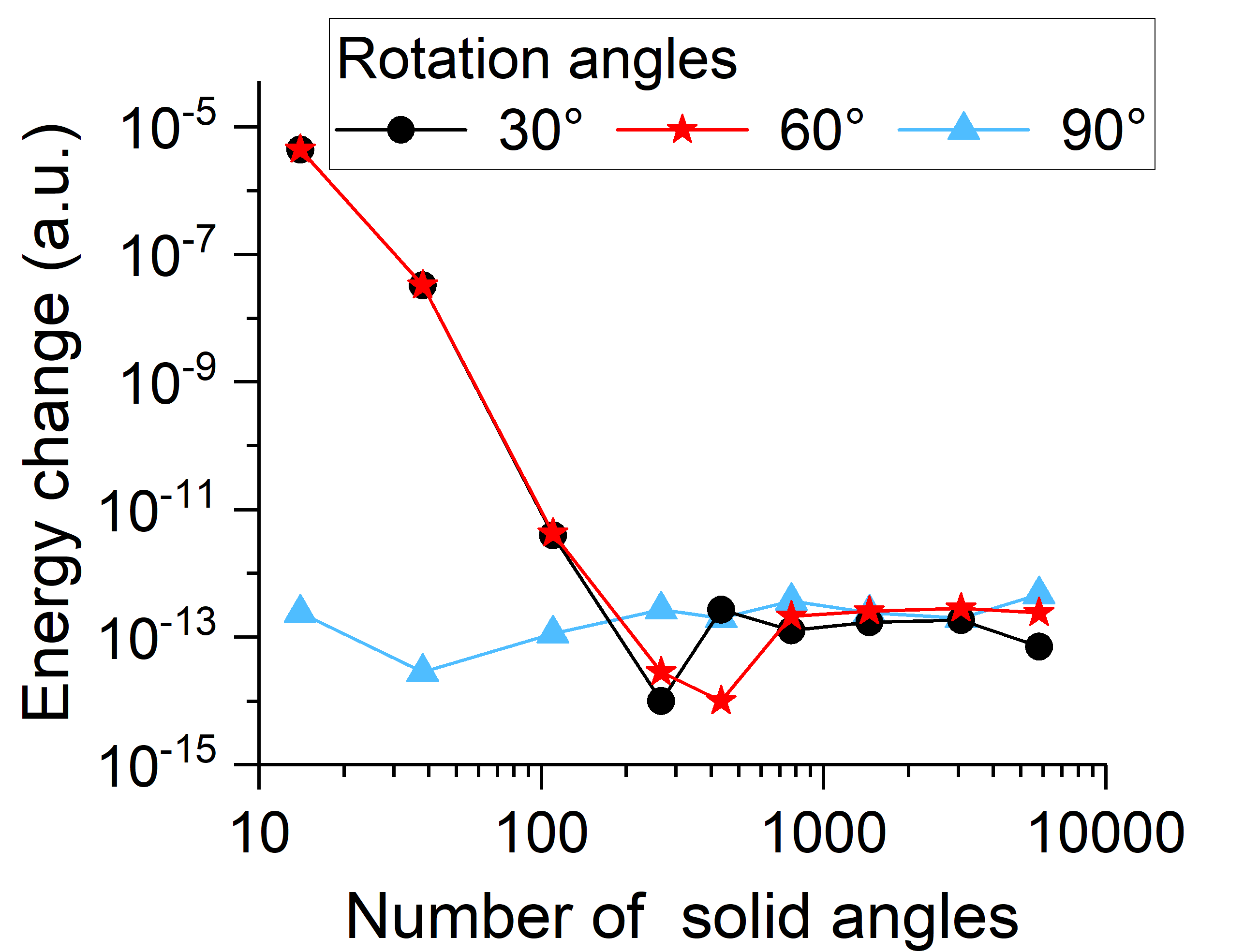}\label{fig4a}}
	\subfigure[Changes of energy under global spin rotations for doublet $\mathrm{H_2O^+}$ using TPSS.]{\includegraphics[width=4.2cm]{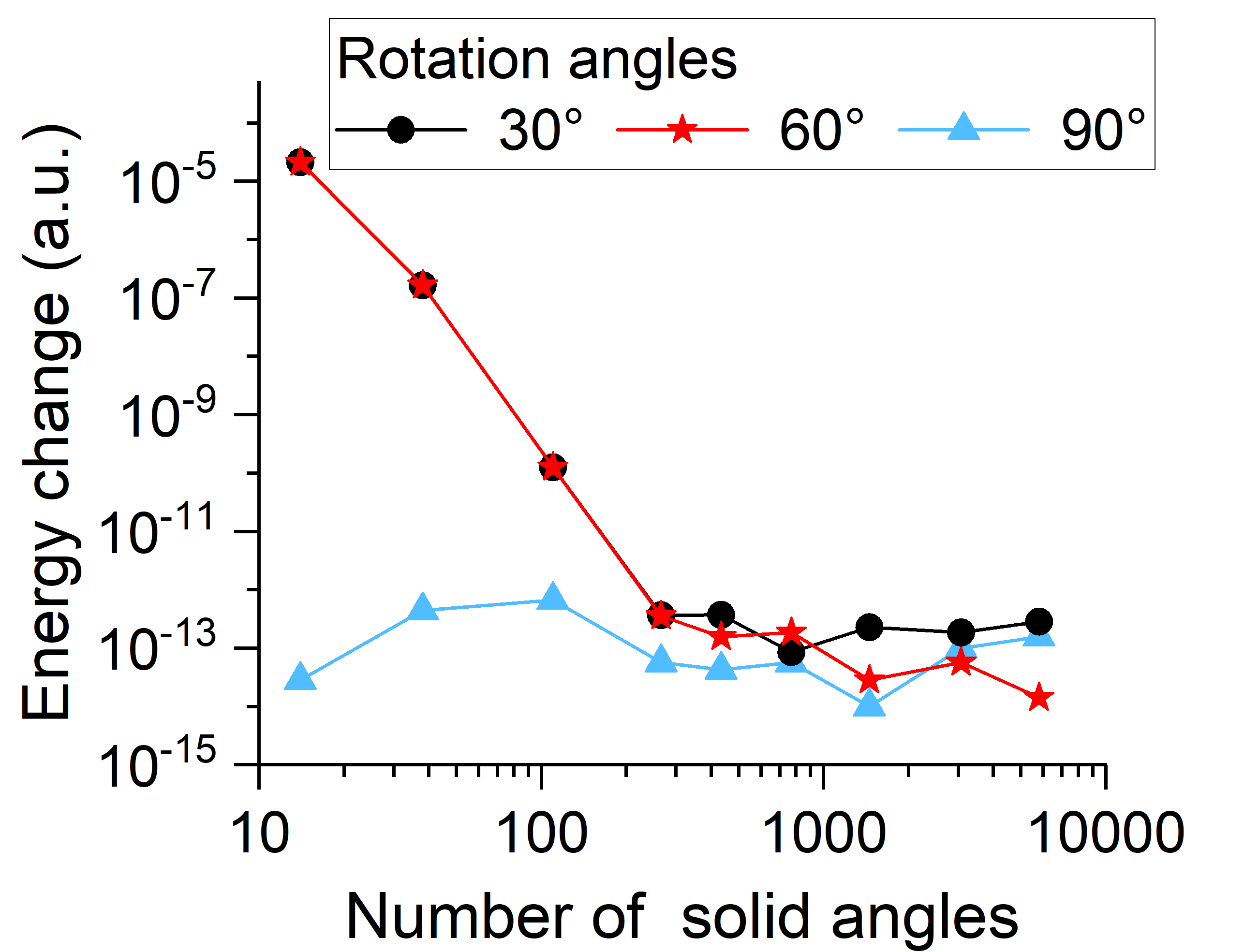}\label{fig4b}}\\
	\caption{Changes of multi-collinear energy under different global spin rotations for a collinear spin system. 
    The doublet $\mathrm{H_2O^+}$ is tested using cc-pVTZ basis set, with $\mathrm{C_{2v}}$ symmetry, bond length 0.9584 Angstroms and bond angle 104.45 degrees, taken from Ref. \citenum{lemberg1975central}. 
    Changes of energy under global spin rotations around the $x$-axis for SVWN5 and TPSS are plotted in (a) and (b), respectively.
    \label{fig4}
	}
\end{figure}

\begin{figure*}
	\centering
	\subfigure[Non-collinear $ \mathrm{Li}_3 $ cluster.]{\includegraphics[width=5.0cm]{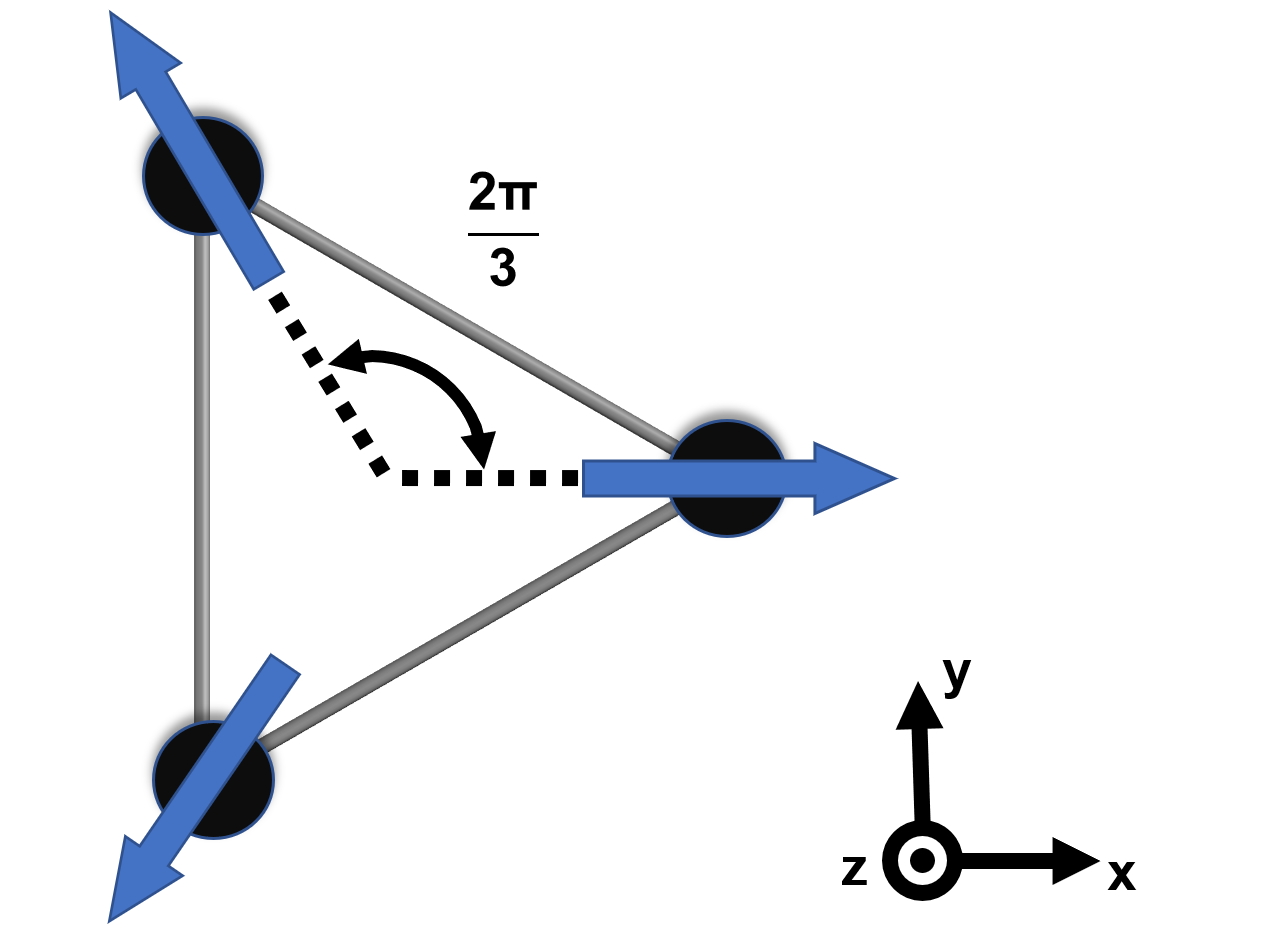}\label{fig5a}} \subfigure[Changes of energy under global spin rotations around $z$-axis.]{\includegraphics[width=5.0cm]{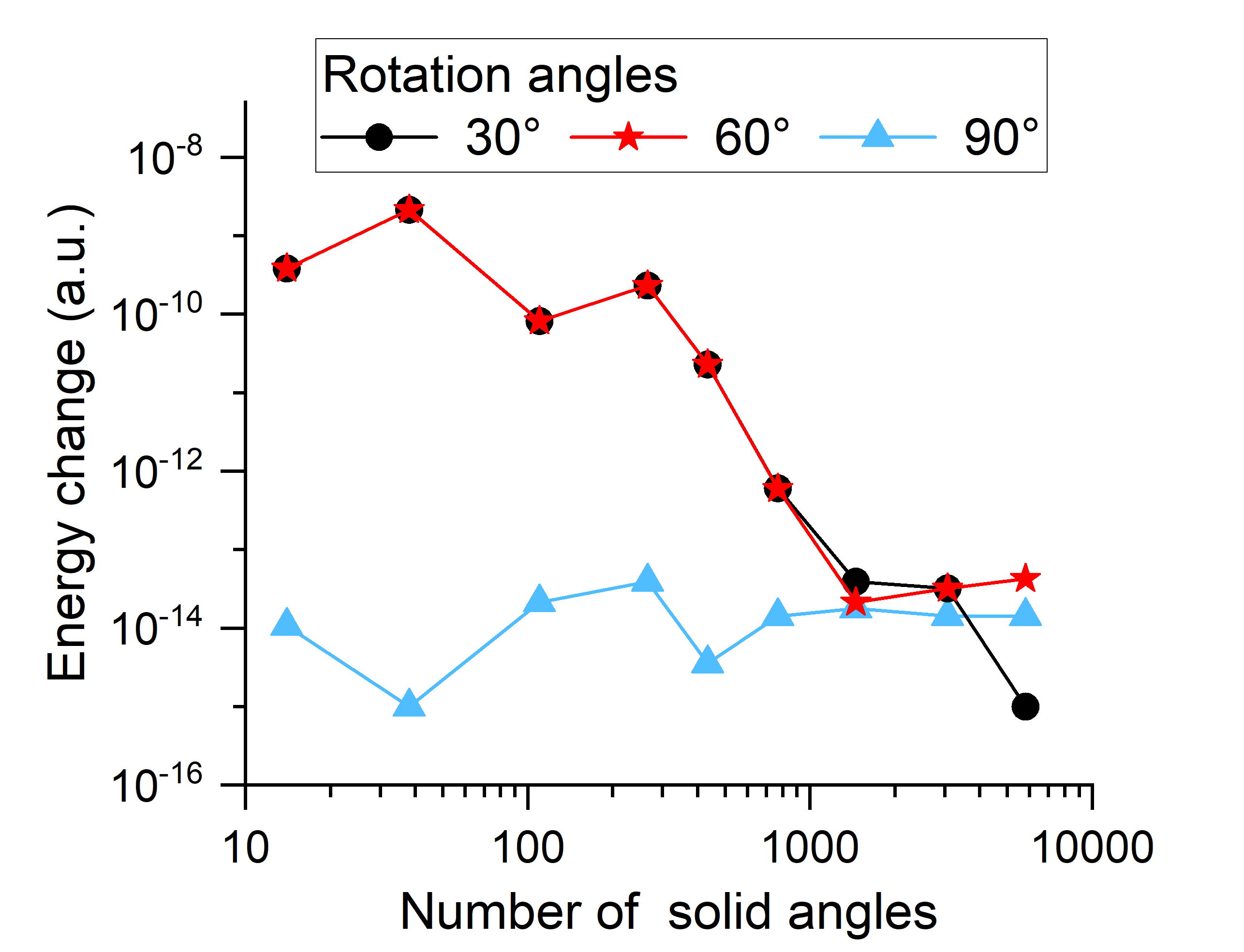}\label{fig5b}}
	\subfigure[Changes of energy under global spin rotations around $x$-axis.]{\includegraphics[width=5.0cm]{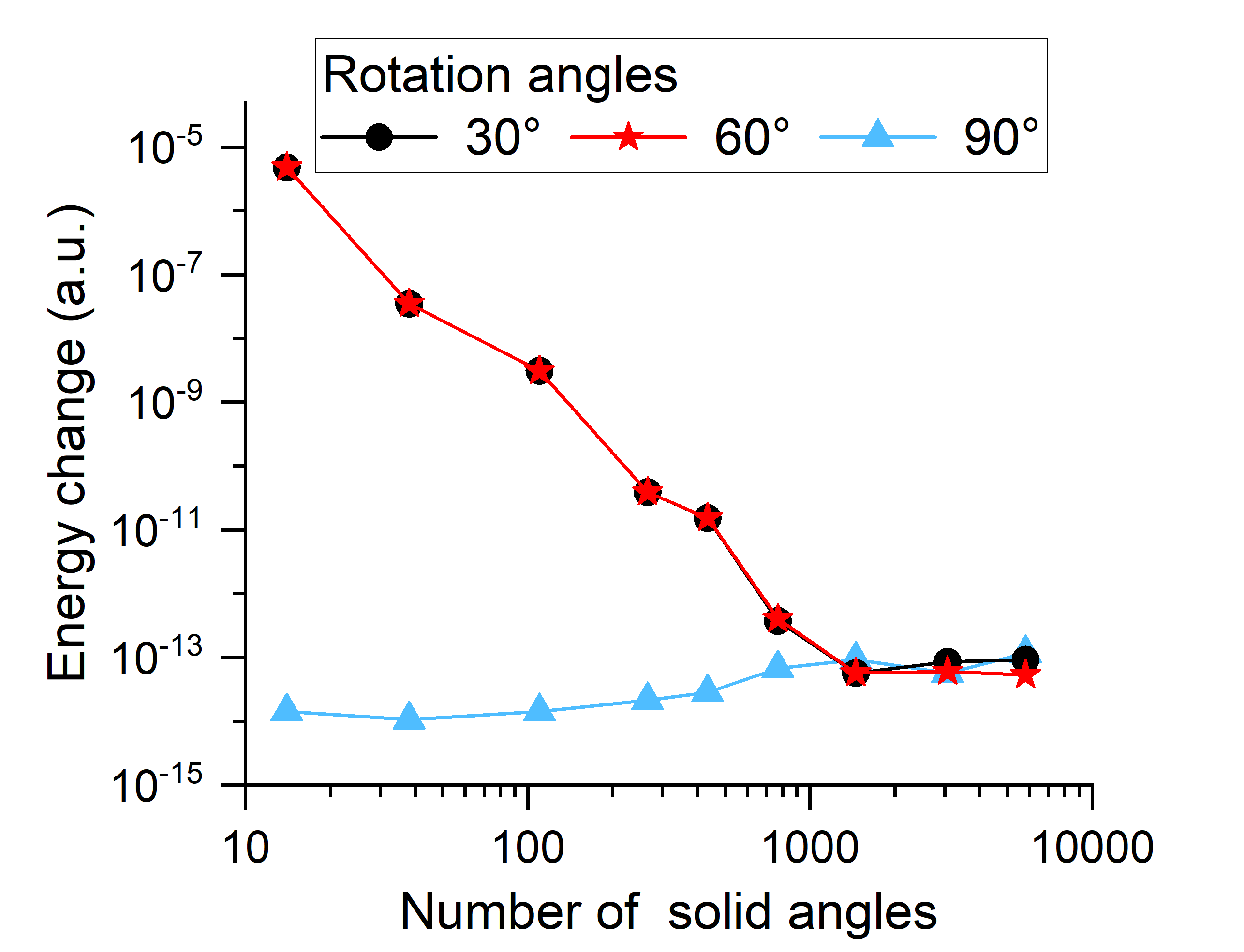}\label{fig5c}}
	\caption{Changes of multi-collinear energy under different global spin rotations for a non-collinear spin system. 
   The $ \mathrm{Li}_3 $ cluster is tested using cc-pVTZ basis set and PBE\cite{perdew1996generalized} functional. 
    (a) The geometry of $ \mathrm{Li}_3 $ cluster has $\mathrm{D_{3h}}$ symmetry with bond length 4.0 Angstroms.
    The arrows represent the magnetization orientation on each atom, exhibiting a non-collinear spin structure. 
    (b) Changes of energy under global spin rotations around $z$-axis. 
    (c) Changes of energy under global spin rotations around $x$-axis.
    \label{fig5}
	}
\end{figure*}

\subsection{Numerical tests for global spin rotation invariance}

Theoretically, the multi-collinear energy is invariant under any global spin rotation. 
However, in our implementation, the average over all directions in spin space is evaluated numerically
using the Lebedev quadrature, which only preserves the $\mathrm{O_h}$ symmetry and  breaks the invariance numerically. 
It is expected that this symmetry breaking will ease with the increase in the number of solid angles.

To see the degree of breaking of invariance, we test a collinear spin system and a non-collinear spin system, by rotating their initial guesses in spin space. The self-consistent field converged energies are shown in FIG. \ref{fig4} and FIG. \ref{fig5}. 
In general, the global spin rotation invariance is preserved satisfactorily, reaching the accuracy of $10^{-12}$ a.u. in energies with about 1000 sample points.

\subsection{Numerical tests for the collinear spin system}

The fact that for collinear spin systems, the multi-collinear approach and the traditional collinear approach share the same energy functional (thanks to the correct collinear limit) and the same potential (shown in Section \ref{sec:potential}), indicates that they provide the same total energy in self-consistent field calculations. Considering that their equivalence holds for arbitrary geometries, their forces are also the same. 

\begin{figure}
	\centering
	\subfigure[Errors of total energy for the doublet $\mathrm{H_2O}^{+}$ cation.]{\includegraphics[width=4.2cm]{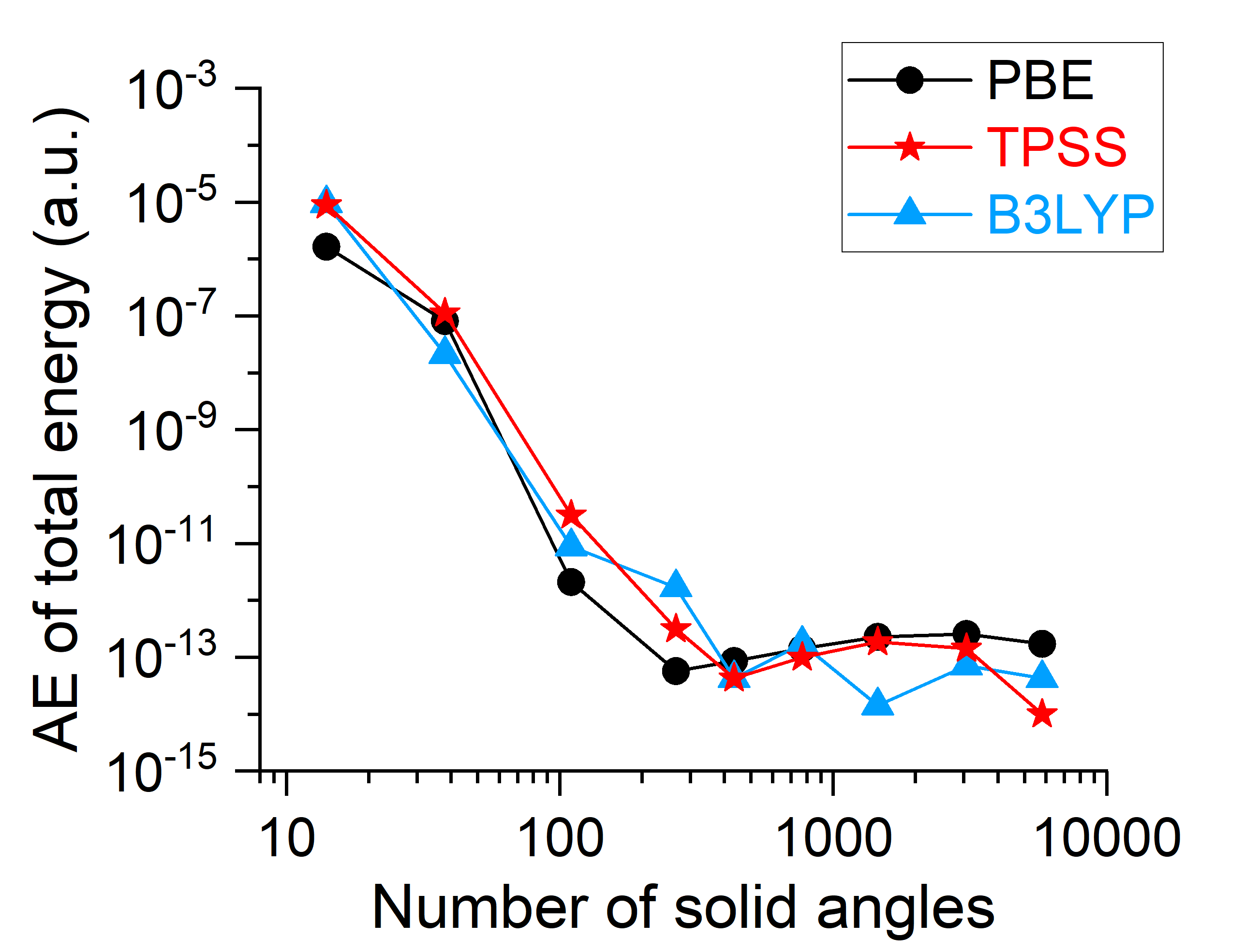}\label{fig6a}} 
	\subfigure[Errors of total energy for the triplet $\mathrm{O_2}$ molecule.]{\includegraphics[width=4.2cm]{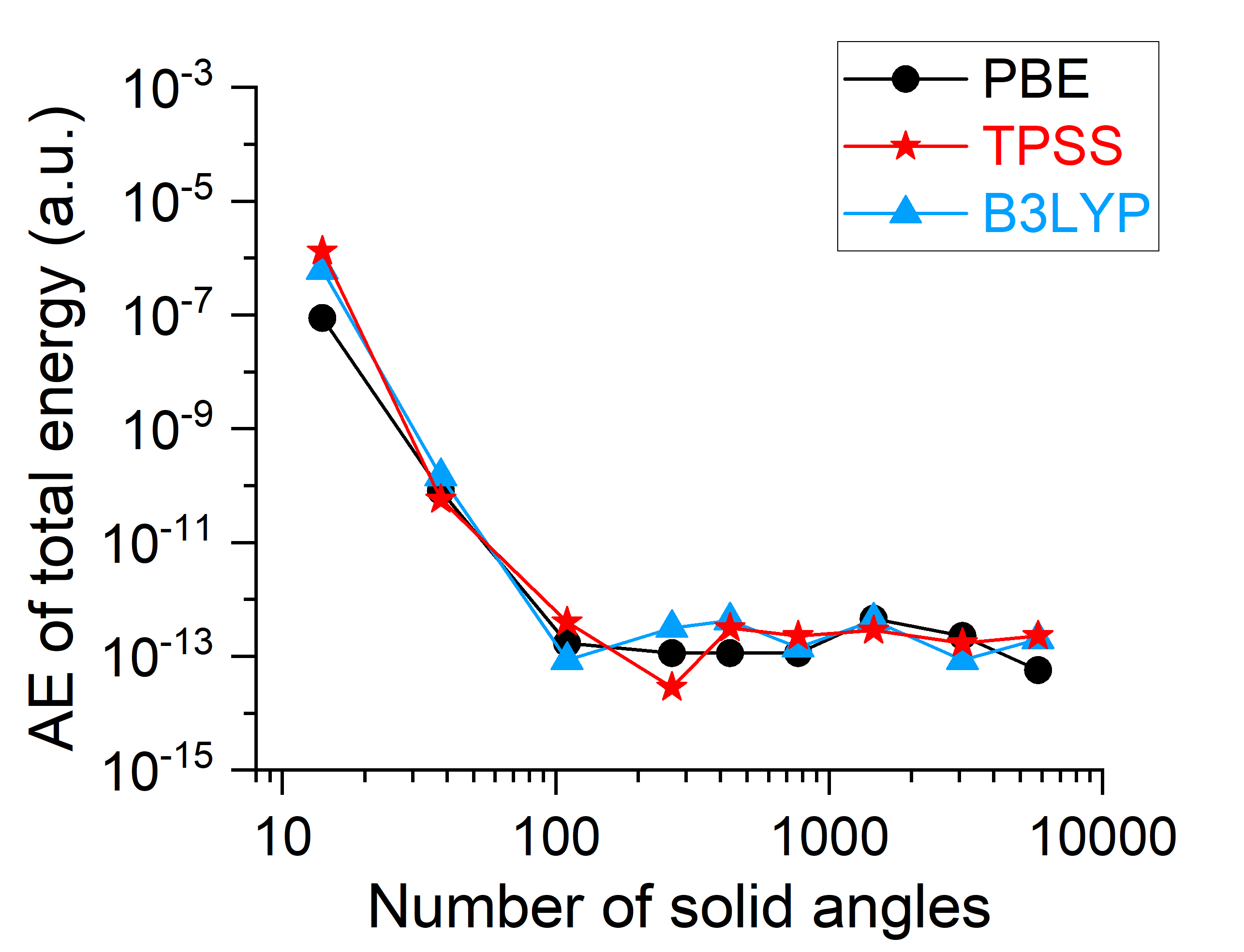}\label{fig6b}}   \\

    \subfigure[Errors of forces for the doublet $\mathrm{H_2O}^{+}$ cation.]{\includegraphics[width=4.2cm]{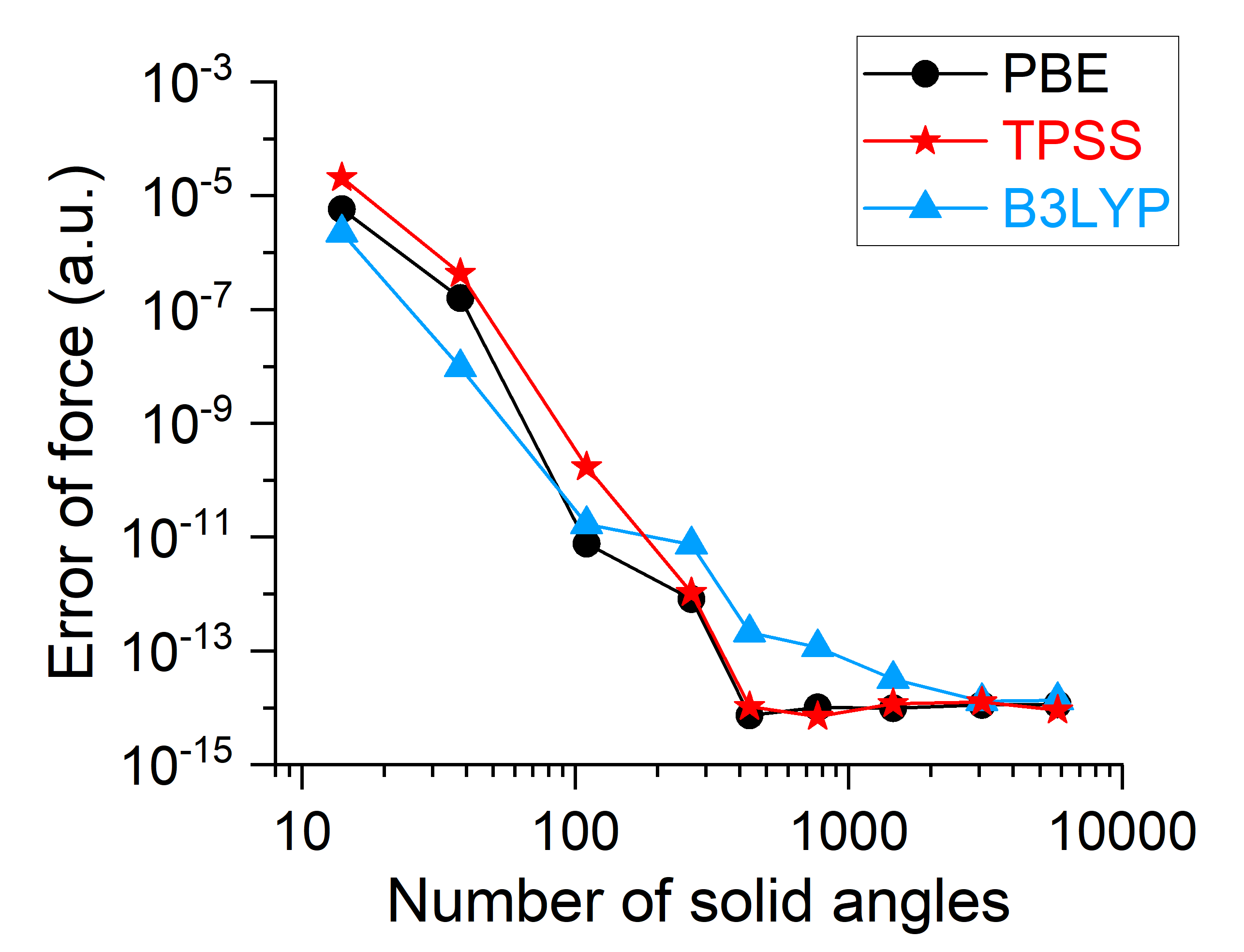}\label{fig6c}} 
	\subfigure[Errors of forces for the triplet $\mathrm{O_2}$ molecule.]{\includegraphics[width=4.2cm]{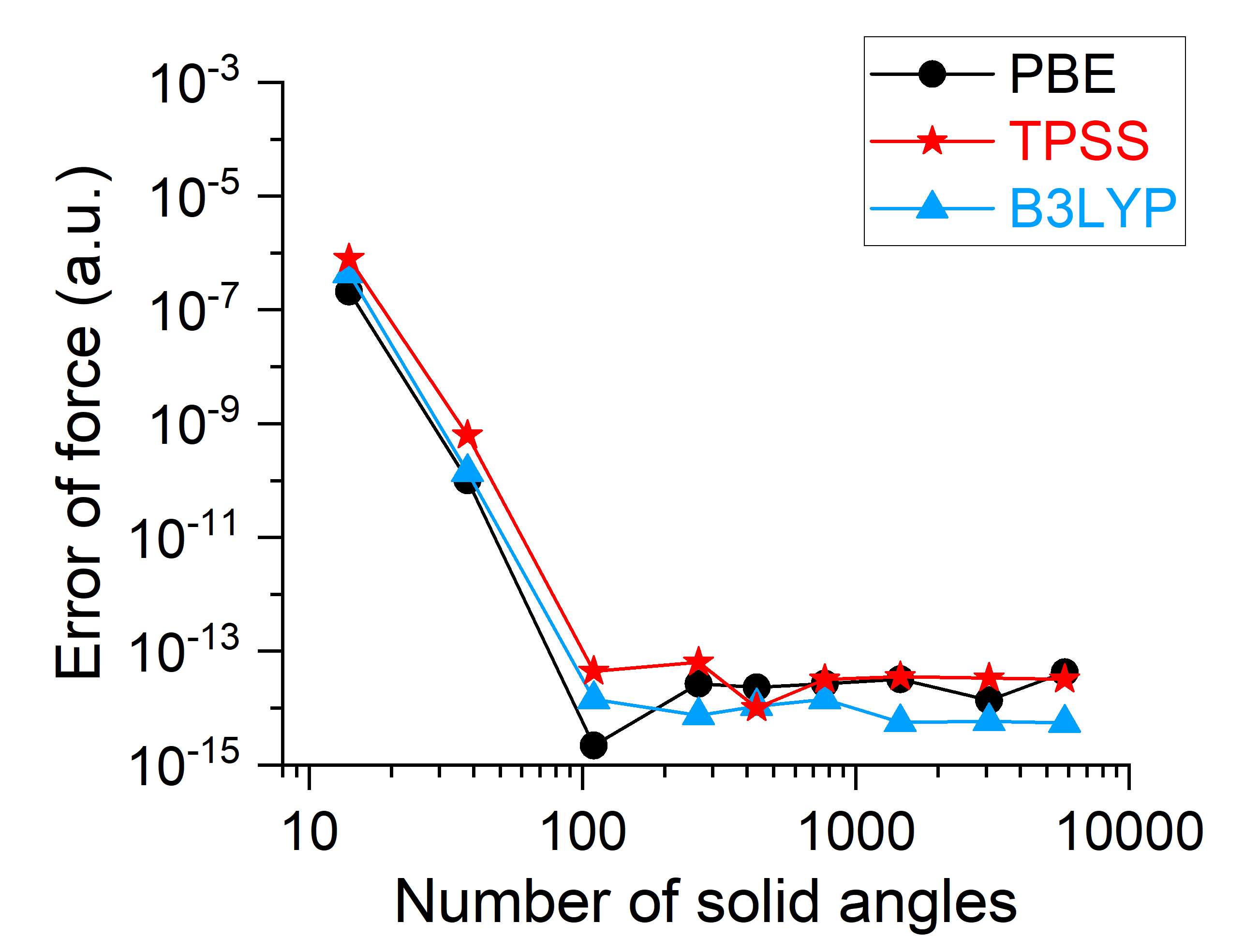}\label{fig6d}} \\
	\caption{
    The errors of multi-collinear energy and forces for collinear spins with respect to the number of solid angles in spin space. The doublet $\mathrm{H_2O}^{+}$ cation with the same geometry as in FIG. \ref{fig4}, and the triplet $\mathrm{O_2}$ molecule with bond length 1.452 Angstroms, are tested.
    Basis set cc-pVTZ and three kinds of functionals, GGA (PBE), meta-GGA (TPSS\cite{tao2003climbing}) and hybrid functional (B3LYP\cite{becke1988density,stephens1994ab}), are used.
    The error of forces is calculated via $ \sum_{K} | \Delta \boldsymbol{F} (K) | $, with $\Delta \boldsymbol{F} (K)$ the error of the force for atom $K$.
    \label{fig6}
	}
\end{figure}

Thus, the traditional collinear approach can be used as the benchmark to test the numerical errors of the energy and forces in the multi-collinear approach for collinear spins.
The doublet $\mathrm{H_2O}^{+}$ cation and triplet $\mathrm{O_2}$ molecule are tested, and the results are shown in FIG. \ref{fig6}, generally reaching the accuracy of $10^{-12}$ a.u. in both energies and forces with about 1000 solid angles in spin space. 

\subsection{Computational time}

In practice, not only the accuracy but also the computational cost is concerned. 
At first glance, the computational cost of the multi-collinear approach, which treats $N$ collinear spin states with $N$ the number of projection directions $\Omega$, seems to be roughly $N$ times the locally collinear approach. 
However, this '$N$ times' computational cost is only for functional or functional derivatives, but independent of the basis set. Thus, the extra cost is a \emph{constant} on each spatial grid and can be ignored for large systems.

\begin{figure}
	\centering
	\includegraphics[width=7.5cm]{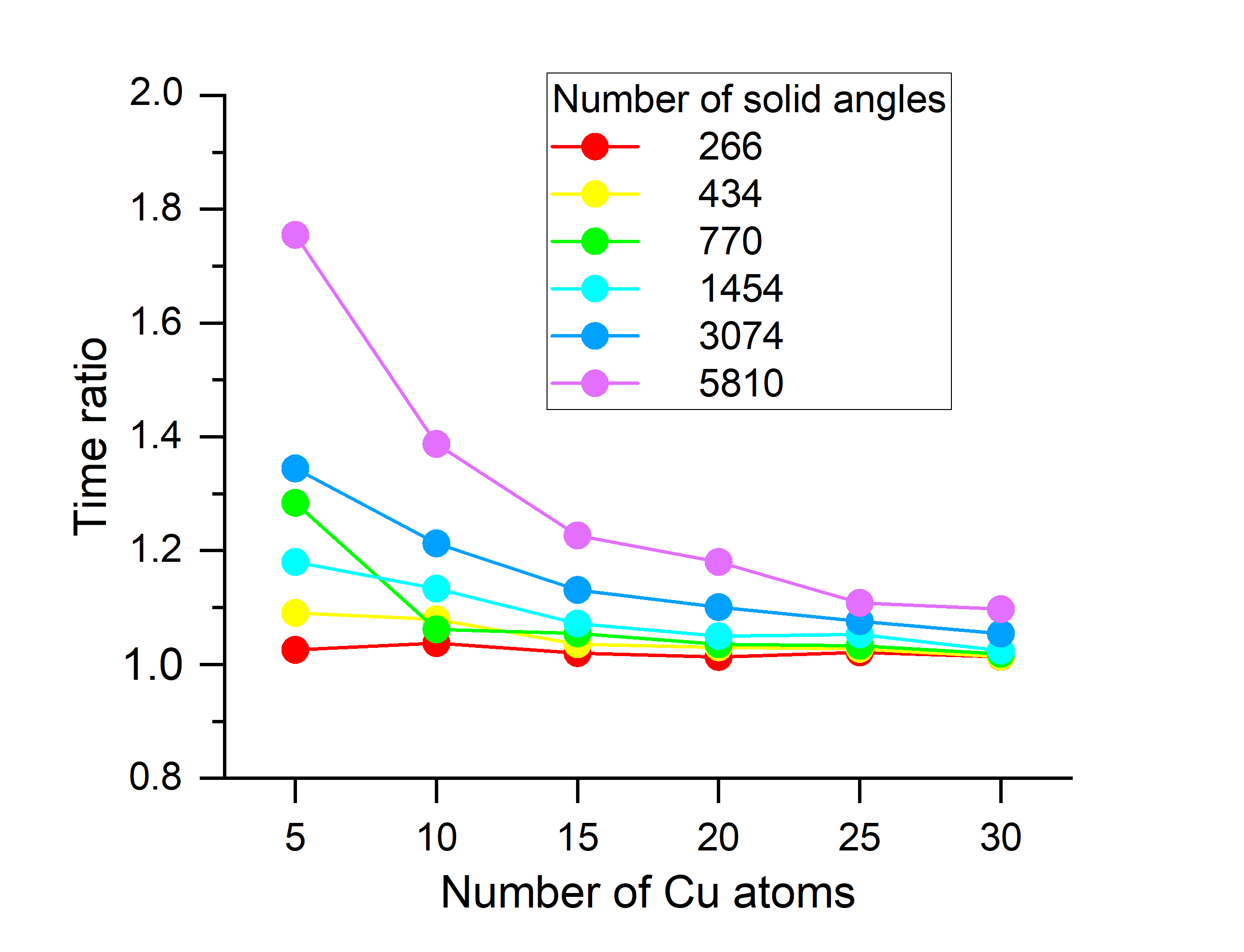}
	\caption{
    The computational time ratio of the multi-collinear approach to the locally collinear approach for Cu chains. One step of self-consistent field calculations is carried out for Cu chains with the interval 5 Angstroms using aug-cc-pVTZ\cite{balabanov2005systematically} basis set, in a moderate computational condition (one computer node with 40 CPU cores and 64G memory with parallelization applied).
    \label{fig7}
}
\end{figure}

Numerical tests are shown in FIG. \ref{fig7}, in which one-dimensional Cu chains with different numbers of Cu atoms are tested. The time ratio of the multi-collinear approach to the locally collinear approach, for one step of self-consistent field calculation using LSDA functional, is plotted. 
The ratios, depending on the number of solid angles, are less than 2 for the chain with five atoms, and decrease with the increase in the number of Cu atoms in the chain, less than 1.2 for Cu chain with more than twenty atoms.

\section{Applications}\label{sec:applications}

Only DFT results will be discussed here, although both the multi-collinear DFT and TDDFT are available in PySCF official version 2.1, supporting LSDA, GGA, meta-GGA and hybrid functionals.
Before the discussions of results, some general remarks on the multi-collinear approach are made.
\begin{enumerate}
    \item The well-definition of functionals and functional derivatives allows the multi-collinear approach to provide a unique and numerical stable result, which does not happen in the locally collinear approach due to the lack of commonly accepted treatment for numerical singular terms \cite{peralta2007noncollinear,desmarais2021spin,komorovsky2019four,bast2009relativistic,egidi2017two,komorovsky2015four,pu2022approach,li2012theoretical,sjostedt2002noncollinear,kurz2004ab} (this numerical instability is not obvious in DFT but severe in TDDFT calculations). 
    Another advantage is providing non-vanishing local torque. To show these two advantages, an application on the $\mathrm{Cr}_3$ cluster is given in Section \ref{sec:Cr3}. 
    \item 
    As a new approach, its universality needs to be tested. 
    Two examples are given, one for the combination with periodic boundary conditions ($\mathrm{Cr}$ monolayer in Section \ref{sec:Cr_mono}) and the other for spin-orbit couplings ($\mathrm{Dy}_3$ cluster using the Dirac-Coulomb Hamiltonian in Section \ref{sec:Dy3}). 
    \item 
    For the currently widely-used collinear functionals, the multi-collinear approach is \emph{not} guaranteed to provide more accurate results than the locally collinear approach, partially because of errors rooted in collinear functionals.
    The quantitative results will be provided only for comparisons, but not to show that the multi-collinear approach is more accurate statistically.
    However, thanks to the correct collinear limit for all kinds of functionals, its accuracy will increase systematically as the accuracy of the collinear functionals increases, which does not happen for approaches without the correct collinear limit.
\end{enumerate}

\subsection{$\boldsymbol{\mathrm{{Cr}_3}}$ cluster \label{sec:Cr3}}

\begin{figure}
	\centering
	\subfigure[$ \boldsymbol{m} $.]{\includegraphics[width=4.2cm]{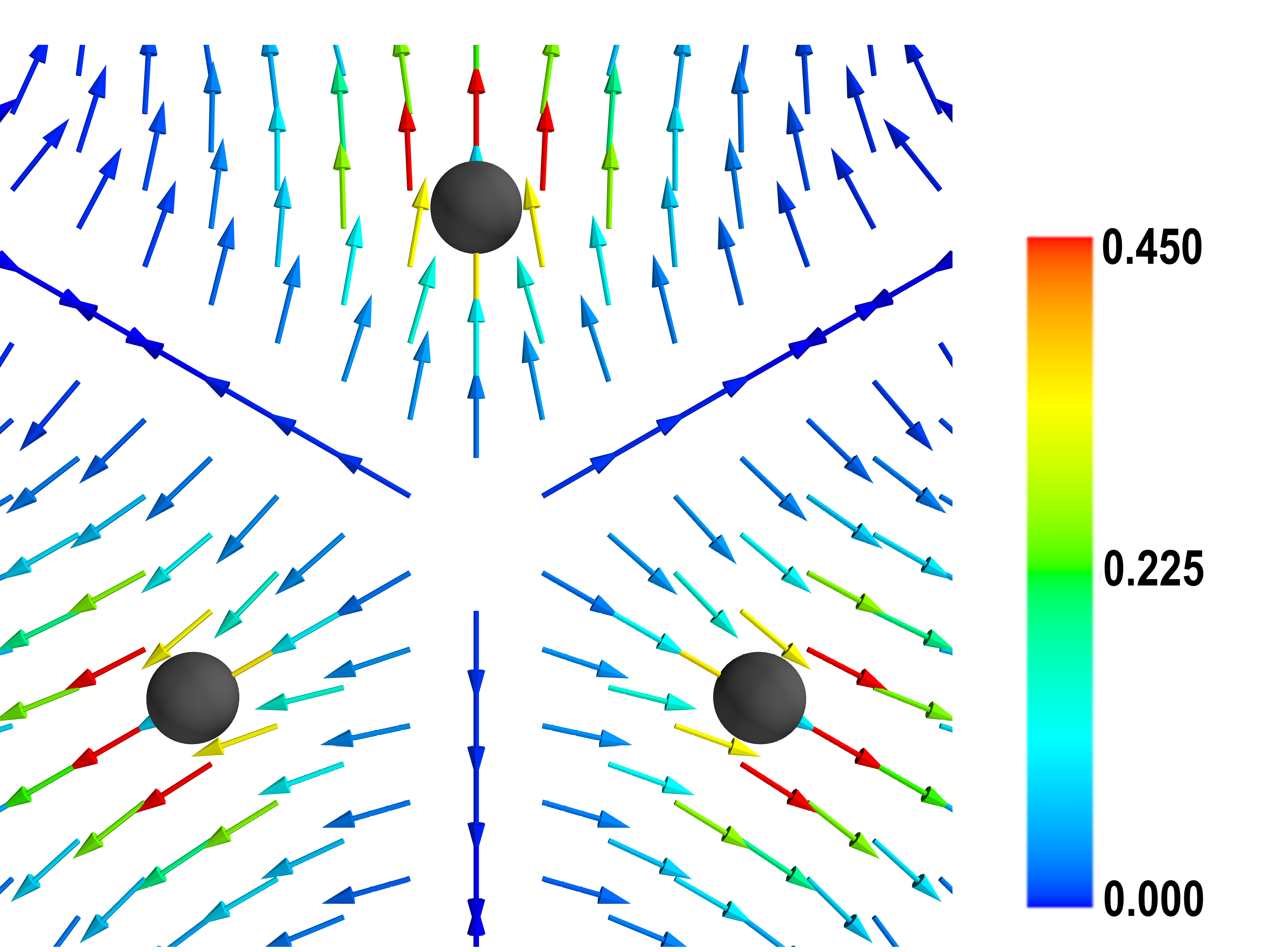}\label{fig8a}}
	\subfigure[$ \boldsymbol{B}^{\mathrm{xc}} $.]{\includegraphics[width=4.2cm]{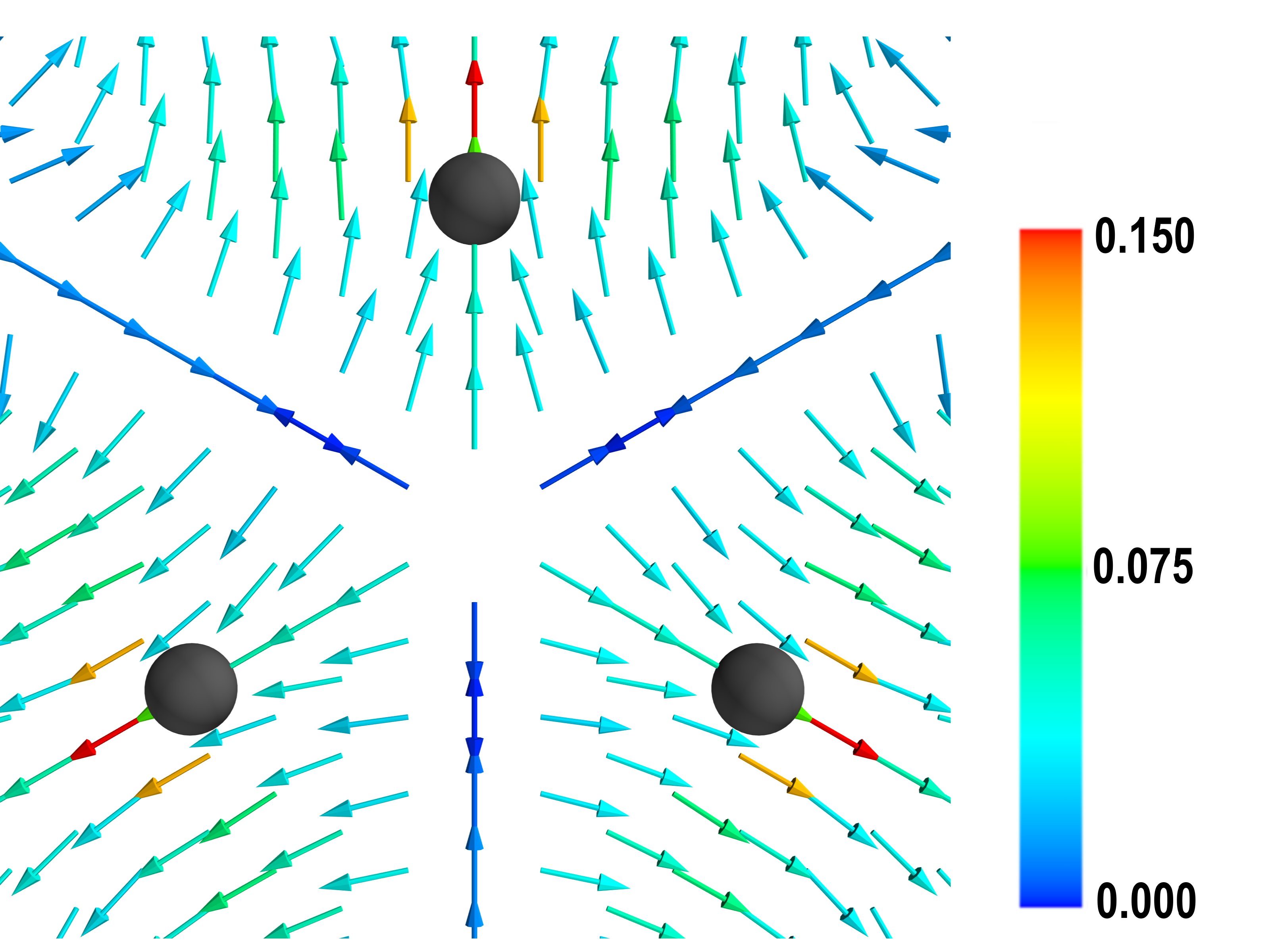}\label{fig8b}} \\
	
	\subfigure[Torque.]{\includegraphics[width=4.2cm]{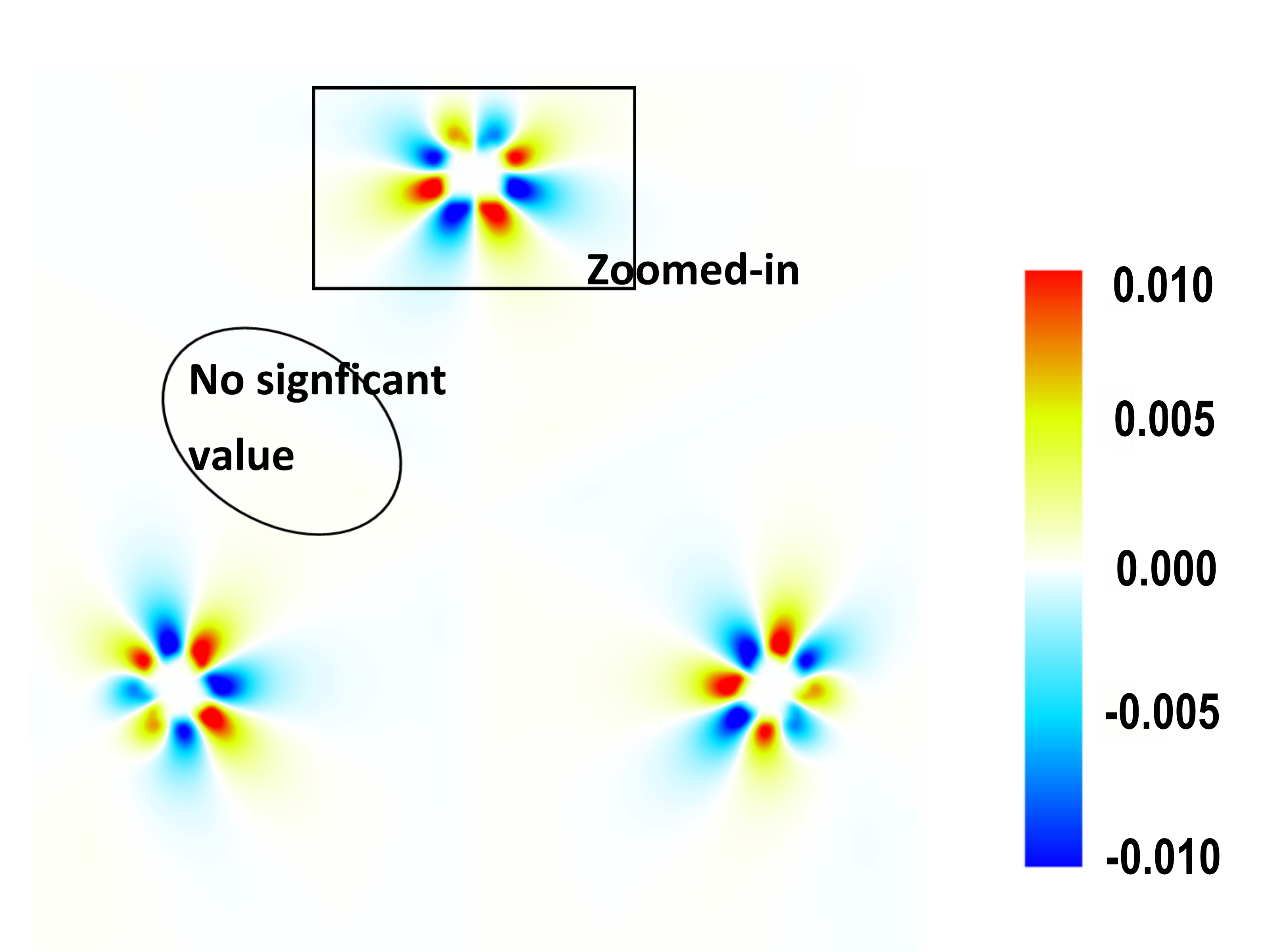}\label{fig8c}}
	\subfigure[Enlarged view of FIG. (c).]{\includegraphics[width=4.2cm]{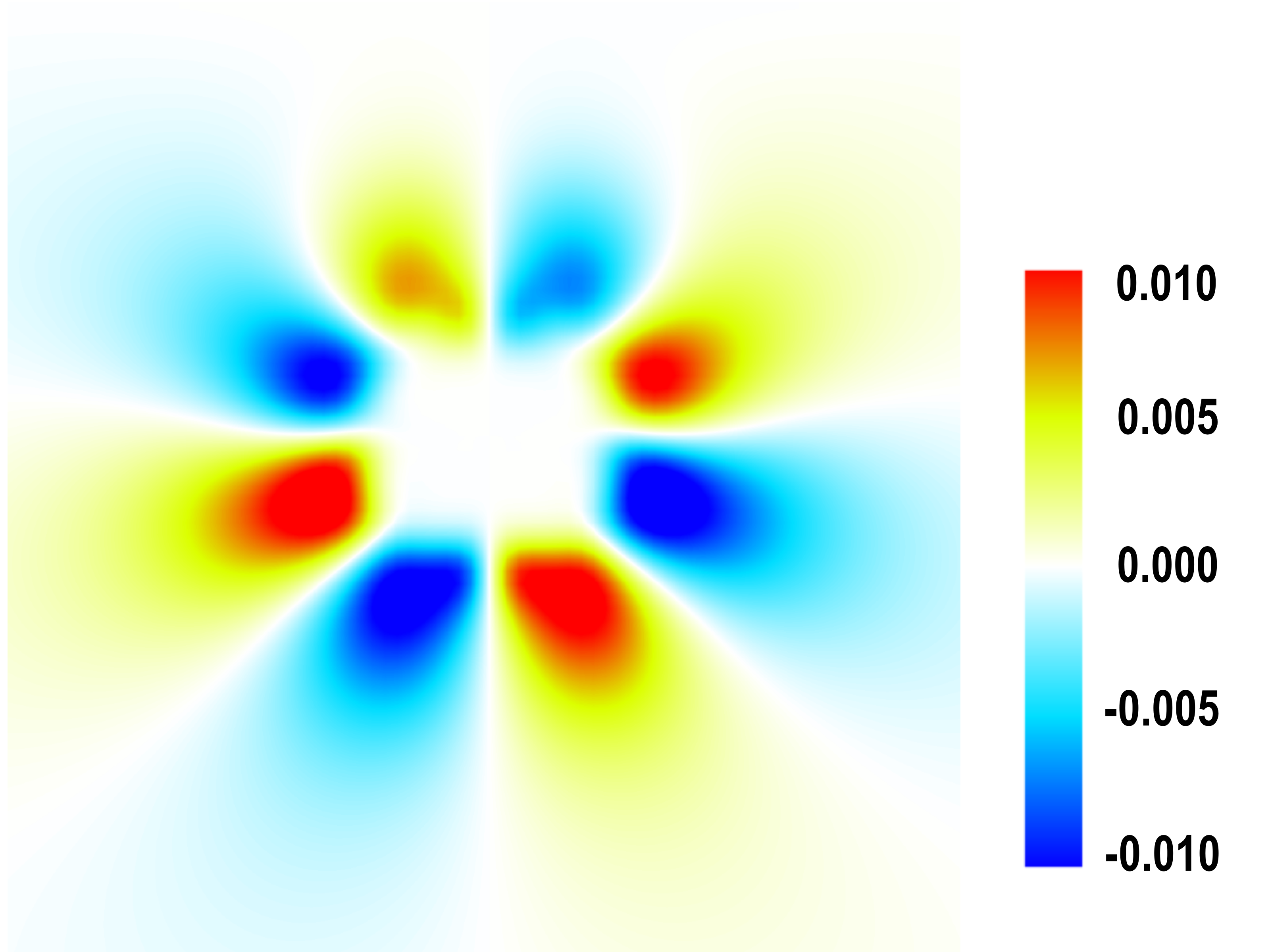}\label{fig8d}}
	\caption{
		Calculation results of $\mathrm{Cr_3}$ using multi-collinear PBE functional, plotted by Mayavi\cite{ramachandran2011mayavi} and Matplotlib \cite{Hunter:2007}.
		(a) $ \boldsymbol{m} $ in the plane of $ \mathrm{Cr_3} $. 
		(b) $ \boldsymbol{B}^{\mathrm{xc}} $ in the plane of $ \mathrm{Cr_3} $.
		(c) The perpendicular component of torque (torque is perpendicular to the $ \mathrm{Cr}_3 $ cluster) in the plane of $ \mathrm{Cr_3} $.
		(d) The enlarged view of the torque around a single atom in FIG. (c). 
		\label{fig8}
	}
\end{figure}

The generalized Kohn-Sham calculation in the multi-collinear approach is applied on $ \mathrm{Cr_3} $ cluster.
The bond length of the $ \mathrm{Cr_3} $ cluster with $ \mathrm{D_{3h}} $ symmetry is $ 3.7000 $ Bohr, the same as in Refs.\citenum{scalmani2012new} and \citenum{peralta2007noncollinear}, for the convenience of comparing. The scalar relativistic effective core potential by Dolg \textit{et al.}\cite{wedig1986energy} is used.
Calculated results using PBE functional are displayed in FIG. \ref{fig8}.

The spin magnetization vector $ \boldsymbol{m}(\boldsymbol{r}) $, shown in FIG. \ref{fig8a}, is mainly around atoms and displays $ \mathrm{D_3} $ symmetry, presenting a similar pattern observed by Peralta \textit{et al.}\cite{peralta2007noncollinear} using PBE functional in the locally collinear approach, where some specific treatments were applied to avoid numerical singularities. For generalized Hartree-Fock (GHF) and more functionals, atomic $\boldsymbol{m}$ using the Mulliken population analysis and the expectation value of the square of the spin operator $\langle S^2 \rangle$ are reported in TABLE \ref{tab:Cr3_S2}, compared with results reported by Peralta \textit{et al.}\cite{peralta2007noncollinear}. 
The GHF results match perfectly with those reported in Ref. \citenum{peralta2007noncollinear} (except for a factor 2 for the atomic magnetic moment, probably caused by whether including the spin-$g$ factor in the definition).
A perfect match is also observed for SVWN5 functional. It is expected because of the equivalence of the multi-collinear approach and the locally collinear approach for LSDA functionals.
For semi-local functionals (PBE, TPSS and PBE0\cite{adamo1999toward}), two approaches provide different but similar results. 

FIG. \ref{fig8b} shows the exchange-correlation magnetic field $ \boldsymbol{B}^{\mathrm{xc}}(\boldsymbol{r}) $, directions deviating from $ \boldsymbol{m}(\boldsymbol{r}) $, allowing non-vanishing local torque. 
The calculations of local torque in the multi-collinear approach are not compared with results of the locally collinear approach, in which local torque is zero, but compared with the results obtained by Scalmani and Frisch in Ref. \citenum{scalmani2012new} using their modified locally collinear approach.
As shown in FIG. \ref{fig8c}, the local torque is perpendicular to the plane of the cluster and centered around $ \mathrm{Cr} $ atoms. As zoomed in in FIG. \ref{fig8d}, it presents a pattern of eight alternating up and down petals, similar to the observation by Scalmani and Frisch\cite{scalmani2012new}. 
However, we do not find significant torque at areas between any two atoms, observed by Scalmani and Frisch\cite{scalmani2012new}.

\begin{table}[H]
  \centering
  \caption{The atomic $\boldsymbol{m}$ and expectation value of the square of spin operator $\langle S^2 \rangle$ (in atomic unit) of $ \mathrm{Cr_3} $ using multi-collinear functionals and locally collinear functionals reported by Peralta \textit{et al.}\cite{peralta2007noncollinear} (in parentheses).} 
    \begin{tabular}{cccccc}
    \toprule
                            & \multicolumn{1}{l}{GHF} & \multicolumn{1}{l}{SVWN5} & \multicolumn{1}{l}{PBE} & \multicolumn{1}{l}{TPSS}  & \multicolumn{1}{l}{PBE0}  \\
    \midrule
    $m$        & 5.90(2.95)              & 2.88(1.44)                & 3.43(1.66)              & 3.95(1.93)                & 4.89(2.40)                \\
    $\langle S^2 \rangle$   & 8.11(8.11)              & 3.25(3.25)                & 3.89(3.87)              & 4.70(4.71)                & 6.36(6.32)                \\
    \bottomrule
    \end{tabular}
  \label{tab:Cr3_S2}
\end{table}

\subsection{Triangular Cr monolayer \label{sec:Cr_mono}}

To show that the multi-collinear approach can be used on systems with the periodic boundary conditions, an application on triangular Cr monolayer is given, with geometry adopted from the Ag(111) surface\cite{owen1933xli} and $46\times46\times1$ k-points, following Ref. \citenum{bulik2013noncollinear} for comparison. The scalar relativistic effective core potential by Dolg \textit{et al.}\cite{wedig1986energy} is used, with exponents below 0.095 $\mathrm{Bohr}^{-2}$ removed. Density fitting with Gaussian functions is used for the Coulomb interaction, with exponents expanded in an even-tempered manner.

\begin{figure} 
	\subfigure[$ \boldsymbol{m} $ of each $ \mathrm{Cr} $ atom.]{\includegraphics[width=4.2cm]{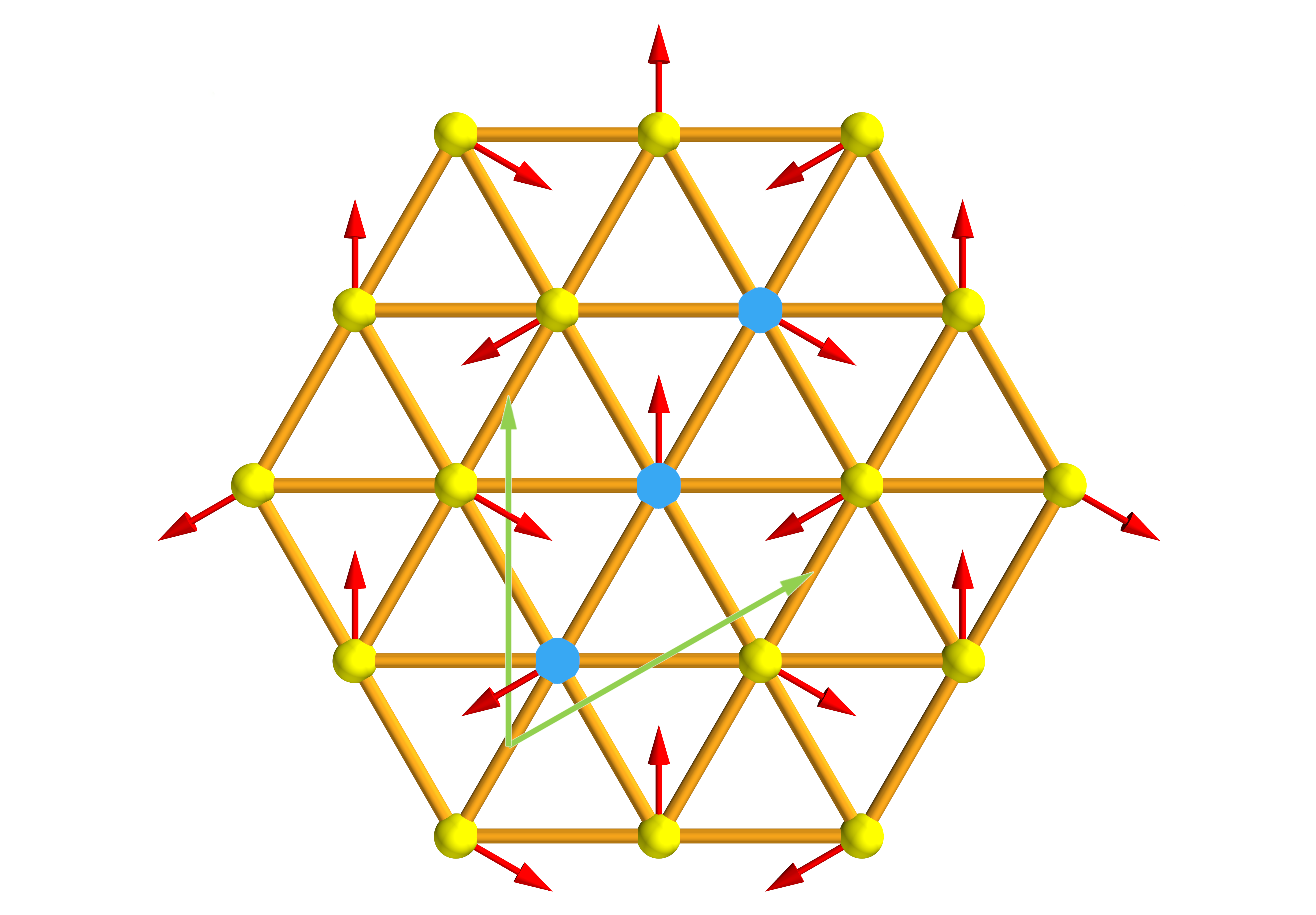}\label{fig9a}}
	\subfigure[$ \boldsymbol{m} $.
	]{\includegraphics[width=4.2cm]{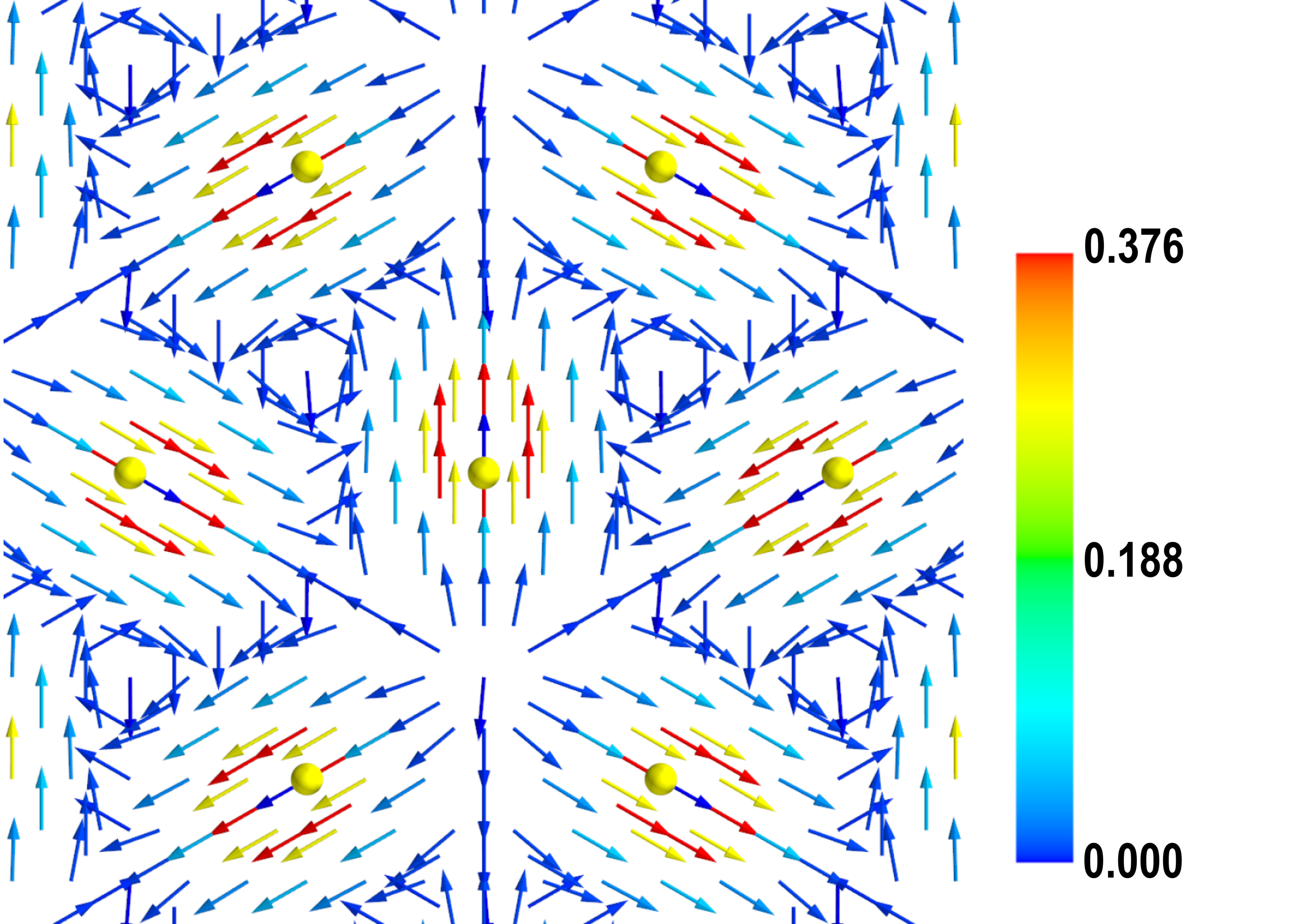}\label{fig9b}} \\
	
	\subfigure[$ \boldsymbol{B}^{\mathrm{xc}} $. ]{\includegraphics[width=4.2cm]{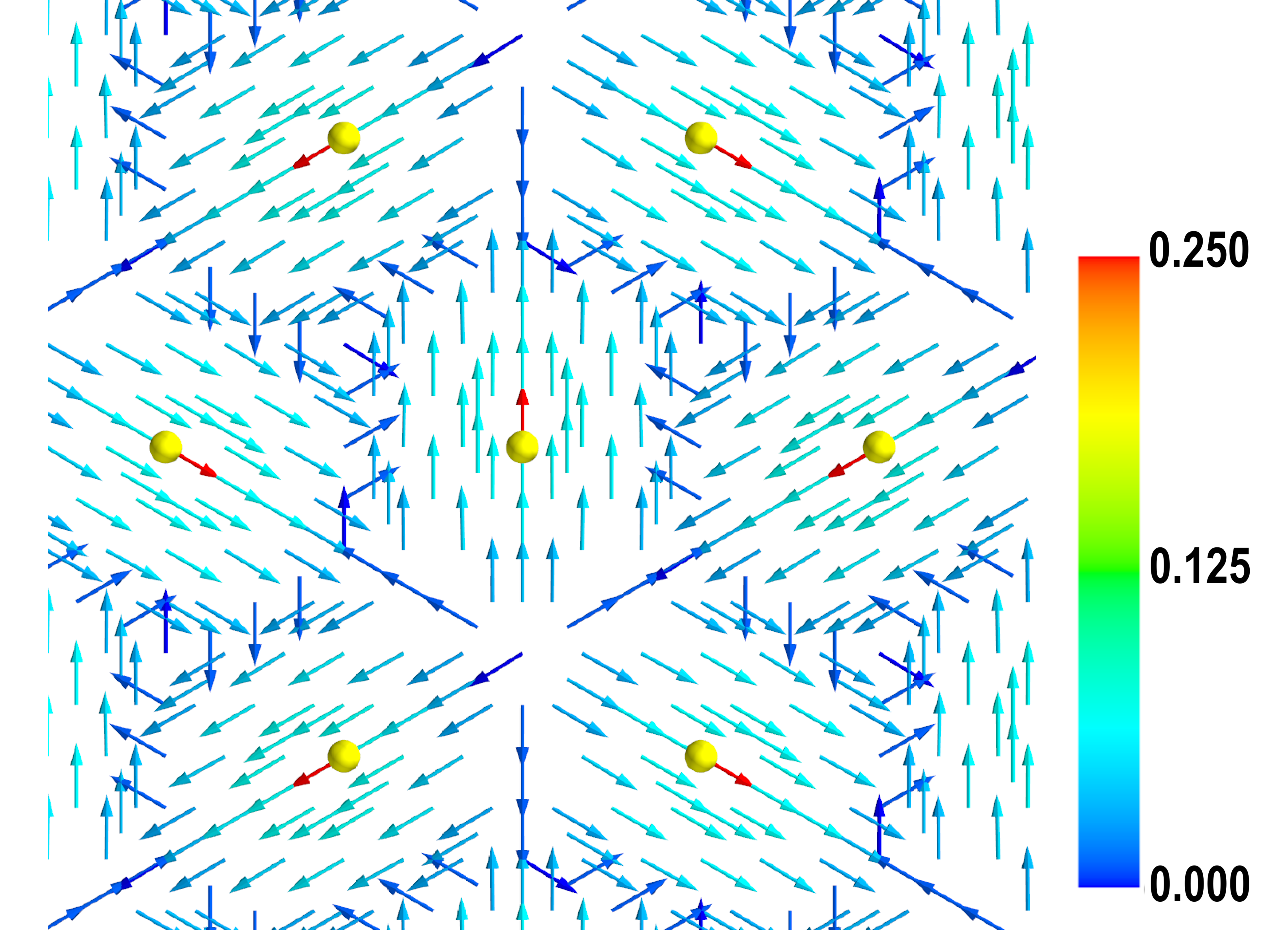}\label{fig9c}}
	\subfigure[Torque.]{\includegraphics[width=4.2cm]{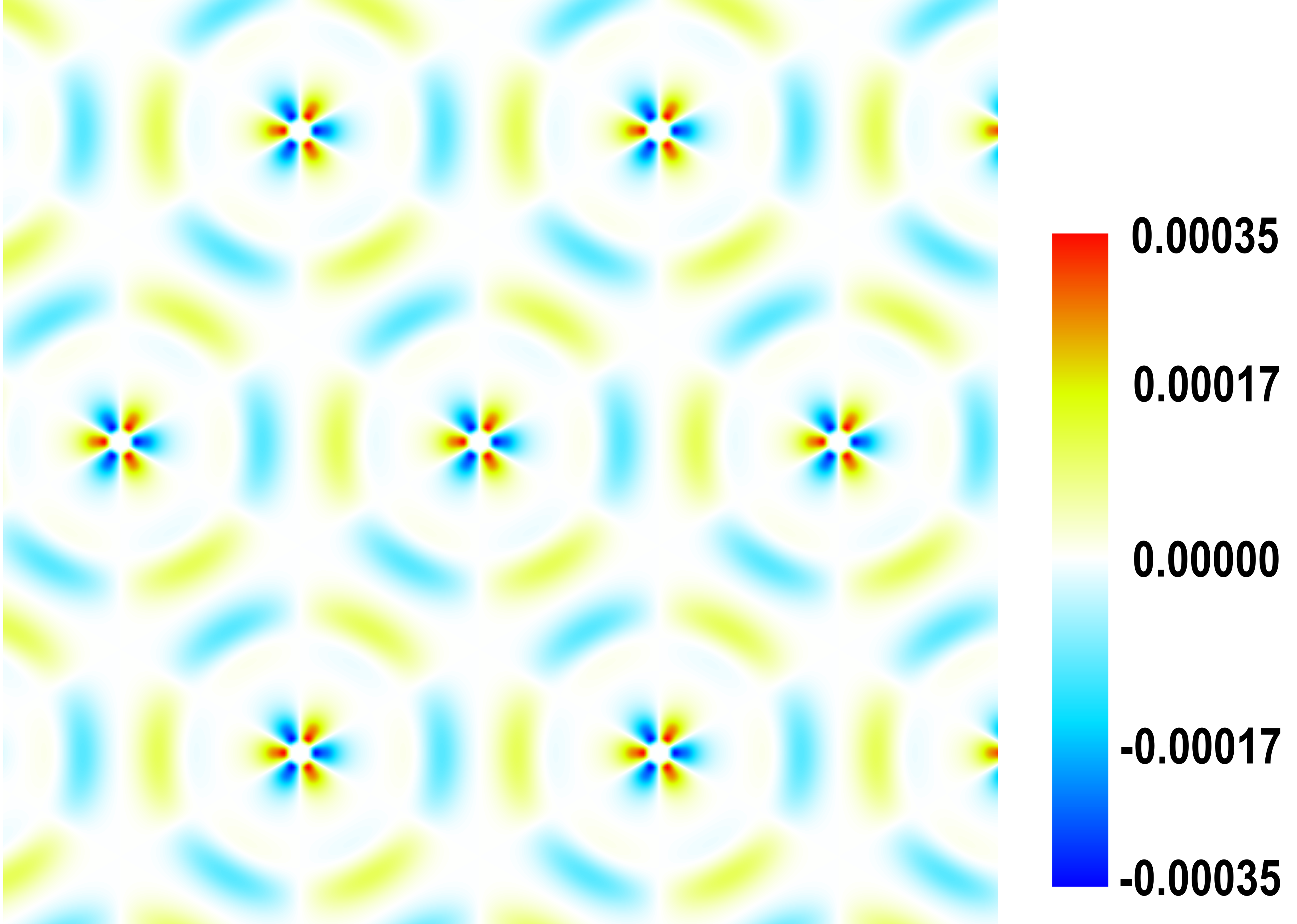}\label{fig9d}}
	\caption{
		Calculation results of the triangular $ \mathrm{Cr} $ monolayer in the multi-collinear approach.
		(a) $ \boldsymbol{m} $ of each $ \mathrm{Cr} $ atom. A magnetic unit cell contains three atoms marked by blue with the translation vectors presented by green arrows. 
		(b) $ \boldsymbol{m} $ in the monolayer. 
		(c) $ \boldsymbol{B}^{\mathrm{xc}} $ in the monolayer. 
		(d) The perpendicular component of torque (torque is perpendicular to the $ \mathrm{Cr} $ monolayer) in the monolayer. \label{fig9}
	}
\end{figure}

Calculation results for PBE functional are displayed in FIG. \ref{fig9}.
FIG. \ref{fig9a} exhibits the calculated non-collinear anti-ferromagnetic N\'{e}el state.
FIG. \ref{fig9b} displays the details of $\boldsymbol{m}$, highly collinear in the core region. The atomic $\boldsymbol{m}$ is 4.36 a.u. using the Hirshfeld partitioning scheme\cite{hirshfeld1977bonded}, close to the result 4.30 a.u. reported in Ref. \citenum{bulik2013noncollinear}, where the modified locally collinear approach by Scalmani and Frisch is used.  
FIG. \ref{fig9c} and FIG. \ref{fig9d} display the $ \boldsymbol{B}^{\mathrm{xc}} $ and the torque, respectively. The torque around an atom presents a pattern of six alternating up and down petals, similar to the pattern observed in Ref. \citenum{bulik2013noncollinear}.

\subsection{$\boldsymbol{\mathrm{Dy}_3}$ cluster with spin-orbit coupling \label{sec:Dy3}}

To illuminate the multi-collinear approach in the presence of spin-orbit coupling, we perform a generalized Kohn-Sham calculation for the $ \mathrm{Dy_3} $ cluster, using the Dirac-Coulomb Hamiltonian.
The bond length of $ \mathrm{Dy_3} $ cluster with $ \mathrm{D_{3h}} $ symmetry is $ 6.6893 $ Bohr, the same as in Ref. \citenum{zhong2020unprecedented}. 

\begin{figure}
	\centering
	\includegraphics[width=6.4cm]{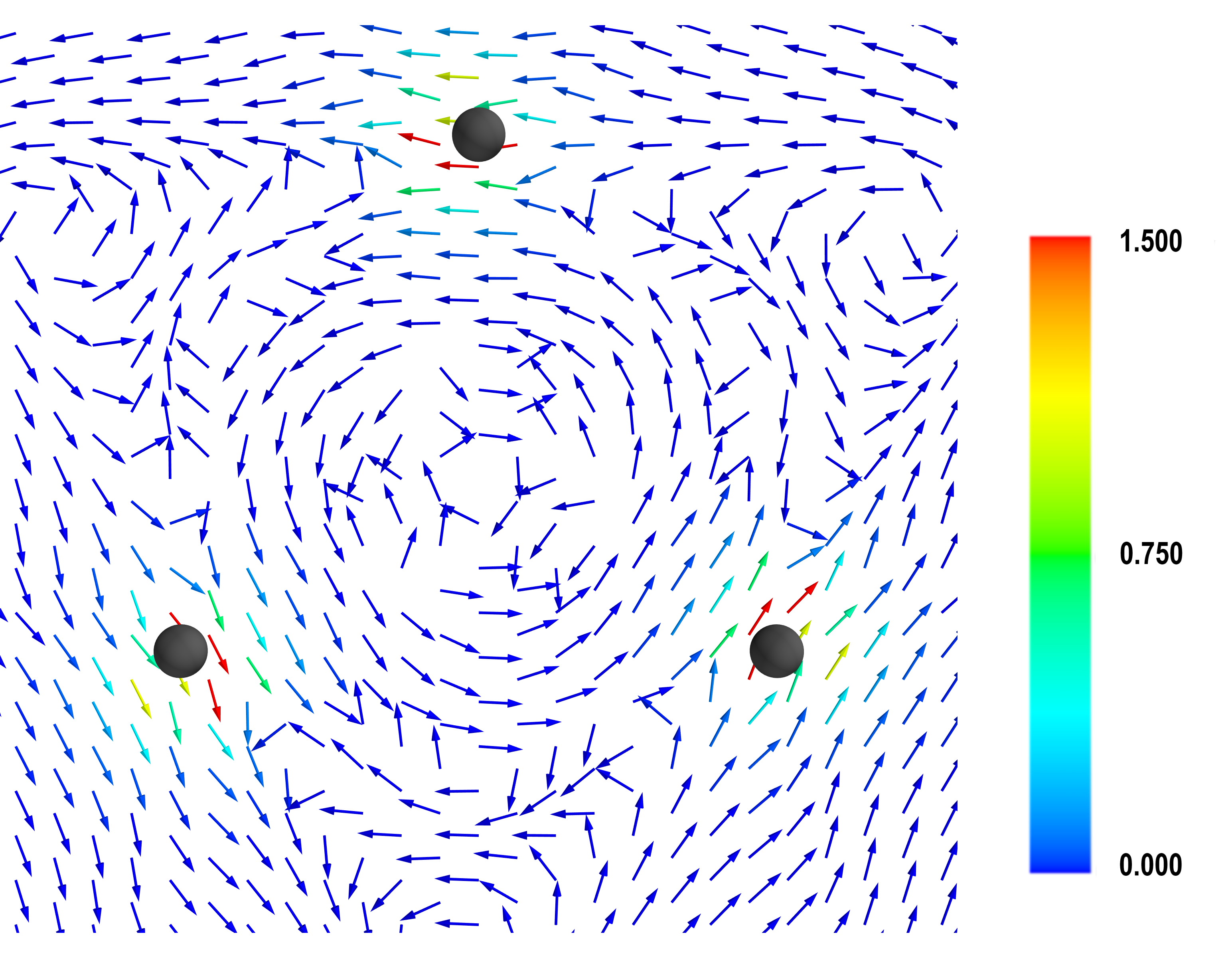}
	\caption{Calculated $ \boldsymbol{m} $ (in atomic units) of $ \mathrm{Dy_3} $ at four-component Dirac-Coulomb level, using multi-collinear PBE and cc-pVDZ-DK\cite{lu2016a} basis set.
    \label{fig10}
    }
\end{figure}

The calculated $\boldsymbol{m}$ is plotted in FIG. \ref{fig10}, exhibiting a toroidal pattern. In 2020, Zhong and co-workers observed such a toroidal pattern in their calculations\cite{zhong2020unprecedented} on a similar system, $ \mathrm{Dy_3} $ but with ligands, at the complete-active-space self-consistent field level.

\section{Summary and outlook}

The multi-collinear approach, satisfying all criteria mentioned in the background section,  treats collinear and non-collinear states in a uniform way. 
Meeting the correct collinear limit generally guarantees the systematic improvement of its accuracy, with the accuracy of collinear functionals increasing in the future.
At present, the multi-collinear approach allows non-collinear DFT to provide non-vanishing local torque, crucial for spin dynamics. Besides, it has well-defined and numerical stable functional derivatives, a desired feature for non-collinear and spin-flip TDDFT.

\begin{acknowledgments}
Y. X thanks Zhendong Li, Chen Li, Jie Xu, Zikuan Wang, Daoling Peng and Hao Chai for valuable discussions.
This work was supported by the National Key Research and Development Program of China (2017YFA0204702) and National Natural Science Foundation of China (21927901, 21821004, 21473002, 11822102).
\end{acknowledgments}

\bibliography{ref}

\begin{thebibliography}{60}%
\makeatletter
\providecommand \@ifxundefined [1]{%
 \@ifx{#1\undefined}
}%
\providecommand \@ifnum [1]{%
 \ifnum #1\expandafter \@firstoftwo
 \else \expandafter \@secondoftwo
 \fi
}%
\providecommand \@ifx [1]{%
 \ifx #1\expandafter \@firstoftwo
 \else \expandafter \@secondoftwo
 \fi
}%
\providecommand \natexlab [1]{#1}%
\providecommand \enquote  [1]{``#1''}%
\providecommand \bibnamefont  [1]{#1}%
\providecommand \bibfnamefont [1]{#1}%
\providecommand \citenamefont [1]{#1}%
\providecommand \href@noop [0]{\@secondoftwo}%
\providecommand \href [0]{\begingroup \@sanitize@url \@href}%
\providecommand \@href[1]{\@@startlink{#1}\@@href}%
\providecommand \@@href[1]{\endgroup#1\@@endlink}%
\providecommand \@sanitize@url [0]{\catcode `\\12\catcode `\$12\catcode
  `\&12\catcode `\#12\catcode `\^12\catcode `\_12\catcode `\%12\relax}%
\providecommand \@@startlink[1]{}%
\providecommand \@@endlink[0]{}%
\providecommand \url  [0]{\begingroup\@sanitize@url \@url }%
\providecommand \@url [1]{\endgroup\@href {#1}{\urlprefix }}%
\providecommand \urlprefix  [0]{URL }%
\providecommand \Eprint [0]{\href }%
\providecommand \doibase [0]{https://doi.org/}%
\providecommand \selectlanguage [0]{\@gobble}%
\providecommand \bibinfo  [0]{\@secondoftwo}%
\providecommand \bibfield  [0]{\@secondoftwo}%
\providecommand \translation [1]{[#1]}%
\providecommand \BibitemOpen [0]{}%
\providecommand \bibitemStop [0]{}%
\providecommand \bibitemNoStop [0]{.\EOS\space}%
\providecommand \EOS [0]{\spacefactor3000\relax}%
\providecommand \BibitemShut  [1]{\csname bibitem#1\endcsname}%
\let\auto@bib@innerbib\@empty
\bibitem [{\citenamefont {Kohn}\ and\ \citenamefont
  {Sham}(1965)}]{kohn1965self}%
  \BibitemOpen
  \bibfield  {author} {\bibinfo {author} {\bibfnamefont {W.}~\bibnamefont
  {Kohn}}\ and\ \bibinfo {author} {\bibfnamefont {L.~J.}\ \bibnamefont
  {Sham}},\ }\bibfield  {title} {\bibinfo {title} {Self-consistent equations
  including exchange and correlation effects},\ }\href@noop {} {\bibfield
  {journal} {\bibinfo  {journal} {Phys. Rev.}\ }\textbf {\bibinfo {volume}
  {140}},\ \bibinfo {pages} {A1133} (\bibinfo {year} {1965})}\BibitemShut
  {NoStop}%
\bibitem [{\citenamefont {Von~Barth}\ and\ \citenamefont
  {Hedin}(1972)}]{von1972local}%
  \BibitemOpen
  \bibfield  {author} {\bibinfo {author} {\bibfnamefont {U.}~\bibnamefont
  {Von~Barth}}\ and\ \bibinfo {author} {\bibfnamefont {L.}~\bibnamefont
  {Hedin}},\ }\bibfield  {title} {\bibinfo {title} {A local
  exchange-correlation potential for the spin polarized case. i},\ }\href@noop
  {} {\bibfield  {journal} {\bibinfo  {journal} {J. phys., C, Solid state
  phys.}\ }\textbf {\bibinfo {volume} {5}},\ \bibinfo {pages} {1629} (\bibinfo
  {year} {1972})}\BibitemShut {NoStop}%
\bibitem [{\citenamefont {K{\"u}bler}\ \emph {et~al.}(1988)\citenamefont
  {K{\"u}bler}, \citenamefont {H{\"o}ck}, \citenamefont {Sticht},\ and\
  \citenamefont {Williams}}]{kubler1988density}%
  \BibitemOpen
  \bibfield  {author} {\bibinfo {author} {\bibfnamefont {J.}~\bibnamefont
  {K{\"u}bler}}, \bibinfo {author} {\bibfnamefont {K.-H.}\ \bibnamefont
  {H{\"o}ck}}, \bibinfo {author} {\bibfnamefont {J.}~\bibnamefont {Sticht}},\
  and\ \bibinfo {author} {\bibfnamefont {A.~R.}\ \bibnamefont {Williams}},\
  }\bibfield  {title} {\bibinfo {title} {Density functional theory of
  non-collinear magnetism},\ }\href@noop {} {\bibfield  {journal} {\bibinfo
  {journal} {J. Phys. F: Met. Phys.}\ }\textbf {\bibinfo {volume} {18}},\
  \bibinfo {pages} {469} (\bibinfo {year} {1988})}\BibitemShut {NoStop}%
\bibitem [{\citenamefont {Nordstr{\"o}m}\ and\ \citenamefont
  {Singh}(1996)}]{nordstrom1996noncollinear}%
  \BibitemOpen
  \bibfield  {author} {\bibinfo {author} {\bibfnamefont {L.}~\bibnamefont
  {Nordstr{\"o}m}}\ and\ \bibinfo {author} {\bibfnamefont {D.~J.}\ \bibnamefont
  {Singh}},\ }\bibfield  {title} {\bibinfo {title} {Noncollinear intra-atomic
  magnetism},\ }\href@noop {} {\bibfield  {journal} {\bibinfo  {journal} {Phys.
  Rev. Lett.}\ }\textbf {\bibinfo {volume} {76}},\ \bibinfo {pages} {4420}
  (\bibinfo {year} {1996})}\BibitemShut {NoStop}%
\bibitem [{\citenamefont {Oda}\ \emph {et~al.}(1998)\citenamefont {Oda},
  \citenamefont {Pasquarello},\ and\ \citenamefont {Car}}]{oda1998fully}%
  \BibitemOpen
  \bibfield  {author} {\bibinfo {author} {\bibfnamefont {T.}~\bibnamefont
  {Oda}}, \bibinfo {author} {\bibfnamefont {A.}~\bibnamefont {Pasquarello}},\
  and\ \bibinfo {author} {\bibfnamefont {R.}~\bibnamefont {Car}},\ }\bibfield
  {title} {\bibinfo {title} {Fully unconstrained approach to noncollinear
  magnetism: application to small fe clusters},\ }\href@noop {} {\bibfield
  {journal} {\bibinfo  {journal} {Phys. Rev. Lett.}\ }\textbf {\bibinfo
  {volume} {80}},\ \bibinfo {pages} {3622} (\bibinfo {year}
  {1998})}\BibitemShut {NoStop}%
\bibitem [{\citenamefont {Van~W{\"u}llen}(2002)}]{van2002spin}%
  \BibitemOpen
  \bibfield  {author} {\bibinfo {author} {\bibfnamefont {C.}~\bibnamefont
  {Van~W{\"u}llen}},\ }\bibfield  {title} {\bibinfo {title} {Spin densities in
  two-component relativistic density functional calculations: Noncollinear
  versus collinear approach},\ }\href@noop {} {\bibfield  {journal} {\bibinfo
  {journal} {J. Comput. Chem.}\ }\textbf {\bibinfo {volume} {23}},\ \bibinfo
  {pages} {779} (\bibinfo {year} {2002})}\BibitemShut {NoStop}%
\bibitem [{\citenamefont {Wang}\ and\ \citenamefont
  {Ziegler}(2004)}]{wang2004time}%
  \BibitemOpen
  \bibfield  {author} {\bibinfo {author} {\bibfnamefont {F.}~\bibnamefont
  {Wang}}\ and\ \bibinfo {author} {\bibfnamefont {T.}~\bibnamefont {Ziegler}},\
  }\bibfield  {title} {\bibinfo {title} {Time-dependent density functional
  theory based on a noncollinear formulation of the exchange-correlation
  potential},\ }\href@noop {} {\bibfield  {journal} {\bibinfo  {journal} {J.
  Chem. Phys.}\ }\textbf {\bibinfo {volume} {121}},\ \bibinfo {pages} {12191}
  (\bibinfo {year} {2004})}\BibitemShut {NoStop}%
\bibitem [{\citenamefont {Gao}\ \emph {et~al.}(2005)\citenamefont {Gao},
  \citenamefont {Zou}, \citenamefont {Liu}, \citenamefont {Xiao}, \citenamefont
  {Peng}, \citenamefont {Song},\ and\ \citenamefont {Liu}}]{gao2005time}%
  \BibitemOpen
  \bibfield  {author} {\bibinfo {author} {\bibfnamefont {J.}~\bibnamefont
  {Gao}}, \bibinfo {author} {\bibfnamefont {W.}~\bibnamefont {Zou}}, \bibinfo
  {author} {\bibfnamefont {W.}~\bibnamefont {Liu}}, \bibinfo {author}
  {\bibfnamefont {Y.}~\bibnamefont {Xiao}}, \bibinfo {author} {\bibfnamefont
  {D.}~\bibnamefont {Peng}}, \bibinfo {author} {\bibfnamefont {B.}~\bibnamefont
  {Song}},\ and\ \bibinfo {author} {\bibfnamefont {C.}~\bibnamefont {Liu}},\
  }\bibfield  {title} {\bibinfo {title} {Time-dependent four-component
  relativistic density-functional theory for excitation energies. ii. the
  exchange-correlation kernel},\ }\href@noop {} {\bibfield  {journal} {\bibinfo
   {journal} {J. Chem. Phys.}\ }\textbf {\bibinfo {volume} {123}},\ \bibinfo
  {pages} {054102} (\bibinfo {year} {2005})}\BibitemShut {NoStop}%
\bibitem [{\citenamefont {Bast}\ \emph {et~al.}(2009)\citenamefont {Bast},
  \citenamefont {Jensen},\ and\ \citenamefont {Saue}}]{bast2009relativistic}%
  \BibitemOpen
  \bibfield  {author} {\bibinfo {author} {\bibfnamefont {R.}~\bibnamefont
  {Bast}}, \bibinfo {author} {\bibfnamefont {H.~J.~A.}\ \bibnamefont
  {Jensen}},\ and\ \bibinfo {author} {\bibfnamefont {T.}~\bibnamefont {Saue}},\
  }\bibfield  {title} {\bibinfo {title} {Relativistic adiabatic time-dependent
  density functional theory using hybrid functionals and noncollinear spin
  magnetization},\ }\href@noop {} {\bibfield  {journal} {\bibinfo  {journal}
  {Int. J. Quantum Chem.}\ }\textbf {\bibinfo {volume} {109}},\ \bibinfo
  {pages} {2091} (\bibinfo {year} {2009})}\BibitemShut {NoStop}%
\bibitem [{\citenamefont {Bulik}\ \emph {et~al.}(2013)\citenamefont {Bulik},
  \citenamefont {Scalmani}, \citenamefont {Frisch},\ and\ \citenamefont
  {Scuseria}}]{bulik2013noncollinear}%
  \BibitemOpen
  \bibfield  {author} {\bibinfo {author} {\bibfnamefont {I.~W.}\ \bibnamefont
  {Bulik}}, \bibinfo {author} {\bibfnamefont {G.}~\bibnamefont {Scalmani}},
  \bibinfo {author} {\bibfnamefont {M.~J.}\ \bibnamefont {Frisch}},\ and\
  \bibinfo {author} {\bibfnamefont {G.~E.}\ \bibnamefont {Scuseria}},\
  }\bibfield  {title} {\bibinfo {title} {Noncollinear density functional theory
  having proper invariance and local torque properties},\ }\href@noop {}
  {\bibfield  {journal} {\bibinfo  {journal} {Phys. Rev. B}\ }\textbf {\bibinfo
  {volume} {87}},\ \bibinfo {pages} {035117} (\bibinfo {year}
  {2013})}\BibitemShut {NoStop}%
\bibitem [{\citenamefont {Egidi}\ \emph {et~al.}(2017)\citenamefont {Egidi},
  \citenamefont {Sun}, \citenamefont {Goings}, \citenamefont {Scalmani},
  \citenamefont {Frisch},\ and\ \citenamefont {Li}}]{egidi2017two}%
  \BibitemOpen
  \bibfield  {author} {\bibinfo {author} {\bibfnamefont {F.}~\bibnamefont
  {Egidi}}, \bibinfo {author} {\bibfnamefont {S.}~\bibnamefont {Sun}}, \bibinfo
  {author} {\bibfnamefont {J.~J.}\ \bibnamefont {Goings}}, \bibinfo {author}
  {\bibfnamefont {G.}~\bibnamefont {Scalmani}}, \bibinfo {author}
  {\bibfnamefont {M.~J.}\ \bibnamefont {Frisch}},\ and\ \bibinfo {author}
  {\bibfnamefont {X.}~\bibnamefont {Li}},\ }\bibfield  {title} {\bibinfo
  {title} {Two-component noncollinear time-dependent spin density functional
  theory for excited state calculations},\ }\href@noop {} {\bibfield  {journal}
  {\bibinfo  {journal} {J. Chem. Theory Comput.}\ }\textbf {\bibinfo {volume}
  {13}},\ \bibinfo {pages} {2591} (\bibinfo {year} {2017})}\BibitemShut
  {NoStop}%
\bibitem [{\citenamefont {Liu}\ and\ \citenamefont
  {Xiao}(2018)}]{liu2018relativistic}%
  \BibitemOpen
  \bibfield  {author} {\bibinfo {author} {\bibfnamefont {W.}~\bibnamefont
  {Liu}}\ and\ \bibinfo {author} {\bibfnamefont {Y.}~\bibnamefont {Xiao}},\
  }\bibfield  {title} {\bibinfo {title} {Relativistic time-dependent density
  functional theories},\ }\href@noop {} {\bibfield  {journal} {\bibinfo
  {journal} {Chem. Soc. Rev.}\ }\textbf {\bibinfo {volume} {47}},\ \bibinfo
  {pages} {4481} (\bibinfo {year} {2018})}\BibitemShut {NoStop}%
\bibitem [{\citenamefont {Li}\ \emph {et~al.}(2020)\citenamefont {Li},
  \citenamefont {Govind}, \citenamefont {Isborn}, \citenamefont
  {DePrince~III},\ and\ \citenamefont {Lopata}}]{li2020real}%
  \BibitemOpen
  \bibfield  {author} {\bibinfo {author} {\bibfnamefont {X.}~\bibnamefont
  {Li}}, \bibinfo {author} {\bibfnamefont {N.}~\bibnamefont {Govind}}, \bibinfo
  {author} {\bibfnamefont {C.}~\bibnamefont {Isborn}}, \bibinfo {author}
  {\bibfnamefont {A.~E.}\ \bibnamefont {DePrince~III}},\ and\ \bibinfo {author}
  {\bibfnamefont {K.}~\bibnamefont {Lopata}},\ }\bibfield  {title} {\bibinfo
  {title} {Real-time time-dependent electronic structure theory},\ }\href@noop
  {} {\bibfield  {journal} {\bibinfo  {journal} {Chem. Rev.}\ }\textbf
  {\bibinfo {volume} {120}},\ \bibinfo {pages} {9951} (\bibinfo {year}
  {2020})}\BibitemShut {NoStop}%
\bibitem [{\citenamefont {Desmarais}\ \emph {et~al.}(2021)\citenamefont
  {Desmarais}, \citenamefont {Komorovsky}, \citenamefont {Flament},\ and\
  \citenamefont {Erba}}]{desmarais2021spin}%
  \BibitemOpen
  \bibfield  {author} {\bibinfo {author} {\bibfnamefont {J.~K.}\ \bibnamefont
  {Desmarais}}, \bibinfo {author} {\bibfnamefont {S.}~\bibnamefont
  {Komorovsky}}, \bibinfo {author} {\bibfnamefont {J.-P.}\ \bibnamefont
  {Flament}},\ and\ \bibinfo {author} {\bibfnamefont {A.}~\bibnamefont
  {Erba}},\ }\bibfield  {title} {\bibinfo {title} {Spin--orbit coupling from a
  two-component self-consistent approach. ii. non-collinear density functional
  theories},\ }\href@noop {} {\bibfield  {journal} {\bibinfo  {journal} {J.
  Chem. Phys}\ }\textbf {\bibinfo {volume} {154}},\ \bibinfo {pages} {204110}
  (\bibinfo {year} {2021})}\BibitemShut {NoStop}%
\bibitem [{\citenamefont {Vosko}\ \emph {et~al.}(1980)\citenamefont {Vosko},
  \citenamefont {Wilk},\ and\ \citenamefont {Nusair}}]{vosko1980accurate}%
  \BibitemOpen
  \bibfield  {author} {\bibinfo {author} {\bibfnamefont {S.~H.}\ \bibnamefont
  {Vosko}}, \bibinfo {author} {\bibfnamefont {L.}~\bibnamefont {Wilk}},\ and\
  \bibinfo {author} {\bibfnamefont {M.}~\bibnamefont {Nusair}},\ }\bibfield
  {title} {\bibinfo {title} {Accurate spin-dependent electron liquid
  correlation energies for local spin density calculations: a critical
  analysis},\ }\href@noop {} {\bibfield  {journal} {\bibinfo  {journal} {Can.
  J. Phys.}\ }\textbf {\bibinfo {volume} {58}},\ \bibinfo {pages} {1200}
  (\bibinfo {year} {1980})}\BibitemShut {NoStop}%
\bibitem [{\citenamefont {Perdew}\ \emph {et~al.}(1992)\citenamefont {Perdew},
  \citenamefont {Chevary}, \citenamefont {Vosko}, \citenamefont {Jackson},
  \citenamefont {Pederson}, \citenamefont {Singh},\ and\ \citenamefont
  {Fiolhais}}]{perdew1992atoms}%
  \BibitemOpen
  \bibfield  {author} {\bibinfo {author} {\bibfnamefont {J.~P.}\ \bibnamefont
  {Perdew}}, \bibinfo {author} {\bibfnamefont {J.~A.}\ \bibnamefont {Chevary}},
  \bibinfo {author} {\bibfnamefont {S.~H.}\ \bibnamefont {Vosko}}, \bibinfo
  {author} {\bibfnamefont {K.~A.}\ \bibnamefont {Jackson}}, \bibinfo {author}
  {\bibfnamefont {M.~R.}\ \bibnamefont {Pederson}}, \bibinfo {author}
  {\bibfnamefont {D.~J.}\ \bibnamefont {Singh}},\ and\ \bibinfo {author}
  {\bibfnamefont {C.}~\bibnamefont {Fiolhais}},\ }\bibfield  {title} {\bibinfo
  {title} {Atoms, molecules, solids, and surfaces: Applications of the
  generalized gradient approximation for exchange and correlation},\
  }\href@noop {} {\bibfield  {journal} {\bibinfo  {journal} {Phys. Rev. B}\
  }\textbf {\bibinfo {volume} {46}},\ \bibinfo {pages} {6671} (\bibinfo {year}
  {1992})}\BibitemShut {NoStop}%
\bibitem [{\citenamefont {Becke}(1988{\natexlab{a}})}]{becke1988density}%
  \BibitemOpen
  \bibfield  {author} {\bibinfo {author} {\bibfnamefont {A.~D.}\ \bibnamefont
  {Becke}},\ }\bibfield  {title} {\bibinfo {title} {Density-functional
  exchange-energy approximation with correct asymptotic behavior},\ }\href@noop
  {} {\bibfield  {journal} {\bibinfo  {journal} {Phys. Rev. A}\ }\textbf
  {\bibinfo {volume} {38}},\ \bibinfo {pages} {3098} (\bibinfo {year}
  {1988}{\natexlab{a}})}\BibitemShut {NoStop}%
\bibitem [{\citenamefont {Gritsenko}\ \emph {et~al.}(1993)\citenamefont
  {Gritsenko}, \citenamefont {Cordero}, \citenamefont {Rubio}, \citenamefont
  {Balb{\'a}s},\ and\ \citenamefont {Alonso}}]{gritsenko1993weighted}%
  \BibitemOpen
  \bibfield  {author} {\bibinfo {author} {\bibfnamefont {O.~V.}\ \bibnamefont
  {Gritsenko}}, \bibinfo {author} {\bibfnamefont {N.~A.}\ \bibnamefont
  {Cordero}}, \bibinfo {author} {\bibfnamefont {A.}~\bibnamefont {Rubio}},
  \bibinfo {author} {\bibfnamefont {L.~C.}\ \bibnamefont {Balb{\'a}s}},\ and\
  \bibinfo {author} {\bibfnamefont {J.~A.}\ \bibnamefont {Alonso}},\ }\bibfield
   {title} {\bibinfo {title} {Weighted-density exchange and local-density
  coulomb correlation energy functionals for finite systems: Application to
  atoms},\ }\href@noop {} {\bibfield  {journal} {\bibinfo  {journal} {Phys.
  Rev. A}\ }\textbf {\bibinfo {volume} {48}},\ \bibinfo {pages} {4197}
  (\bibinfo {year} {1993})}\BibitemShut {NoStop}%
\bibitem [{\citenamefont {Casida}(1995)}]{casida1995time}%
  \BibitemOpen
  \bibfield  {author} {\bibinfo {author} {\bibfnamefont {M.~E.}\ \bibnamefont
  {Casida}},\ }\bibfield  {title} {\bibinfo {title} {Time-dependent density
  functional response theory for molecules},\ }in\ \href@noop {} {\emph
  {\bibinfo {booktitle} {Recent Advances In Density Functional Methods: (Part
  I)}}}\ (\bibinfo  {publisher} {World Scientific},\ \bibinfo {year} {1995})\
  pp.\ \bibinfo {pages} {155--192}\BibitemShut {NoStop}%
\bibitem [{\citenamefont {Runge}\ and\ \citenamefont
  {Gross}(1984)}]{TDDFT_1984}%
  \BibitemOpen
  \bibfield  {author} {\bibinfo {author} {\bibfnamefont {E.}~\bibnamefont
  {Runge}}\ and\ \bibinfo {author} {\bibfnamefont {E.~K.~U.}\ \bibnamefont
  {Gross}},\ }\bibfield  {title} {\bibinfo {title} {Density-functional theory
  for time-dependent systems},\ }\href
  {https://doi.org/10.1103/PhysRevLett.52.997} {\bibfield  {journal} {\bibinfo
  {journal} {Phys. Rev. Lett.}\ }\textbf {\bibinfo {volume} {52}},\ \bibinfo
  {pages} {997} (\bibinfo {year} {1984})}\BibitemShut {NoStop}%
\bibitem [{\citenamefont {Komorovsky}\ \emph {et~al.}(2019)\citenamefont
  {Komorovsky}, \citenamefont {Cherry},\ and\ \citenamefont
  {Repisky}}]{komorovsky2019four}%
  \BibitemOpen
  \bibfield  {author} {\bibinfo {author} {\bibfnamefont {S.}~\bibnamefont
  {Komorovsky}}, \bibinfo {author} {\bibfnamefont {P.~J.}\ \bibnamefont
  {Cherry}},\ and\ \bibinfo {author} {\bibfnamefont {M.}~\bibnamefont
  {Repisky}},\ }\bibfield  {title} {\bibinfo {title} {Four-component
  relativistic time-dependent density-functional theory using a stable
  noncollinear dft ansatz applicable to both closed-and open-shell systems},\
  }\href@noop {} {\bibfield  {journal} {\bibinfo  {journal} {J. Chem. Phys.}\
  }\textbf {\bibinfo {volume} {151}},\ \bibinfo {pages} {184111} (\bibinfo
  {year} {2019})}\BibitemShut {NoStop}%
\bibitem [{\citenamefont {Li}\ and\ \citenamefont
  {Liu}(2012)}]{li2012theoretical}%
  \BibitemOpen
  \bibfield  {author} {\bibinfo {author} {\bibfnamefont {Z.}~\bibnamefont
  {Li}}\ and\ \bibinfo {author} {\bibfnamefont {W.}~\bibnamefont {Liu}},\
  }\bibfield  {title} {\bibinfo {title} {Theoretical and numerical assessments
  of spin-flip time-dependent density functional theory},\ }\href@noop {}
  {\bibfield  {journal} {\bibinfo  {journal} {J. Chem. Phys.}\ }\textbf
  {\bibinfo {volume} {136}},\ \bibinfo {pages} {024107} (\bibinfo {year}
  {2012})}\BibitemShut {NoStop}%
\bibitem [{\citenamefont {Capelle}\ \emph {et~al.}(2001)\citenamefont
  {Capelle}, \citenamefont {Vignale},\ and\ \citenamefont
  {Gy{\"o}rffy}}]{capelle2001spin}%
  \BibitemOpen
  \bibfield  {author} {\bibinfo {author} {\bibfnamefont {K.}~\bibnamefont
  {Capelle}}, \bibinfo {author} {\bibfnamefont {G.}~\bibnamefont {Vignale}},\
  and\ \bibinfo {author} {\bibfnamefont {B.~L.}\ \bibnamefont {Gy{\"o}rffy}},\
  }\bibfield  {title} {\bibinfo {title} {Spin currents and spin dynamics in
  time-dependent density-functional theory},\ }\href@noop {} {\bibfield
  {journal} {\bibinfo  {journal} {Phys. Rev. Lett.}\ }\textbf {\bibinfo
  {volume} {87}},\ \bibinfo {pages} {206403} (\bibinfo {year}
  {2001})}\BibitemShut {NoStop}%
\bibitem [{\citenamefont {Sharma}\ \emph {et~al.}(2007)\citenamefont {Sharma},
  \citenamefont {Dewhurst}, \citenamefont {Ambrosch-Draxl}, \citenamefont
  {Kurth}, \citenamefont {Helbig}, \citenamefont {Pittalis}, \citenamefont
  {Shallcross}, \citenamefont {Nordstr{\"o}m},\ and\ \citenamefont
  {Gross}}]{sharma2007first}%
  \BibitemOpen
  \bibfield  {author} {\bibinfo {author} {\bibfnamefont {S.}~\bibnamefont
  {Sharma}}, \bibinfo {author} {\bibfnamefont {J.~K.}\ \bibnamefont
  {Dewhurst}}, \bibinfo {author} {\bibfnamefont {C.}~\bibnamefont
  {Ambrosch-Draxl}}, \bibinfo {author} {\bibfnamefont {S.}~\bibnamefont
  {Kurth}}, \bibinfo {author} {\bibfnamefont {N.}~\bibnamefont {Helbig}},
  \bibinfo {author} {\bibfnamefont {S.}~\bibnamefont {Pittalis}}, \bibinfo
  {author} {\bibfnamefont {S.}~\bibnamefont {Shallcross}}, \bibinfo {author}
  {\bibfnamefont {L.}~\bibnamefont {Nordstr{\"o}m}},\ and\ \bibinfo {author}
  {\bibfnamefont {E.~K.~U.}\ \bibnamefont {Gross}},\ }\bibfield  {title}
  {\bibinfo {title} {First-principles approach to noncollinear magnetism:
  Towards spin dynamics},\ }\href@noop {} {\bibfield  {journal} {\bibinfo
  {journal} {Phys. Rev. Lett.}\ }\textbf {\bibinfo {volume} {98}},\ \bibinfo
  {pages} {196405} (\bibinfo {year} {2007})}\BibitemShut {NoStop}%
\bibitem [{\citenamefont {Scalmani}\ and\ \citenamefont
  {Frisch}(2012)}]{scalmani2012new}%
  \BibitemOpen
  \bibfield  {author} {\bibinfo {author} {\bibfnamefont {G.}~\bibnamefont
  {Scalmani}}\ and\ \bibinfo {author} {\bibfnamefont {M.~J.}\ \bibnamefont
  {Frisch}},\ }\bibfield  {title} {\bibinfo {title} {A new approach to
  noncollinear spin density functional theory beyond the local density
  approximation},\ }\href@noop {} {\bibfield  {journal} {\bibinfo  {journal}
  {J. Chem. Theory Comput.}\ }\textbf {\bibinfo {volume} {8}},\ \bibinfo
  {pages} {2193} (\bibinfo {year} {2012})}\BibitemShut {NoStop}%
\bibitem [{\citenamefont {Eich}\ and\ \citenamefont
  {Gross}(2013)}]{eich2013transverse}%
  \BibitemOpen
  \bibfield  {author} {\bibinfo {author} {\bibfnamefont {F.~G.}\ \bibnamefont
  {Eich}}\ and\ \bibinfo {author} {\bibfnamefont {E.~K.~U.}\ \bibnamefont
  {Gross}},\ }\bibfield  {title} {\bibinfo {title} {Transverse spin-gradient
  functional for noncollinear spin-density-functional theory},\ }\href@noop {}
  {\bibfield  {journal} {\bibinfo  {journal} {Phys. Rev. Lett.}\ }\textbf
  {\bibinfo {volume} {111}},\ \bibinfo {pages} {156401} (\bibinfo {year}
  {2013})}\BibitemShut {NoStop}%
\bibitem [{\citenamefont {Eich}\ \emph {et~al.}(2013)\citenamefont {Eich},
  \citenamefont {Pittalis},\ and\ \citenamefont
  {Vignale}}]{eich2013transverse2}%
  \BibitemOpen
  \bibfield  {author} {\bibinfo {author} {\bibfnamefont {F.~G.}\ \bibnamefont
  {Eich}}, \bibinfo {author} {\bibfnamefont {S.}~\bibnamefont {Pittalis}},\
  and\ \bibinfo {author} {\bibfnamefont {G.}~\bibnamefont {Vignale}},\
  }\bibfield  {title} {\bibinfo {title} {Transverse and longitudinal gradients
  of the spin magnetization in spin-density-functional theory},\ }\href@noop {}
  {\bibfield  {journal} {\bibinfo  {journal} {Phys. Rev. B}\ }\textbf {\bibinfo
  {volume} {88}},\ \bibinfo {pages} {245102} (\bibinfo {year}
  {2013})}\BibitemShut {NoStop}%
\bibitem [{\citenamefont {Talman}\ and\ \citenamefont
  {Shadwick}(1976)}]{talman1976optimized}%
  \BibitemOpen
  \bibfield  {author} {\bibinfo {author} {\bibfnamefont {J.~D.}\ \bibnamefont
  {Talman}}\ and\ \bibinfo {author} {\bibfnamefont {W.~F.}\ \bibnamefont
  {Shadwick}},\ }\bibfield  {title} {\bibinfo {title} {Optimized effective
  atomic central potential},\ }\href@noop {} {\bibfield  {journal} {\bibinfo
  {journal} {Phys. Rev. A}\ }\textbf {\bibinfo {volume} {14}},\ \bibinfo
  {pages} {36} (\bibinfo {year} {1976})}\BibitemShut {NoStop}%
\bibitem [{\citenamefont {Becke}(1988{\natexlab{b}})}]{becke1988correlation}%
  \BibitemOpen
  \bibfield  {author} {\bibinfo {author} {\bibfnamefont {A.~D.}\ \bibnamefont
  {Becke}},\ }\bibfield  {title} {\bibinfo {title} {Correlation energy of an
  inhomogeneous electron gas: A coordinate-space model},\ }\href@noop {}
  {\bibfield  {journal} {\bibinfo  {journal} {J. Chem. Phys.}\ }\textbf
  {\bibinfo {volume} {88}},\ \bibinfo {pages} {1053} (\bibinfo {year}
  {1988}{\natexlab{b}})}\BibitemShut {NoStop}%
\bibitem [{\citenamefont {Tschinke}\ and\ \citenamefont
  {Ziegler}(1989)}]{tschinke1989shape}%
  \BibitemOpen
  \bibfield  {author} {\bibinfo {author} {\bibfnamefont {V.}~\bibnamefont
  {Tschinke}}\ and\ \bibinfo {author} {\bibfnamefont {T.}~\bibnamefont
  {Ziegler}},\ }\bibfield  {title} {\bibinfo {title} {On the shape of
  spherically averaged fermi-hole correlation functions in density functional
  theory. 1. atomic systems},\ }\href@noop {} {\bibfield  {journal} {\bibinfo
  {journal} {Can. J. Chem.}\ }\textbf {\bibinfo {volume} {67}},\ \bibinfo
  {pages} {460} (\bibinfo {year} {1989})}\BibitemShut {NoStop}%
\bibitem [{\citenamefont {Neumann}\ and\ \citenamefont
  {Handy}(1997)}]{neumann1997higher}%
  \BibitemOpen
  \bibfield  {author} {\bibinfo {author} {\bibfnamefont {R.}~\bibnamefont
  {Neumann}}\ and\ \bibinfo {author} {\bibfnamefont {N.~C.}\ \bibnamefont
  {Handy}},\ }\bibfield  {title} {\bibinfo {title} {Higher-order gradient
  corrections for exchange-correlation functionals},\ }\href@noop {} {\bibfield
   {journal} {\bibinfo  {journal} {Chem. Phys. Lett.}\ }\textbf {\bibinfo
  {volume} {266}},\ \bibinfo {pages} {16} (\bibinfo {year} {1997})}\BibitemShut
  {NoStop}%
\bibitem [{\citenamefont {Becke}(1993)}]{becke1993new}%
  \BibitemOpen
  \bibfield  {author} {\bibinfo {author} {\bibfnamefont {A.~D.}\ \bibnamefont
  {Becke}},\ }\bibfield  {title} {\bibinfo {title} {A new mixing of
  hartree--fock and local density-functional theories},\ }\href@noop {}
  {\bibfield  {journal} {\bibinfo  {journal} {J. Chem. Phys.}\ }\textbf
  {\bibinfo {volume} {98}},\ \bibinfo {pages} {1372} (\bibinfo {year}
  {1993})}\BibitemShut {NoStop}%
\bibitem [{\citenamefont {Kurz}\ \emph {et~al.}(2004)\citenamefont {Kurz},
  \citenamefont {F{\"o}rster}, \citenamefont {Nordstr{\"o}m}, \citenamefont
  {Bihlmayer},\ and\ \citenamefont {Bl{\"u}gel}}]{kurz2004ab}%
  \BibitemOpen
  \bibfield  {author} {\bibinfo {author} {\bibfnamefont {P.}~\bibnamefont
  {Kurz}}, \bibinfo {author} {\bibfnamefont {F.}~\bibnamefont {F{\"o}rster}},
  \bibinfo {author} {\bibfnamefont {L.}~\bibnamefont {Nordstr{\"o}m}}, \bibinfo
  {author} {\bibfnamefont {G.}~\bibnamefont {Bihlmayer}},\ and\ \bibinfo
  {author} {\bibfnamefont {S.}~\bibnamefont {Bl{\"u}gel}},\ }\bibfield  {title}
  {\bibinfo {title} {Ab initio treatment of noncollinear magnets with the
  full-potential linearized augmented plane wave method},\ }\href@noop {}
  {\bibfield  {journal} {\bibinfo  {journal} {Phys. Rev. B}\ }\textbf {\bibinfo
  {volume} {69}},\ \bibinfo {pages} {024415} (\bibinfo {year}
  {2004})}\BibitemShut {NoStop}%
\bibitem [{\citenamefont {Sj{\"o}stedt}\ and\ \citenamefont
  {Nordstr{\"o}m}(2002)}]{sjostedt2002noncollinear}%
  \BibitemOpen
  \bibfield  {author} {\bibinfo {author} {\bibfnamefont {E.}~\bibnamefont
  {Sj{\"o}stedt}}\ and\ \bibinfo {author} {\bibfnamefont {L.}~\bibnamefont
  {Nordstr{\"o}m}},\ }\bibfield  {title} {\bibinfo {title} {Noncollinear
  full-potential studies of $\gamma$- fe},\ }\href@noop {} {\bibfield
  {journal} {\bibinfo  {journal} {Phys. Rev. B}\ }\textbf {\bibinfo {volume}
  {66}},\ \bibinfo {pages} {014447} (\bibinfo {year} {2002})}\BibitemShut
  {NoStop}%
\bibitem [{\citenamefont {Peralta}\ \emph {et~al.}(2007)\citenamefont
  {Peralta}, \citenamefont {Scuseria},\ and\ \citenamefont
  {Frisch}}]{peralta2007noncollinear}%
  \BibitemOpen
  \bibfield  {author} {\bibinfo {author} {\bibfnamefont {J.~E.}\ \bibnamefont
  {Peralta}}, \bibinfo {author} {\bibfnamefont {G.~E.}\ \bibnamefont
  {Scuseria}},\ and\ \bibinfo {author} {\bibfnamefont {M.~J.}\ \bibnamefont
  {Frisch}},\ }\bibfield  {title} {\bibinfo {title} {Noncollinear magnetism in
  density functional calculations},\ }\href@noop {} {\bibfield  {journal}
  {\bibinfo  {journal} {Phys. Rev. B}\ }\textbf {\bibinfo {volume} {75}},\
  \bibinfo {pages} {125119} (\bibinfo {year} {2007})}\BibitemShut {NoStop}%
\bibitem [{\citenamefont {Kn{\"o}pfle}\ \emph {et~al.}(2000)\citenamefont
  {Kn{\"o}pfle}, \citenamefont {Sandratskii},\ and\ \citenamefont
  {K{\"u}bler}}]{knopfle2000spin}%
  \BibitemOpen
  \bibfield  {author} {\bibinfo {author} {\bibfnamefont {K.}~\bibnamefont
  {Kn{\"o}pfle}}, \bibinfo {author} {\bibfnamefont {L.~M.}\ \bibnamefont
  {Sandratskii}},\ and\ \bibinfo {author} {\bibfnamefont {J.}~\bibnamefont
  {K{\"u}bler}},\ }\bibfield  {title} {\bibinfo {title} {Spin spiral ground
  state of $\gamma$-iron},\ }\href@noop {} {\bibfield  {journal} {\bibinfo
  {journal} {Phys. Rev. B}\ }\textbf {\bibinfo {volume} {62}},\ \bibinfo
  {pages} {5564} (\bibinfo {year} {2000})}\BibitemShut {NoStop}%
\bibitem [{\citenamefont {Overhauser}(1962)}]{overhauser1962spin}%
  \BibitemOpen
  \bibfield  {author} {\bibinfo {author} {\bibfnamefont {A.~W.}\ \bibnamefont
  {Overhauser}},\ }\bibfield  {title} {\bibinfo {title} {Spin density waves in
  an electron gas},\ }\href@noop {} {\bibfield  {journal} {\bibinfo  {journal}
  {Phys. Rev.}\ }\textbf {\bibinfo {volume} {128}},\ \bibinfo {pages} {1437}
  (\bibinfo {year} {1962})}\BibitemShut {NoStop}%
\bibitem [{\citenamefont {Skyrme}(1962)}]{skyrme1962unified}%
  \BibitemOpen
  \bibfield  {author} {\bibinfo {author} {\bibfnamefont {T.~H.~R.}\
  \bibnamefont {Skyrme}},\ }\bibfield  {title} {\bibinfo {title} {A unified
  field theory of mesons and baryons},\ }\href@noop {} {\bibfield  {journal}
  {\bibinfo  {journal} {Nucl. Phys.}\ }\textbf {\bibinfo {volume} {31}},\
  \bibinfo {pages} {556} (\bibinfo {year} {1962})}\BibitemShut {NoStop}%
\bibitem [{\citenamefont {Sun}(2015)}]{sun2015libcint}%
  \BibitemOpen
  \bibfield  {author} {\bibinfo {author} {\bibfnamefont {Q.}~\bibnamefont
  {Sun}},\ }\bibfield  {title} {\bibinfo {title} {Libcint: An efficient general
  integral library for gaussian basis functions},\ }\href@noop {} {\bibfield
  {journal} {\bibinfo  {journal} {J. Comput. Chem.}\ }\textbf {\bibinfo
  {volume} {36}},\ \bibinfo {pages} {1664} (\bibinfo {year}
  {2015})}\BibitemShut {NoStop}%
\bibitem [{\citenamefont {Sun}\ \emph {et~al.}(2018)\citenamefont {Sun},
  \citenamefont {Berkelbach}, \citenamefont {Blunt}, \citenamefont {Booth},
  \citenamefont {Guo}, \citenamefont {Li}, \citenamefont {Liu}, \citenamefont
  {McClain}, \citenamefont {Sayfutyarova}, \citenamefont {Sharma} \emph
  {et~al.}}]{sun2018pyscf}%
  \BibitemOpen
  \bibfield  {author} {\bibinfo {author} {\bibfnamefont {Q.}~\bibnamefont
  {Sun}}, \bibinfo {author} {\bibfnamefont {T.~C.}\ \bibnamefont {Berkelbach}},
  \bibinfo {author} {\bibfnamefont {N.~S.}\ \bibnamefont {Blunt}}, \bibinfo
  {author} {\bibfnamefont {G.~H.}\ \bibnamefont {Booth}}, \bibinfo {author}
  {\bibfnamefont {S.}~\bibnamefont {Guo}}, \bibinfo {author} {\bibfnamefont
  {Z.}~\bibnamefont {Li}}, \bibinfo {author} {\bibfnamefont {J.}~\bibnamefont
  {Liu}}, \bibinfo {author} {\bibfnamefont {J.~D.}\ \bibnamefont {McClain}},
  \bibinfo {author} {\bibfnamefont {E.~R.}\ \bibnamefont {Sayfutyarova}},
  \bibinfo {author} {\bibfnamefont {S.}~\bibnamefont {Sharma}}, \emph
  {et~al.},\ }\bibfield  {title} {\bibinfo {title} {Pyscf: the python-based
  simulations of chemistry framework},\ }\href@noop {} {\bibfield  {journal}
  {\bibinfo  {journal} {Wiley Interdiscip. Rev.: Comput. Mol. Sci.}\ }\textbf
  {\bibinfo {volume} {8}},\ \bibinfo {pages} {e1340} (\bibinfo {year}
  {2018})}\BibitemShut {NoStop}%
\bibitem [{\citenamefont {Sun}\ \emph {et~al.}(2020)\citenamefont {Sun},
  \citenamefont {Zhang}, \citenamefont {Banerjee}, \citenamefont {Bao},
  \citenamefont {Barbry}, \citenamefont {Blunt}, \citenamefont {Bogdanov},
  \citenamefont {Booth}, \citenamefont {Chen}, \citenamefont {Cui} \emph
  {et~al.}}]{sun2020recent}%
  \BibitemOpen
  \bibfield  {author} {\bibinfo {author} {\bibfnamefont {Q.}~\bibnamefont
  {Sun}}, \bibinfo {author} {\bibfnamefont {X.}~\bibnamefont {Zhang}}, \bibinfo
  {author} {\bibfnamefont {S.}~\bibnamefont {Banerjee}}, \bibinfo {author}
  {\bibfnamefont {P.}~\bibnamefont {Bao}}, \bibinfo {author} {\bibfnamefont
  {M.}~\bibnamefont {Barbry}}, \bibinfo {author} {\bibfnamefont {N.~S.}\
  \bibnamefont {Blunt}}, \bibinfo {author} {\bibfnamefont {N.~A.}\ \bibnamefont
  {Bogdanov}}, \bibinfo {author} {\bibfnamefont {G.~H.}\ \bibnamefont {Booth}},
  \bibinfo {author} {\bibfnamefont {J.}~\bibnamefont {Chen}}, \bibinfo {author}
  {\bibfnamefont {Z.-H.}\ \bibnamefont {Cui}}, \emph {et~al.},\ }\bibfield
  {title} {\bibinfo {title} {Recent developments in the pyscf program
  package},\ }\href@noop {} {\bibfield  {journal} {\bibinfo  {journal} {J.
  Chem. Phys.}\ }\textbf {\bibinfo {volume} {153}},\ \bibinfo {pages} {024109}
  (\bibinfo {year} {2020})}\BibitemShut {NoStop}%
\bibitem [{\citenamefont {Lebedev}(1975)}]{Lebedev1975}%
  \BibitemOpen
  \bibfield  {author} {\bibinfo {author} {\bibfnamefont {V.~I.}\ \bibnamefont
  {Lebedev}},\ }\bibfield  {title} {\bibinfo {title} {{Values of the nodes and
  weights of ninth to seventeenth order gauss-markov quadrature formulae
  invariant under the octahedron group with inversion}},\ }\href
  {https://doi.org/10.1016/0041-5553(75)90133-0} {\bibfield  {journal}
  {\bibinfo  {journal} {USSR Computational Mathematics and Mathematical
  Physics}\ }\textbf {\bibinfo {volume} {15}},\ \bibinfo {pages} {44} (\bibinfo
  {year} {1975})}\BibitemShut {NoStop}%
\bibitem [{\citenamefont {Lebedev}\ and\ \citenamefont
  {Laikov}(1999)}]{lebedev1999quadrature}%
  \BibitemOpen
  \bibfield  {author} {\bibinfo {author} {\bibfnamefont {V.~I.}\ \bibnamefont
  {Lebedev}}\ and\ \bibinfo {author} {\bibfnamefont {D.}~\bibnamefont
  {Laikov}},\ }\bibfield  {title} {\bibinfo {title} {A quadrature formula for
  the sphere of the 131st algebraic order of accuracy},\ }in\ \href@noop {}
  {\emph {\bibinfo {booktitle} {Doklady Mathematics}}},\ Vol.~\bibinfo {volume}
  {59}\ (\bibinfo {year} {1999})\ pp.\ \bibinfo {pages} {477--481}\BibitemShut
  {NoStop}%
\bibitem [{\citenamefont {Golub}\ and\ \citenamefont
  {Welsch}(1969)}]{golub1969calculation}%
  \BibitemOpen
  \bibfield  {author} {\bibinfo {author} {\bibfnamefont {G.~H.}\ \bibnamefont
  {Golub}}\ and\ \bibinfo {author} {\bibfnamefont {J.~H.}\ \bibnamefont
  {Welsch}},\ }\bibfield  {title} {\bibinfo {title} {Calculation of gauss
  quadrature rules},\ }\href@noop {} {\bibfield  {journal} {\bibinfo  {journal}
  {Math. Comput.}\ }\textbf {\bibinfo {volume} {23}},\ \bibinfo {pages} {221}
  (\bibinfo {year} {1969})}\BibitemShut {NoStop}%
\bibitem [{\citenamefont {Gonz{\'a}lez}(2010)}]{gonzalez2010measurement}%
  \BibitemOpen
  \bibfield  {author} {\bibinfo {author} {\bibfnamefont {{\'A}.}~\bibnamefont
  {Gonz{\'a}lez}},\ }\bibfield  {title} {\bibinfo {title} {Measurement of areas
  on a sphere using fibonacci and latitude--longitude lattices},\ }\href@noop
  {} {\bibfield  {journal} {\bibinfo  {journal} {Math. Geosci.}\ }\textbf
  {\bibinfo {volume} {42}},\ \bibinfo {pages} {49} (\bibinfo {year}
  {2010})}\BibitemShut {NoStop}%
\bibitem [{\citenamefont {Balabanov}\ and\ \citenamefont
  {Peterson}(2005)}]{balabanov2005systematically}%
  \BibitemOpen
  \bibfield  {author} {\bibinfo {author} {\bibfnamefont {N.~B.}\ \bibnamefont
  {Balabanov}}\ and\ \bibinfo {author} {\bibfnamefont {K.~A.}\ \bibnamefont
  {Peterson}},\ }\bibfield  {title} {\bibinfo {title} {Systematically
  convergent basis sets for transition metals. i. all-electron correlation
  consistent basis sets for the 3 d elements sc--zn},\ }\href@noop {}
  {\bibfield  {journal} {\bibinfo  {journal} {J. Chem. Phys.}\ }\textbf
  {\bibinfo {volume} {123}},\ \bibinfo {pages} {064107} (\bibinfo {year}
  {2005})}\BibitemShut {NoStop}%
\bibitem [{\citenamefont {Lemberg}\ and\ \citenamefont
  {Stillinger}(1975)}]{lemberg1975central}%
  \BibitemOpen
  \bibfield  {author} {\bibinfo {author} {\bibfnamefont {H.~L.}\ \bibnamefont
  {Lemberg}}\ and\ \bibinfo {author} {\bibfnamefont {F.~H.}\ \bibnamefont
  {Stillinger}},\ }\bibfield  {title} {\bibinfo {title} {Central-force model
  for liquid water},\ }\href@noop {} {\bibfield  {journal} {\bibinfo  {journal}
  {J. Chem. Phys}\ }\textbf {\bibinfo {volume} {62}},\ \bibinfo {pages} {1677}
  (\bibinfo {year} {1975})}\BibitemShut {NoStop}%
\bibitem [{\citenamefont {Perdew}\ \emph {et~al.}(1996)\citenamefont {Perdew},
  \citenamefont {Burke},\ and\ \citenamefont
  {Ernzerhof}}]{perdew1996generalized}%
  \BibitemOpen
  \bibfield  {author} {\bibinfo {author} {\bibfnamefont {J.~P.}\ \bibnamefont
  {Perdew}}, \bibinfo {author} {\bibfnamefont {K.}~\bibnamefont {Burke}},\ and\
  \bibinfo {author} {\bibfnamefont {M.}~\bibnamefont {Ernzerhof}},\ }\bibfield
  {title} {\bibinfo {title} {Generalized gradient approximation made simple},\
  }\href@noop {} {\bibfield  {journal} {\bibinfo  {journal} {Phys. Rev. Lett.}\
  }\textbf {\bibinfo {volume} {77}},\ \bibinfo {pages} {3865} (\bibinfo {year}
  {1996})}\BibitemShut {NoStop}%
\bibitem [{\citenamefont {Tao}\ \emph {et~al.}(2003)\citenamefont {Tao},
  \citenamefont {Perdew}, \citenamefont {Staroverov},\ and\ \citenamefont
  {Scuseria}}]{tao2003climbing}%
  \BibitemOpen
  \bibfield  {author} {\bibinfo {author} {\bibfnamefont {J.}~\bibnamefont
  {Tao}}, \bibinfo {author} {\bibfnamefont {J.~P.}\ \bibnamefont {Perdew}},
  \bibinfo {author} {\bibfnamefont {V.~N.}\ \bibnamefont {Staroverov}},\ and\
  \bibinfo {author} {\bibfnamefont {G.~E.}\ \bibnamefont {Scuseria}},\
  }\bibfield  {title} {\bibinfo {title} {Climbing the density functional
  ladder: Nonempirical meta--generalized gradient approximation designed for
  molecules and solids},\ }\href@noop {} {\bibfield  {journal} {\bibinfo
  {journal} {Phys. Rev. Lett.}\ }\textbf {\bibinfo {volume} {91}},\ \bibinfo
  {pages} {146401} (\bibinfo {year} {2003})}\BibitemShut {NoStop}%
\bibitem [{\citenamefont {Stephens}\ \emph {et~al.}(1994)\citenamefont
  {Stephens}, \citenamefont {Devlin}, \citenamefont {Chabalowski},\ and\
  \citenamefont {Frisch}}]{stephens1994ab}%
  \BibitemOpen
  \bibfield  {author} {\bibinfo {author} {\bibfnamefont {P.~J.}\ \bibnamefont
  {Stephens}}, \bibinfo {author} {\bibfnamefont {F.~J.}\ \bibnamefont
  {Devlin}}, \bibinfo {author} {\bibfnamefont {C.~F.}\ \bibnamefont
  {Chabalowski}},\ and\ \bibinfo {author} {\bibfnamefont {M.~J.}\ \bibnamefont
  {Frisch}},\ }\bibfield  {title} {\bibinfo {title} {Ab initio calculation of
  vibrational absorption and circular dichroism spectra using density
  functional force fields},\ }\href@noop {} {\bibfield  {journal} {\bibinfo
  {journal} {J. Phys. Chem.}\ }\textbf {\bibinfo {volume} {98}},\ \bibinfo
  {pages} {11623} (\bibinfo {year} {1994})}\BibitemShut {NoStop}%
\bibitem [{\citenamefont {Komorovsky}\ \emph {et~al.}(2015)\citenamefont
  {Komorovsky}, \citenamefont {Repisky}, \citenamefont {Malkin}, \citenamefont
  {Demissie},\ and\ \citenamefont {Ruud}}]{komorovsky2015four}%
  \BibitemOpen
  \bibfield  {author} {\bibinfo {author} {\bibfnamefont {S.}~\bibnamefont
  {Komorovsky}}, \bibinfo {author} {\bibfnamefont {M.}~\bibnamefont {Repisky}},
  \bibinfo {author} {\bibfnamefont {E.}~\bibnamefont {Malkin}}, \bibinfo
  {author} {\bibfnamefont {T.~B.}\ \bibnamefont {Demissie}},\ and\ \bibinfo
  {author} {\bibfnamefont {K.}~\bibnamefont {Ruud}},\ }\bibfield  {title}
  {\bibinfo {title} {Four-component relativistic density-functional theory
  calculations of nuclear spin--rotation constants: Relativistic effects in
  p-block hydrides},\ }\href@noop {} {\bibfield  {journal} {\bibinfo  {journal}
  {J. Chem. Theory Comput.}\ }\textbf {\bibinfo {volume} {11}},\ \bibinfo
  {pages} {3729} (\bibinfo {year} {2015})}\BibitemShut {NoStop}%
\bibitem [{\citenamefont {Pu}\ \emph {et~al.}(2022)\citenamefont {Pu},
  \citenamefont {Zhang}, \citenamefont {Jiang},\ and\ \citenamefont
  {Xiao}}]{pu2022approach}%
  \BibitemOpen
  \bibfield  {author} {\bibinfo {author} {\bibfnamefont {Z.}~\bibnamefont
  {Pu}}, \bibinfo {author} {\bibfnamefont {N.}~\bibnamefont {Zhang}}, \bibinfo
  {author} {\bibfnamefont {H.}~\bibnamefont {Jiang}},\ and\ \bibinfo {author}
  {\bibfnamefont {Y.}~\bibnamefont {Xiao}},\ }\bibfield  {title} {\bibinfo
  {title} {Approach for noncollinear gga kernels in closed-shell systems},\
  }\href@noop {} {\bibfield  {journal} {\bibinfo  {journal} {Phys. Rev. B}\
  }\textbf {\bibinfo {volume} {105}},\ \bibinfo {pages} {035114} (\bibinfo
  {year} {2022})}\BibitemShut {NoStop}%
\bibitem [{\citenamefont {Wedig}\ \emph {et~al.}(1986)\citenamefont {Wedig},
  \citenamefont {Dolg}, \citenamefont {Stoll},\ and\ \citenamefont
  {Preuss}}]{wedig1986energy}%
  \BibitemOpen
  \bibfield  {author} {\bibinfo {author} {\bibfnamefont {U.}~\bibnamefont
  {Wedig}}, \bibinfo {author} {\bibfnamefont {M.}~\bibnamefont {Dolg}},
  \bibinfo {author} {\bibfnamefont {H.}~\bibnamefont {Stoll}},\ and\ \bibinfo
  {author} {\bibfnamefont {H.}~\bibnamefont {Preuss}},\ }\bibfield  {title}
  {\bibinfo {title} {Energy-adjusted pseudopotentials for transition-metal
  elements},\ }in\ \href@noop {} {\emph {\bibinfo {booktitle} {Quantum
  chemistry: the challenge of transition metals and coordination chemistry}}}\
  (\bibinfo  {publisher} {Springer},\ \bibinfo {year} {1986})\ pp.\ \bibinfo
  {pages} {79--89}\BibitemShut {NoStop}%
\bibitem [{\citenamefont {Adamo}\ and\ \citenamefont
  {Barone}(1999)}]{adamo1999toward}%
  \BibitemOpen
  \bibfield  {author} {\bibinfo {author} {\bibfnamefont {C.}~\bibnamefont
  {Adamo}}\ and\ \bibinfo {author} {\bibfnamefont {V.}~\bibnamefont {Barone}},\
  }\bibfield  {title} {\bibinfo {title} {Toward reliable density functional
  methods without adjustable parameters: The pbe0 model},\ }\href@noop {}
  {\bibfield  {journal} {\bibinfo  {journal} {J. Chem. Phys.}\ }\textbf
  {\bibinfo {volume} {110}},\ \bibinfo {pages} {6158} (\bibinfo {year}
  {1999})}\BibitemShut {NoStop}%
\bibitem [{\citenamefont {Ramachandran}\ and\ \citenamefont
  {Varoquaux}(2011)}]{ramachandran2011mayavi}%
  \BibitemOpen
  \bibfield  {author} {\bibinfo {author} {\bibfnamefont {P.}~\bibnamefont
  {Ramachandran}}\ and\ \bibinfo {author} {\bibfnamefont {G.}~\bibnamefont
  {Varoquaux}},\ }\bibfield  {title} {\bibinfo {title} {{Mayavi: 3D
  Visualization of Scientific Data}},\ }\href@noop {} {\bibfield  {journal}
  {\bibinfo  {journal} {Comput. Sci. Eng.}\ }\textbf {\bibinfo {volume} {13}},\
  \bibinfo {pages} {40} (\bibinfo {year} {2011})}\BibitemShut {NoStop}%
\bibitem [{\citenamefont {Hunter}(2007)}]{Hunter:2007}%
  \BibitemOpen
  \bibfield  {author} {\bibinfo {author} {\bibfnamefont {J.~D.}\ \bibnamefont
  {Hunter}},\ }\bibfield  {title} {\bibinfo {title} {Matplotlib: A 2d graphics
  environment},\ }\href {https://doi.org/10.1109/MCSE.2007.55} {\bibfield
  {journal} {\bibinfo  {journal} {Comput. Sci. Eng.}\ }\textbf {\bibinfo
  {volume} {9}},\ \bibinfo {pages} {90} (\bibinfo {year} {2007})}\BibitemShut
  {NoStop}%
\bibitem [{\citenamefont {Owen}\ and\ \citenamefont
  {Yates}(1933)}]{owen1933xli}%
  \BibitemOpen
  \bibfield  {author} {\bibinfo {author} {\bibfnamefont {E.}~\bibnamefont
  {Owen}}\ and\ \bibinfo {author} {\bibfnamefont {E.}~\bibnamefont {Yates}},\
  }\bibfield  {title} {\bibinfo {title} {Xli. precision measurements of crystal
  parameters},\ }\href@noop {} {\bibfield  {journal} {\bibinfo  {journal}
  {Philos. Mag.}\ }\textbf {\bibinfo {volume} {15}},\ \bibinfo {pages} {472}
  (\bibinfo {year} {1933})}\BibitemShut {NoStop}%
\bibitem [{\citenamefont {Hirshfeld}(1977)}]{hirshfeld1977bonded}%
  \BibitemOpen
  \bibfield  {author} {\bibinfo {author} {\bibfnamefont {F.~L.}\ \bibnamefont
  {Hirshfeld}},\ }\bibfield  {title} {\bibinfo {title} {Bonded-atom fragments
  for describing molecular charge densities},\ }\href@noop {} {\bibfield
  {journal} {\bibinfo  {journal} {Theor. Chem. Acc.}\ }\textbf {\bibinfo
  {volume} {44}},\ \bibinfo {pages} {129} (\bibinfo {year} {1977})}\BibitemShut
  {NoStop}%
\bibitem [{\citenamefont {Zhong}\ \emph {et~al.}(2020)\citenamefont {Zhong},
  \citenamefont {Chen}, \citenamefont {OuYang}, \citenamefont {Yang},
  \citenamefont {Zhang}, \citenamefont {Gao}, \citenamefont {Schulze},
  \citenamefont {Wernsdorfer},\ and\ \citenamefont
  {Dong}}]{zhong2020unprecedented}%
  \BibitemOpen
  \bibfield  {author} {\bibinfo {author} {\bibfnamefont {L.}~\bibnamefont
  {Zhong}}, \bibinfo {author} {\bibfnamefont {W.-B.}\ \bibnamefont {Chen}},
  \bibinfo {author} {\bibfnamefont {Z.-J.}\ \bibnamefont {OuYang}}, \bibinfo
  {author} {\bibfnamefont {M.}~\bibnamefont {Yang}}, \bibinfo {author}
  {\bibfnamefont {Y.-Q.}\ \bibnamefont {Zhang}}, \bibinfo {author}
  {\bibfnamefont {S.}~\bibnamefont {Gao}}, \bibinfo {author} {\bibfnamefont
  {M.}~\bibnamefont {Schulze}}, \bibinfo {author} {\bibfnamefont
  {W.}~\bibnamefont {Wernsdorfer}},\ and\ \bibinfo {author} {\bibfnamefont
  {W.}~\bibnamefont {Dong}},\ }\bibfield  {title} {\bibinfo {title}
  {Unprecedented one-dimensional chain and two-dimensional network dysprosium
  (iii) single-molecule toroics with white-light emission},\ }\href@noop {}
  {\bibfield  {journal} {\bibinfo  {journal} {Chem. Commun.}\ }\textbf
  {\bibinfo {volume} {56}},\ \bibinfo {pages} {2590} (\bibinfo {year}
  {2020})}\BibitemShut {NoStop}%
\bibitem [{\citenamefont {Lu}\ and\ \citenamefont {Peterson}(2016)}]{lu2016a}%
  \BibitemOpen
  \bibfield  {author} {\bibinfo {author} {\bibfnamefont {Q.}~\bibnamefont
  {Lu}}\ and\ \bibinfo {author} {\bibfnamefont {K.~A.}\ \bibnamefont
  {Peterson}},\ }\bibfield  {title} {\bibinfo {title} {Correlation consistent
  basis sets for lanthanides: The atoms la-lu},\ }\href
  {https://doi.org/10.1063/1.4959280} {\bibfield  {journal} {\bibinfo
  {journal} {J. Chem. Phys.}\ }\textbf {\bibinfo {volume} {145}},\ \bibinfo
  {pages} {054111} (\bibinfo {year} {2016})}\BibitemShut {NoStop}%
\end{thebibliography}%

\end{document}